# SNAKE AND SNAKE ROBOT LOCOMOTION IN COMPLEX, 3-D TERRAIN

by
Qiyuan Fu

A dissertation submitted to Johns Hopkins University in conformity with the requirements for the degree of Doctor of Philosophy

Baltimore, Maryland
February 2023

# Abstract

Snakes are able to traverse almost all types of environments by bending their elongate bodies in three dimensions to interact with the terrain. Similarly, a snake robot is a promising platform to perform critical tasks in various environments. Understanding how 3-D body bending effectively interacts with the terrain for propulsion and stability can not only inform how snakes move through natural environments, but also inspire snake robots to achieve similar performance to assist humans.

How snakes and snake robots move on flat surfaces has been understood relatively well in previous studies. However, such ideal terrain is rare in natural environments and little was understood about how to generate propulsion and maintain stability when large height variations occur, except for some qualitative descriptions of arboreal snake locomotion and a few robots using geometric planning. To bridge this knowledge gap, in this dissertation research we integrated animal experiments and robotic studies in three representative environments: a large smooth step, an uneven arena of blocks of large height variation, and large bumps.

We discovered that vertical body bending induces stability challenges but can generate large propulsion. When traversing a large smooth step, a snake robot is challenged by roll instability that increases with larger vertical body bending because of a higher center of mass. The instability can be reduced by body compliance that statistically increases surface contact. Despite the stability challenge, vertical body bending can potentially allow snakes to push against terrain for propulsion similar to lateral body bending, as demonstrated by corn snakes traversing an uneven arena. This ability to generate large propulsion was confirmed on a robot if body-terrain contact is well maintained. Contact feedback control can help the strategy accommodate perturbations such as novel terrain geometry or excessive external forces by helping the body regain lost contact. Our findings provide insights into how snakes and snake robots can use vertical body bending for efficient and versatile traversal of the three-dimensional world while maintaining stability.

**Primary Reader and Advisor:** Chen Li

**Secondary Readers:** Noah J. Cowan, Henry C. Astley

# Acknowledgments

I feel extremely lucky to have worked with my advisor, Prof. Chen Li, during my Ph.D. study. I cannot thank him enough for training me on rigorous scientific research, guidance on becoming an independent person, and help during the pandemic and my career development. His optimism and dedication to research and mentoring have made him a role model for me in my journey toward independent research in the future.

I am also deeply grateful to Prof. Henry Astley at the University of Akron for his vital suggestions on animal care and on research, which significantly contribute to the formation of my research direction and the design of some studies, and for his support in my career development and dissertation reading. I would like to thank Prof. Noah Cowan for the two fundamental courses I took from him, his suggestions on my research, and his support in my career development and dissertation reading. I would also like to thank my other collaborators, Dr. Jin Seob Kim and Prof. Gregory Chirikjian, for their help with the development of several methods that laid a solid foundation for some of the studies in this dissertation.

It is my pleasure to have worked with the members of the Terradynamics Lab and I am grateful for their help both in research and in life: Dr. Thomas Mitchel, Dr. Sean Gart, Qihan Xuan, Yifeng Zhang, Divya Ramesh, Yaqing Wang, Dr. Ratan Othayoth, Dr. Yuanfeng Han, Nansong Yi, Dr. Huidong Gao, Changxing Yan, Dr. Hongtao Wu, Zhiyi Ren, Xiangyu Peng, Yulong Wang, Kaiwen Wang, Daniel Deng, Eric Lara, Zihao Rao, Gargi Sadalgekar, Eugene Lin, Kangxin Wang, and many others. Special thanks to Tommy and Sean, who led the initial exploration of snake locomotion in our lab and passed on their expertise to me. Special thanks to Qihan and Yifeng for their discussions with me and their work on modeling and simulation that provided many insights into the physics of snake locomotion. Special thanks to Divya and Yaqing for help with animal care, efforts in developing a sensorized snake robot, and enlightening discussions in the lab. Special thanks to Ratan and Yuanfeng for the development of many fundamental experimental setups and procedures in the lab and helpful discussions. Special thanks to Nansong, Huidong, Changxing, Hongtao, Zhiyi, Xiangyu, Kaiwen, and Daniel for their help with robot design, construction of experimental platforms, and preliminary experiments.

I would like to thank Prof. Noah Cowan, Prof. Gregory Chirikjian, Prof. Louis Whitcomb, Prof. Gretar Tryggvason, Prof. Pablo Iglesias, and Prof. Avanti Athreya for examining me during my qualification exams and support of my Ph.D. candidacy.

During my Ph.D. studies, I also received many helpful suggestions and comments from my colleagues. I would especially like to thank Dr. Derek Jurestovsky at the University of Akron for sharing their snake experimental data and Laura Paez at the École Polytechnique Fédérale de Lausanne for suggestions on servo motor control. I would like to especially thank Prof. Daniel Goldman, Prof. David Hu, Prof. Jake Socha, Dr. Baxi Chong, Tianyu Wang, Dr. Jessica Tingle, and Dr. Perrin Schiebel for helpful discussions that provided many inspirations to improve my work. I had the privilege of attending conferences to present our work and receive feedback from the scientific communities, especially the Society for Integrative and Comparative Biology and the American Physical Society.

I would like to especially thank Prof. Guolei Wang, Prof. Auke Ijspeert and his group members, Prof. Daniel Goldman and his group members, Prof. Kohei Nakajima, Prof. Shai Revzen, Prof. Edward Ionides, Prof. Matt McHenry, Prof. Eva Kanso, Prof. Andrew Spence, Prof. Jianyu Chen, Dr. Guanya Shi, Prof. Jiaoyang Li, Dr. Zhiyuan Li, and Xiangyu Peng for their help in my post-doc applications and suggestions on my career development. The application process provided me with not only a post-doc position to explore new directions, but also an opportunity to synthesize my past contributions and shape my scientific vision, training on fellowship application, and motivation to improve myself toward an independent researcher.

I would like to thank the administrative staff at the Department of Mechanical Engineering and the Laboratory for Computational Sensing and Robotics for their support of my academic program and laboratory research that saved me much time and effort to focus on coursework and research, especially Mike Bernard, Kevin Adams, John Soos, Patrick Caufield, Lorrie Dodd, Jordan Card, Ashley Moriarty, and Alison Morrow. Thanks to Rohit Agrawal at the Johns Hopkins University Counseling center who helped me out of my depression during the pandemic. Thanks to the funding resources that enabled my Ph.D. studies: the Burroughs Wellcome Fund, the Arnold and Mabel Beckman Foundation, the Whiting School of Engineering, and the Johns Hopkins University.

I would like to thank my awesome friends who provided emotional support that kept me optimistic along the way and helped me get through my struggles. Special thanks to my roommates Decheng Hou and Zhenhui Liu for creating harmonic environments in my apartments, where I can relax after my course and lab work. Special thanks to Shufan Wang, Cheng Qian, and Ziyan Wang for the casual chats and entertainment that reduced my stress. Special thanks to my cats Nana and Banana for their companionship and emotional support.

Finally, I would like to thank my family, especially my parents Hailin Fu and Xiaoli Guo, for their unconditional love and support. It would be impossible for me to come to this point as a Ph.D. without their unremitting efforts that supported me both financially and emotionally.

# Table of contents

# List of Tables

# List of Figures

# Chapter 1

# Introduction

## 1.1 Motivation and overview

Snakes can traverse a wide diversity of complex, 3-D environments efficiently and versatilely using various body bending patterns (**Figure 1**). Thanks to an elongate body, a snake can form multiple contact points with the environment. By modulating body bending, it can coordinate the contact forces on these points for propulsion, braking, or maneuvering. The large number of contact points can provide a large base of support for stability. The small cross-section allows them to enter the most confined spaces such as rock crevices. Similarly, a snake robot is a promising platform to perform critical tasks in complex environments, such as search and rescue in earthquake rubbles, examination of factory plants, monitoring environments, and exploration of subterranean or extraterrestrial environments (**Figure 2**). Understanding how a snake or snake robot bends a limbless body to move through complex environments can not only inform understanding of the vastly abundant limbless animals in nature (Astley, 2020a) but also inspire engineering applications that can assist humans in these critical tasks.

How snakes move on flat surfaces with or without vertical extrusions has been well understood and allowed snake robots to achieve similar performance in such environments (Sanfilippo et al., 2017; Walker et al., 2016). However, such terrain rarely exists in natural environments, and little is understood about how snakes adapt to height variations and maintain propulsion and stability when using 3-D body bending. This lack of understanding results in a significant performance gap between snake robots and snakes in efficiency, stability, and versatility in complex, 3-D environments. Most of the existing robots traversing 3-D terrain either experience large slipping and instability or heavily depend on human operators to plan the bending patterns dedicated to limited environments such as steps and pipes.

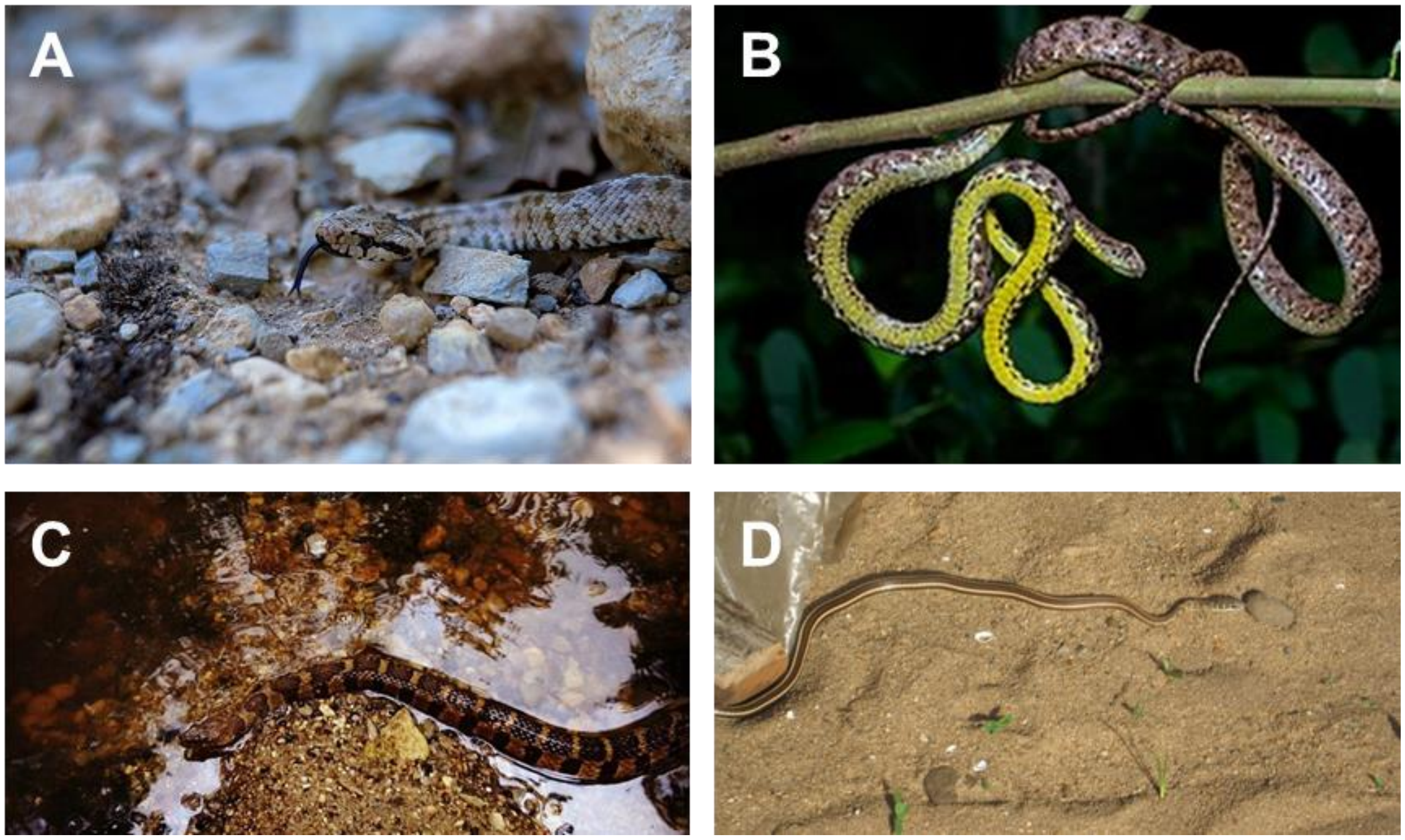

**Figure 1. Snakes traversing various complex environments. (A)** Over a pile of rocks. **(B)** On tree branches. **(C)** Transition from water to land. **(D)** On loose sand. Images courtesy of (A) Mostafameraji, (B) Rushenb, (C) Tom VanderVelde, and (D) Phillip1949 from Wikimedia Commons.

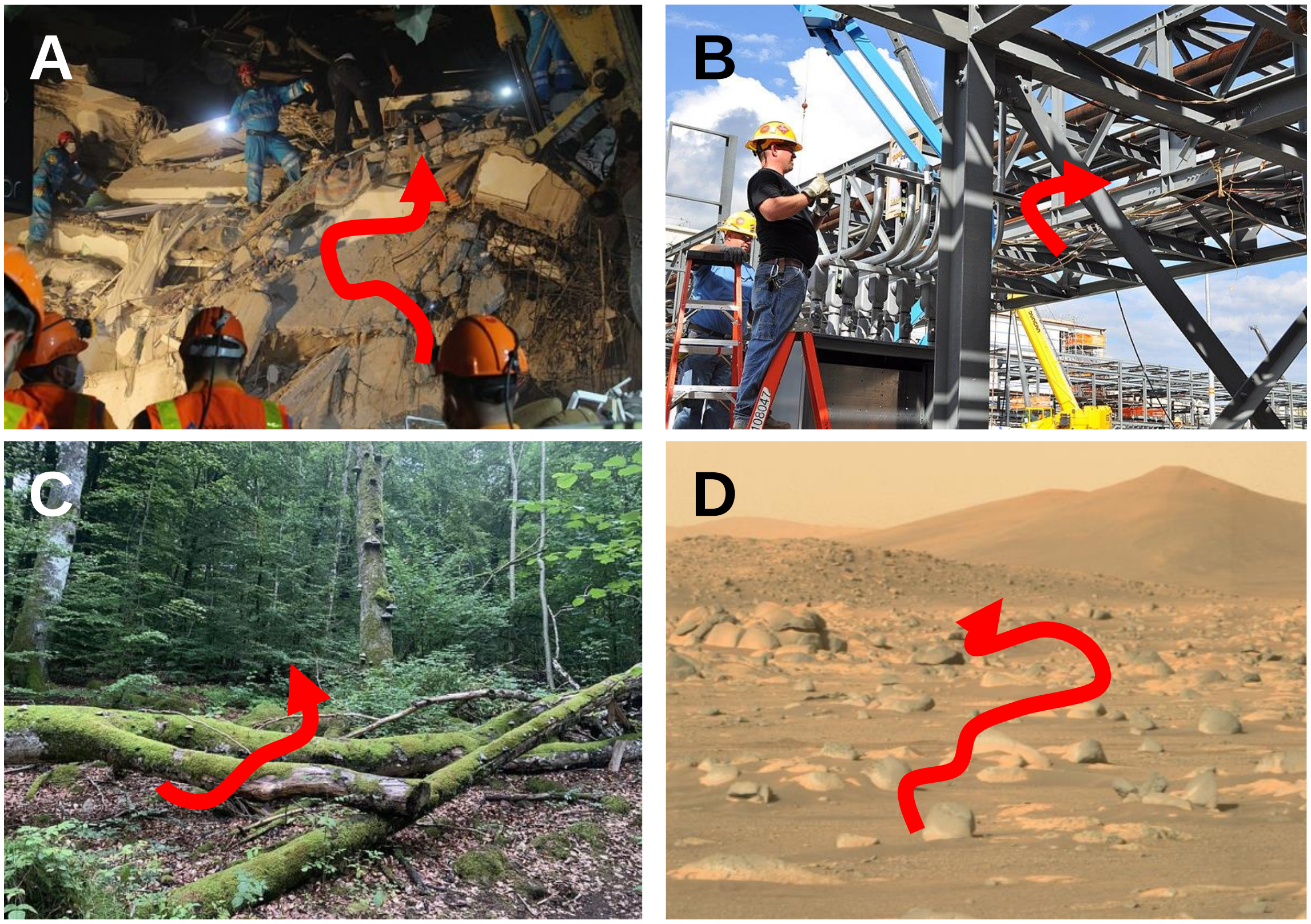

**Figure 2. Potential applications of snake robots. (A)** Search and rescue. **(B)** Industrial inspection. **(C)** Environmental monitoring. **(D)** Extra-terrestrial exploration. Image courtesy of (A) Oğulcan Bakiler, (B) PEO ACWA, (C) Vitaly Repin, and (D) NASA from Wikimedia Commons.

To bridge this knowledge gap, in this dissertation we studied how 3-D body bending of a snake or a snake robot relates to body-terrain contact and how changes in contact affect performance such as stability and propulsion generation. We integrated biological and robotic experiments in three representative 3-D environments: (1) a large, smooth step, (2) an uneven terrain with blocks of random heights, and (3) large bumps. The studies revealed principles of using vertical body bending for propulsion when traversing 3-D terrain and how body compliance and contact feedback control can further improve the stability and robustness of this strategy by improving contact. The principles of how they affect performance by modulating physical interaction not only advance our understanding of how snakes traverse 3-D environments versatilely and efficiently, but also provide inspiration to design snake robots for better traversal performance in similar environments. Below, we will first elaborate on relevant background and knowledge gaps, then summarize the specific questions we sought to answer and how we answered them.

## 1.2 Background

### 1.2.1 Snake and snake robot locomotion on flat surfaces

#### 1.2.1.1 Locomotion on flat surfaces without vertical extrusions

On a flat surface without large vertical extrusions, snakes can use various periodic movement modes (gaits) to move forward, which vary between species and environmental conditions such as slope and substrate strength. Traditionally, four major gaits are classified: lateral undulation, sidewinding, rectilinear, and concertina (**Figure 3**) (Jayne, 2020).

When using lateral undulation (Gray, 1946; Gray and Lissmann, 1950; Hu et al., 2009; Jayne, 1986; Walton et al., 1990), a snake propagates a wave of lateral bending down the body (**Figure 3**A). Propulsion is generated by the body pushing against the terrain. A snake can generate a larger contact force along the normal direction of a local body section on the ground than along the longitudinal (along the body axis) direction (hereafter referred to as force anisotropy). This anisotropy can result from anisotropic friction between the scales and the ground (Hu et al., 2009) or contact forces against terrain asperities such as depressed sand (Schiebel et al., 2020a), artificial turf (Gerald and Wass, 2019; Jayne and Bennett, 1989; Jayne and Bennett, 1990; Walton et al., 1990), and large vertical extrusions (see Chapter 1.2.1.2). The sum of contact forces distributed along the body is thus along the desired movement direction (Hu et al., 2009). If the anisotropy is insignificant, such as on a smooth glass plate, a snake slips sideways much when using lateral undulation (**Figure 3**A, bottom). To further improve the efficiency, a snake can lift its body sections that produce much larger drag than propulsion (sinus-lifting) (Hirose, 1993; Hu et al., 2009). While always accompanied by large slipping along the body axis, lateral undulation does not cost more energy than running by legged animals of similar sizes (Walton et al., 1990).

Different from when using lateral undulation to generate propulsion while the body always slips against contact points, a snake often forms static contact points with the terrain when using the other three modes. When using sidewinding (Gray, 1946; Jayne, 1986; Jayne, 1988; Marvi et al., 2014; Mosauer, 1932; Secor et al., 1992), a snake forms static contact with the terrain along several disconnected tracks oblique to the direction of movement (**Figure 3**B). The body sections in between these tracks are lifted and transferred between the tracks. When using concertina (Jayne, 2020), a snake bends a part of its body to form static contact with the terrain while moving the other parts forward and exchanges the role of each part alternatively (**Figure 3**C). When a snake uses rectilinear (Lissmann, 1950; Newman and Jayne, 2018), the skin deforms periodically such that the ventral scales oscillate longitudinally and form alternating contact points with the terrain (**Figure 3**D).

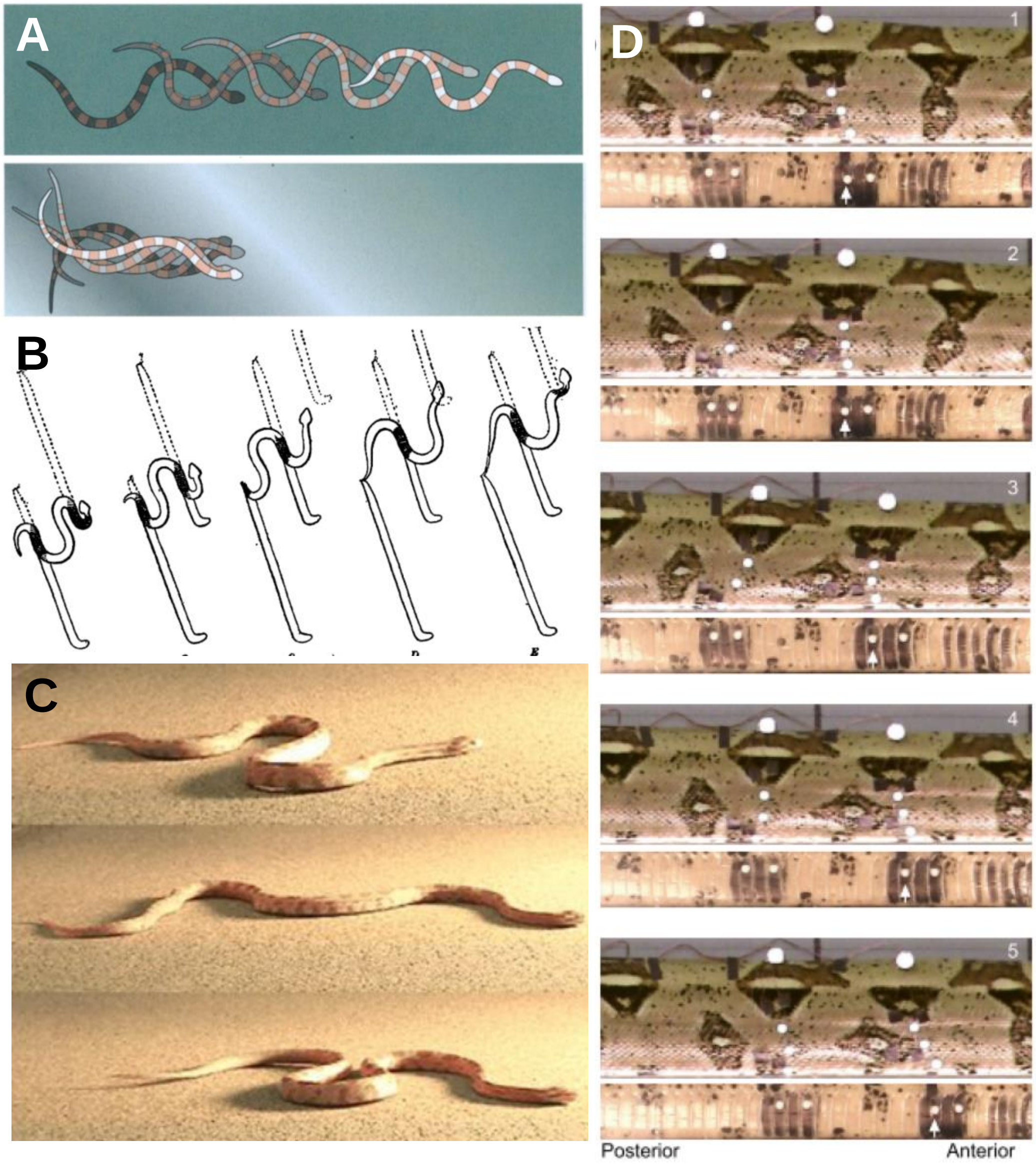

**Figure 3. Snakes moving on flat surfaces without vertical extrusions using four major gaits.** **(A)** Lateral undulation on rough (top) and smooth (below) surfaces. Reproduced from (Goldman and Hu, 2010). **(B)** Sidewinding. A snake forms static contact (black regions) with obliquely oriented, disconnected tracks. Reproduced from (Mosauer, 1932). **(C)** Concertina. Reproduced from (Marvi et al., 2011). **(D)** Rectilinear. Note how white markers on snake move with skin deformation. Reproduced from (Newman and Jayne, 2018).

Inspired by snakes, many snake robots can use these cyclic modes to move forward on flat surfaces without large vertical extrusions. Snake robots that use lateral undulation to provide propulsion in such environments need the force anisotropy similar to snakes. This is typically realized by adding passive wheels to reduce longitudinal resistance (**Figure 4**A) (Hirose, 1993; Klaassen and Paap, 1999; Liljebäck et al., 2012a; Mori and Hirose, 2001; Nakajima et al., 2018; Togawa et al., 2000) or by moving in deformable substrates such as sand (Maladen et al., 2011; Schiebel et al., 2020a). Rectilinear requires fine control of skin deformation, which is hard to realize on robots. Instead, some robots combine mechanisms with larger backward friction than forward friction, such as one-way wheels or tilted scales, and axial sliding to realize movements similar to a combination of rectilinear and concertina (**Figure 4**C) (Marvi et al., 2011; Tang et al., 2015; Yim et al., 2001). Sidewinding (**Figure 4**B) (Astley et al., 2015; Chong et al., 2021; Gong et al., 2016; Marvi et al., 2014) and concertina (**Figure 4**D) (Manzoor et al., 2019) can be realized by body bending only and do not need additional structures.

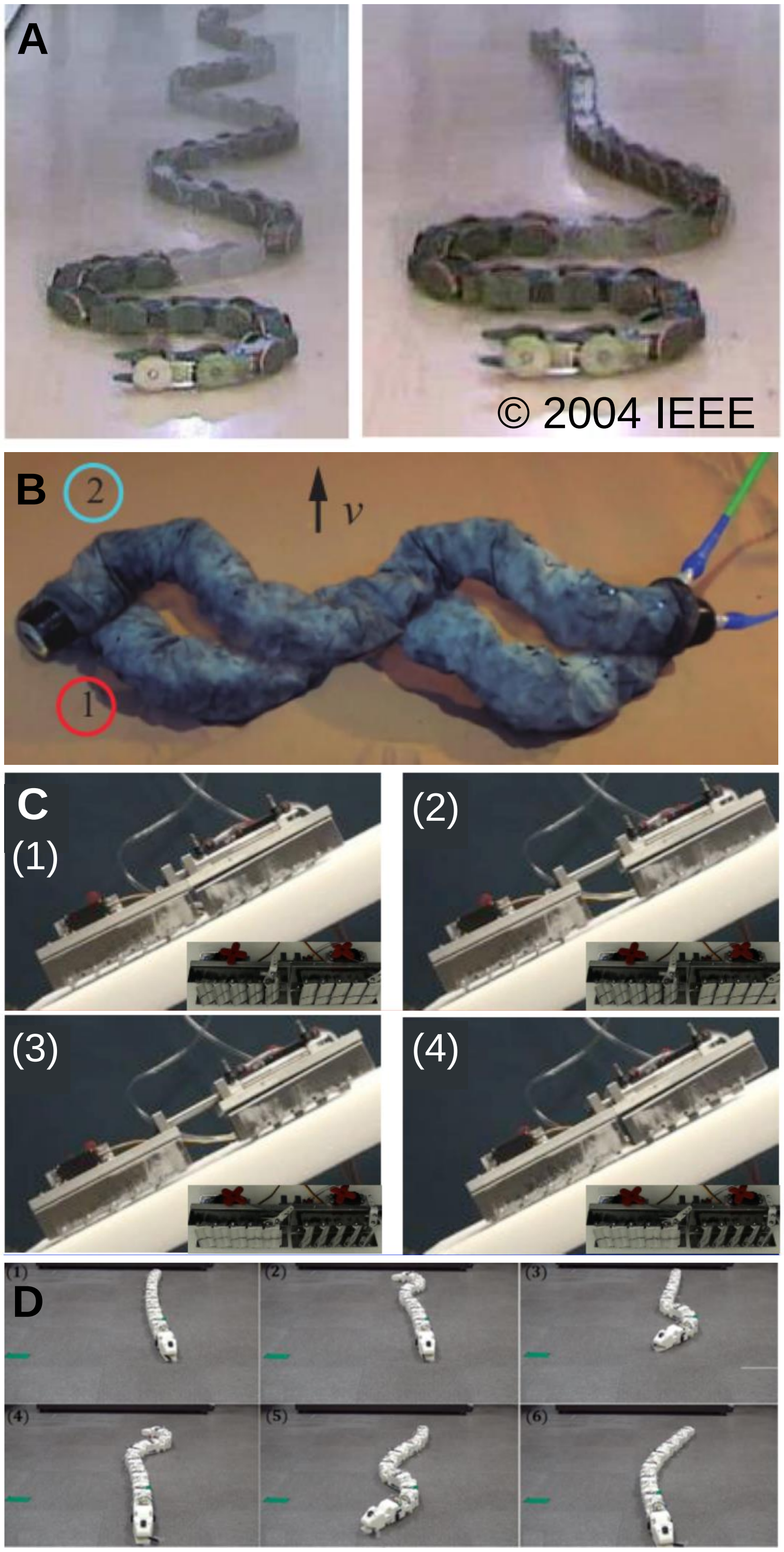

**Figure 4. Robot locomotion on flat surfaces using modes inspired by animals. (A)** Lateral undulation using variable wave shapes. Robot is equipped with passive wheels around body. **(B)**

Sidewinding[1]. **(C)** Concertina-inspired gait using axial sliding and controllable mechanical scales similar to a snake deforming its belly during rectilinear. **(D)** Concertina using body bending. Reproduced from (Hirose and Mori, 2004; Manzoor et al., 2019; Marvi et al., 2011; Marvi et al., 2014).

Aside from these bio-inspired gaits, some snake robots have also used artificial gaits to traverse a smooth flat surface (**Figure 5**). For example, by forming a rolling loop (**Figure 5**A), a snake robot can continuously move forward like a tread (Ohashi et al., 2010; Sastra et al., 2009). A snake robot can also roll forward using a crawler gait (**Figure 5**B) by forming static contact with the terrain (Takemori et al., 2018b), similar to sidewinding. By continuously rolling around its centerline while bending into an arc (**Figure 5**C), a robot can move laterally and obliquely (Hirose and Mori, 2004; Tesch et al., 2009). By vertically propagating a wave (**Figure 5**D), a snake robot can rely on the static contact points that travel down the body to move forward (Hirose and Mori, 2004; Koopaee, 2019; Yim et al., 2001). In addition, many robots use actuated wheels or treads (**Figure 5**E) to generate propulsion without the need for proper body shape control (Walker et al., 2016). However, the additional actuators can occupy a considerable amount of weight and internal space of the robot, which can reduce the ability to bend its body into complex shapes. In addition, they are more difficult to seal for applications in dusty or wet environments compared to robots that use body bending for locomotion.

[1] From H. Marvi et al., “Sidewinding with minimal slip: Snake and robot ascent of sandy slopes,” *Science*, vol. 346, no. 6206, pp. 224--229, Oct. 2014. Reprinted with permission from AAAS.

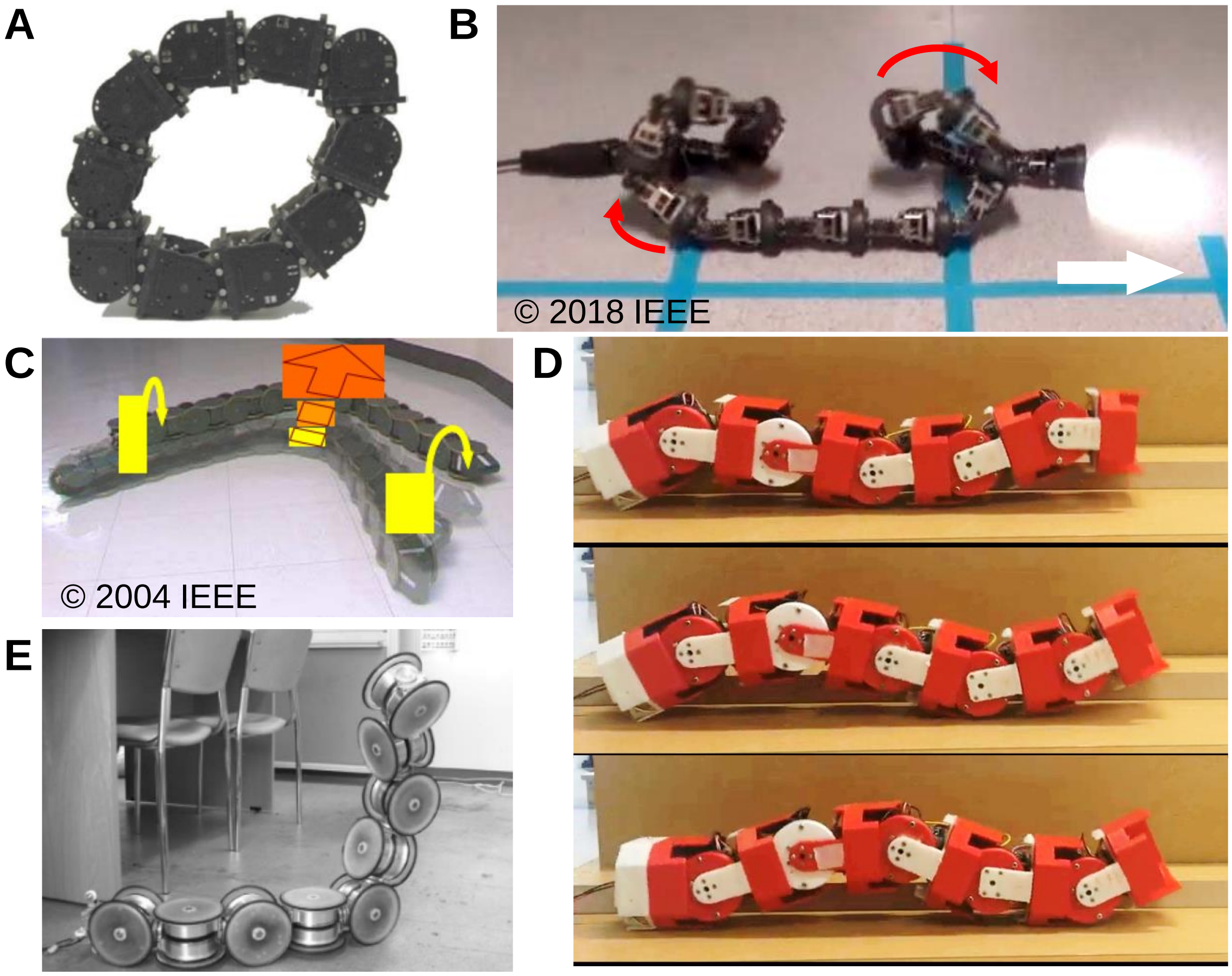

**Figure 5. Snake robots moving on flat surfaces using artificial modes. (A)** Loop rolling. **(B)** Crawler gait. **(C)** Lateral rolling. **(D)** Vertical undulation. **(E)** Using active wheels. Reproduced from (Hirose and Mori, 2004; Koopaee, 2019; Sastra et al., 2009; Takemori et al., 2018b; Yamada and Hirose, 2006).

#### 1.2.1.2 Locomotion on flat surfaces with vertical extrusions

When large vertical extrusions such as plants or rocks are available on a flat surface, snakes can utilize them during locomotion by body bending. The most well-understood strategy used by snakes in such environments is using lateral undulation to push against lateral contact points (**Figure 6**A) (Gasc et al., 1989; Gray, 1946; Gray and Lissmann, 1950; Jayne, 1986; Kano et al., 2012; Moon and Gans, 1998; Schiebel et al., 2020b). To allow a substantially large component of the contact force on a contact point to convert to propulsion, these contact surfaces need to be properly oriented facing the movement direction. Another common strategy to utilize these vertical extrusions is by bracing against opposite-facing structures to form static contacts and moving the other part of the body forward (**Figure 6**H), which is also classified as concertina (Gray, 1946; Jayne, 1986; Jayne, 1988; Jayne and Davis, 1991; Marvi and Hu, 2012; Walton et al., 1990).

To fully exploit these contact points, generalist snakes can adjust their lateral body bending to maintain pushing against vertical structures at different locations (**Figure 6**A) (Gray, 1946; Gray and Lissmann, 1950; Jayne, 1988; Jayne and Davis, 1991; Kano et al., 2012; Schiebel et al., 2020b), presumably using sensory information such as vision (Gans, 1975; Gripshover and Jayne, 2020) or mechanosensation (Crowe, 1992; Von Düring and Miller, 1979). Some snakes, especially specialists that reside in habitats lacking variation of terrain asperities in nature, tend to maintain their bending patterns similar to that used on smooth surfaces. However, their traversal performance can still be affected by the collision between the compliant body and the vertical extrusions (Rieser et al., 2019; Schiebel et al., 2020c).

The exploitation of lateral contact points for propulsion has inspired many snake robots to use similar lateral undulation or concertina strategies to traverse similar environments. Some robots can use open-loop control with passive mechanical structures to accommodate the uncertainties of push point locations (**Figure 6**B) (Kojouharov et al., 2023; Schiebel et al., 2020c; Wang et al., 2023). Some robotic studies investigated using knowledge of accurate environment geometry to compute a desired path or joint torque profile (**Figure 6**C) (Akbarzadeh et al., 2011; Hanssen et al., 2020; Holden and Stavdahl, 2013; Holden et al., 2014; Sanfilippo et al., 2016; Transeth et al., 2007).

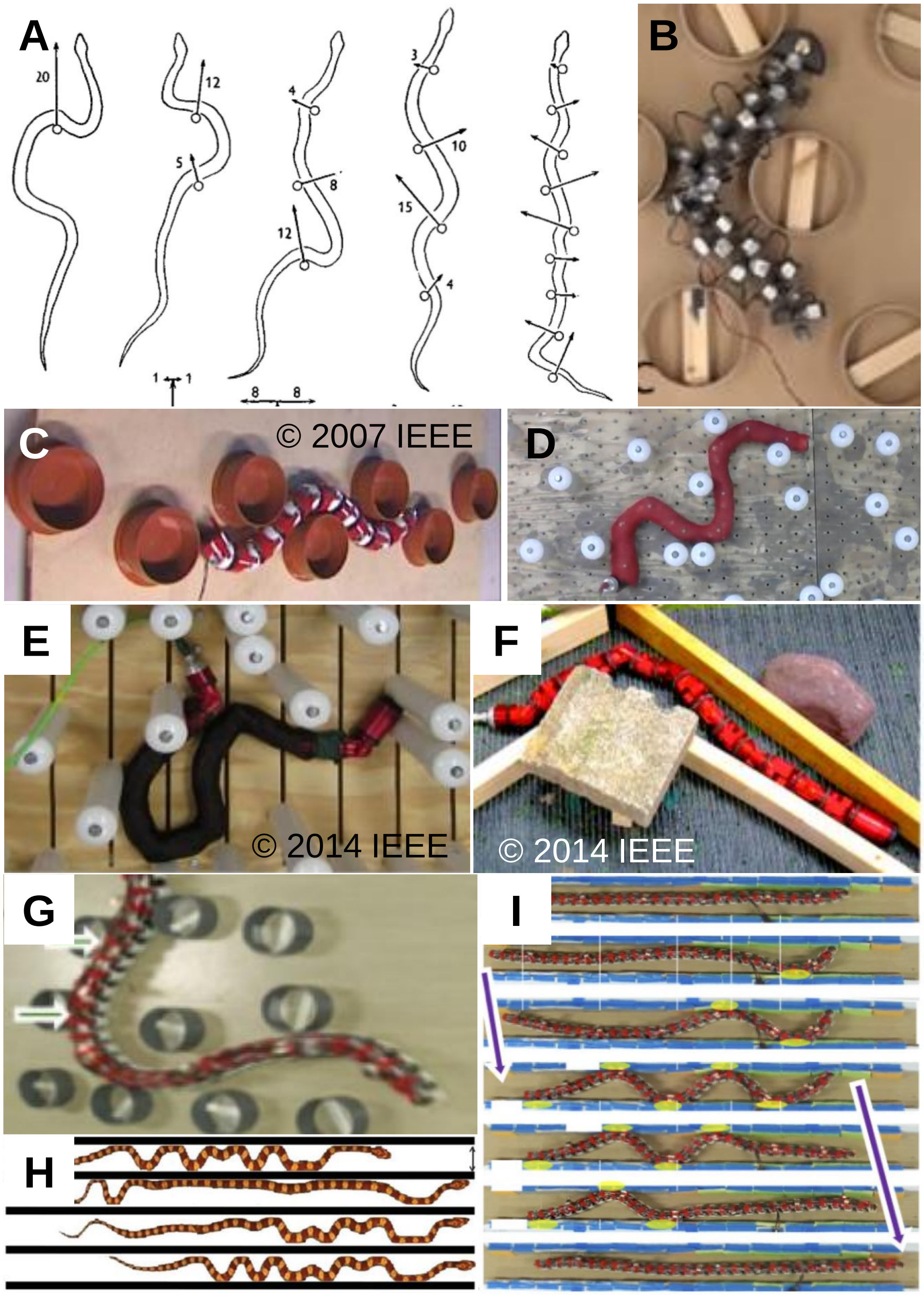

**Figure 6. Representative snake and snake robot locomotion on flat surfaces with vertical extrusions. (A-G)** Pushing against pillars for propulsion using lateral undulation. Robots are controlled in open loop with mechanical compliance (B), using feedforward patterns matching

terrain geometry (C), using proprioceptive feedback (D, E), using feedforward joint torque profiles (F), or using both proprioceptive and exteroceptive feedback (G). **(H-I)** Bracing against parallel walls using concertina. Robot is controlled using both proprioceptive and exteroceptive feedback (I). Purple arrows in (I) show temporal sequence. Reproduced from (Gray and Lissmann, 1950; Kano et al., 2018; Marvi and Hu, 2012; Rollinson et al., 2014; Schiebel et al., 2020c; Transeth et al., 2007; Travers et al., 2018).

Most of the robots that push against the lateral contact points modulate their bending patterns using feedback control based on proprioception (joint angle/torque) or exteroception (contact force) (for a review, see (Sanfilippo et al., 2017)). Some rely on parameterized functions to control the entire body shape modulated by proprioception-based feedback control (**Figure 6**D) (Gong et al., 2016; Rollinson and Choset, 2013; Travers et al., 2015; Travers et al., 2018; Wang et al., 2020a; Whitman et al., 2016). Some use decentralized joint control with proprioceptive feedback (**Figure 6**E) (Boyle et al., 2013; Date and Takita, 2005; Rollinson et al., 2014) or feedforward joint torque control (**Figure 6**F) (Rollinson et al., 2014) to propagate a shape for automatic adaptation to the environment. Some combine proprioceptive and exteroceptive feedback in control to propagate a lateral bending shape and react to contact (**Figure 6**G, I) (Hirose, 1993; Kamegawa et al., 2014; Kano and Ishiguro, 2013; Kano and Ishiguro, 2020; Kano et al., 2012; Liljebäck et al., 2011; Liljebäck et al., 2014b).

### 1.2.2 Snake and snake robot locomotion in 3-D environments

When traversing 3-D environments, snakes often bend their bodies dorsoventrally to accommodate height variations (**Figure 7**). Only in limited cases similar to inclined flat surfaces with vertical extrusions, such as inside a vertical gap (Shapiro et al., 2007) or on tree barks with asperities that can be gripped by scales (Marvi and Hu, 2012), they can reorient the body to achieve vertical displacement using only lateral body bending. Unlike *C. elegans* that lie on their sides during movement, snakes and snake robots rarely move with perfect 90° rolling. Thus, dorsoventral body bending almost always involves a change of height in the vertical direction. Below, we do not distinguish vertical body bending and dorsoventral body bending.

Many of the previous snake studies that investigated large vertical body bending focused on arboreal environments (**Figure 7**A, green circles), such as a single cylindrical perch with different slopes and pegs extruded from the perch (Astley and Jayne, 2007; Astley and Jayne, 2009; Crotty and Jayne, 2015; Jayne and Byrnes, 2015; Jayne and Herrmann, 2011; Jayne et al., 2015), a large gap between two such perches (**Figure 7**D) (Byrnes and Jayne, 2012; Graham and Socha, 2021; Jayne and Riley, 2007; Jayne et al., 2014; Jorgensen and Jayne, 2017; Mansfield and Jayne, 2011), or flying snakes gliding in the air (Socha et al., 2005; Socha et al., 2011; Yeaton et al., 2020). Snakes bend their bodies on the perches in 3-D to surmount or wrap the perches, push against or grip the pegs and move along the perches using concertina or lateral undulation (Jayne, 2020). When bridging a gap, the body in between the perches either cantilevers in the air quasi-statically or lunges dynamically for longer displacement, both of which can involve large vertical bending produced by muscle activation or gravity (Jorgensen and Jayne, 2017). When gliding in the air, vertical bending can be used to control pitch motion (Yeaton et al., 2020).

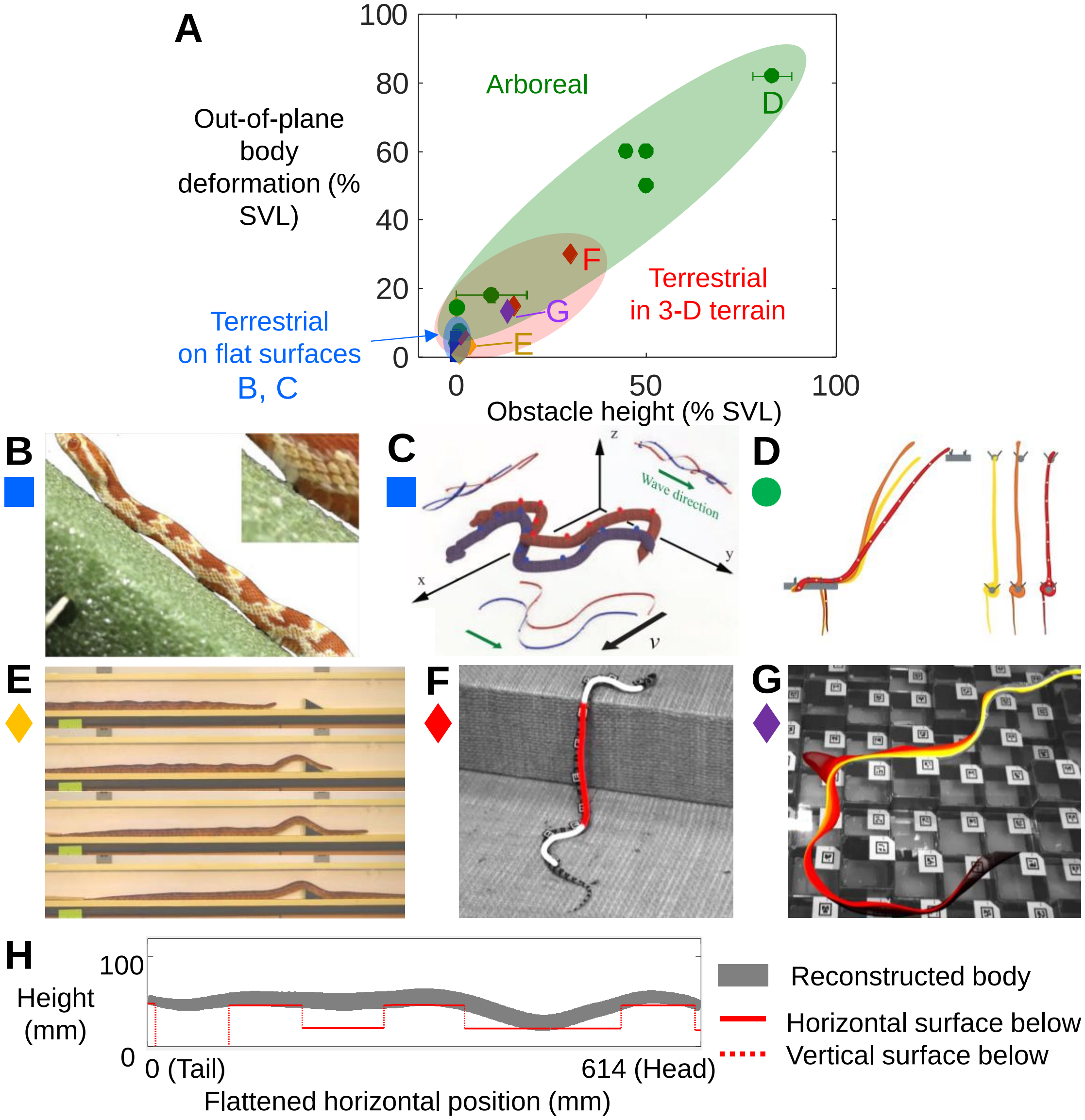

**Figure 7. Representative snake locomotion studies with quantified out-of-plane body deformation. (A)** Out-of-plane body deformation and obstacle height. Circles show arboreal locomotion, squares show terrestrial locomotion on flat surfaces, and diamonds show terrestrial locomotion in 3-D terrain. SVL stands for snout-vent length. **(B-C)** Snakes traversing flat surfaces with out-of-plane body deformation using concertina (B) or sidewinding[2] (C). **(D)** Arboreal snake

---

[2] From H. Marvi et al., "Sidewinding with minimal slip: Snake and robot ascent of sandy slopes," *Science*, vol. 346, no. 6206, pp. 224--229, Oct. 2014. Reprinted with permission from AAAS.

bridging a vertical gap. **(E)** Snake traversing a triangular wedge inside a narrow tunnel involving vertical body bending. **(F)** Snake traversing a large smooth step using partitioned gait. **(G-H)** Snake traversing uneven terrain with quantified kinematics (G) and body-terrain contact (H) in this dissertation. (G) shows timelapse with midline overlaid at different time instances in a representative trial, (H) shows contact between reconstructed body (gray) and terrain surfaces below (red) in flattened sagittal view at a representative time instance. Reproduced from (Byrnes and Jayne, 2012; Fu et al., 2022; Gart et al., 2019; Jurestovsky et al., 2021; Marvi and Hu, 2012; Marvi et al., 2014).

Previous studies on terrestrial snake locomotion have rarely used 3-D terrain surfaces with significant height variations. Vertical body bending observed in terrestrial snake locomotion was mostly from gaits used on flat surfaces (**Figure 7**A, blue squares) such as sidewinding (Astley et al., 2015; Marvi et al., 2014) (**Figure 7**C) or concertina (Marvi and Hu, 2012) (**Figure 7**B). Our lab previously (Gart et al., 2019) quantified how a generalist kingsnake climbs up a large, smooth step by partitioning its body into three sections performing different patterns (**Figure 7**A, red diamonds; **Figure 7**F). A recent study from our collaborators found that a generalist corn snake can use vertical bending only to traverse terrain with large height variations, such as a row of horizontal cylinders or a triangular wedge (Jurestovsky et al., 2021) (**Figure 7**A, dark yellow diamonds; **Figure 7**E).

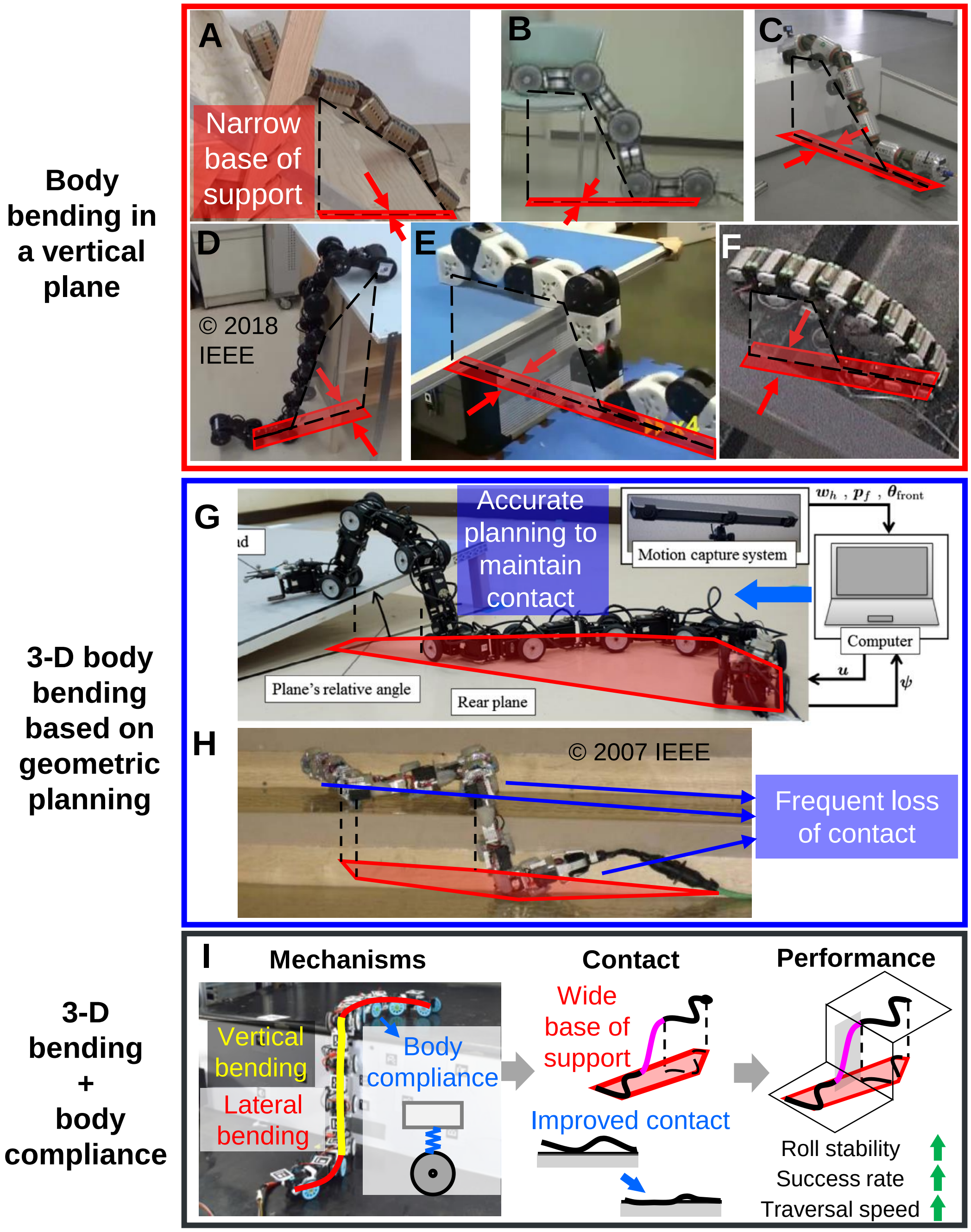

**Figure 8. Snake robots climbing steps. (A-F)** Representative snake robots climbing steps with a narrow base of support (red polygon). (A-D) use active propellers and are controlled by human operators (A-B) or geometric shape planning (C-D). (E-F) use only vertical body bending for propulsion and are controlled with prescribed shape planning (E, F) and compliant joint control (F).

**(G-H)** Snake robots climbing a step combining lateral and vertical body bending for a larger base of support. However, they need accurate control or lose contact frequently. They are controlled by geometric shape planning based on knowledge of terrain geometry. **(I)** Our study on stability principles during step traversal. Although 3-D body bending is also controlled by geometric shape planning, we discovered that passive body compliance (blue) improves body-terrain contact without accurate control, which increases roll stability and allows faster and more robust traversal. Reproduced from (Borenstein and Hansen, 2007; Fu and Li, 2020; Kurokawa et al., 2008; Lipkin et al., 2007; Nakajima et al., 2018; Pfotzer et al., 2017; Tanaka et al., 2018; Yamada and Hirose, 2006; Yim et al., 2006).

Snake robots are far less stable, versatile and efficient than animals when traversing 3-D terrain, in contrast to their well-studied locomotion on flat surfaces. Many snake robots have attempted climbing up steps or stairs. The majority of them form a narrow base of ground support without lateral body bending (**Figure 8**A-F) and can be unstable when there are lateral perturbations such as on uneven terrain. Some rely on active propellers such as actuated treads or wheels for propulsion and rely on human operators (**Figure 8**A, B) or geometric shape planning (**Figure 8**C, D) to control body bending (Birkenhofer, 2010; Jing et al., 2017; Komura et al., 2015; Kurokawa et al., 2008; Nakajima et al., 2018; Scholl et al., 2000; Takaoka et al., 2011; Takayama and Hirose, 2000; Tanaka and Tanaka, 2013; Tanaka et al., 2018; Yamada and Hirose, 2006). Some (**Figure 8**E, F) rely on prescribed body bending patterns to move forward without using active propellers (Kurokawa et al., 2008; Yim et al., 2001). Only two robots (**Figure 8**G, H) deliberately use lateral body bending for a wide base of support but still rely on careful geometric shape planning and accurate feedback control or will lose contact frequently (Lipkin et al., 2007; Nakajima et al., 2018; Tanaka and Tanaka, 2015a). The instability and the consequent demand for careful control resulted in slow traversal speeds, which can be improved if we better understand the stability principles.

Aside from the instability, previous snake robots cannot adapt to diverse terrain versatilely. Many robots use prescribed motion patterns specifically designed for limited environments such as steps (**Figure 8**A-E, G, H), pipes, and ladders (**Figure 9**) (Inazawa et al., 2021; Jurestovsky et al., 2021; Lipkin et al., 2007; Melo and Paez, 2012; Nakajima et al., 2018; Rollinson and Choset, 2013; Takemori et al., 2018a; Takemori et al., 2018b; Tanaka and Tanaka, 2013; Walker et al., 2016; Wang et al., 2022). Although some are able to use torque control (Inazawa et al., 2021; Takemori et al., 2018b), proprioception-based feedback control (Rollinson and Choset, 2013), path planning based on laser sensing (Pfotzer et al., 2017), or elastic joints (Vespignani et al., 2015) to automatically adapt to terrain with similar geometry at different scales, such as pipes of different dimensions, they typically lack the versatility to adapt to more diverse terrain geometry.

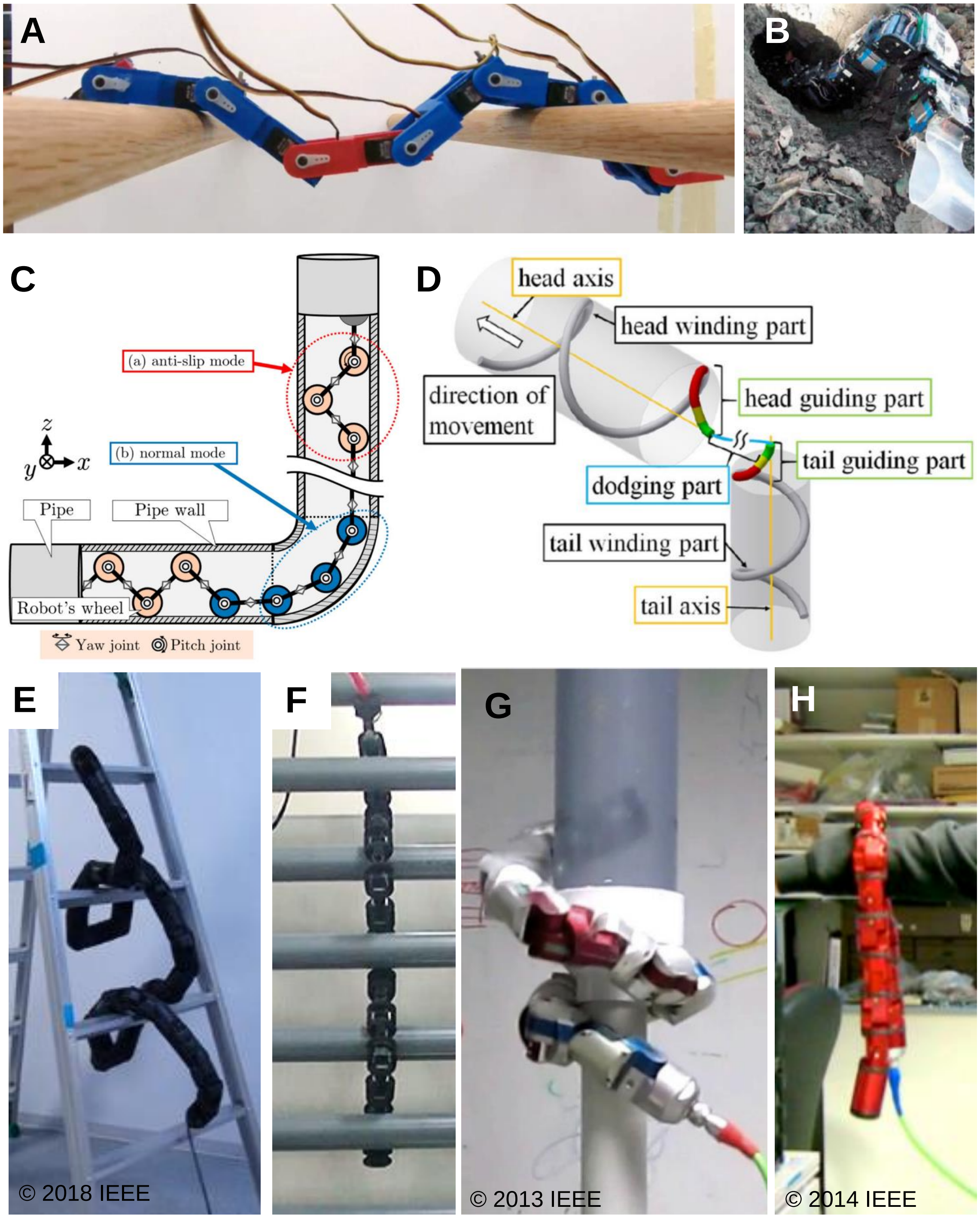

**Figure 9. Representative snake robots using prescribed motion patterns specifically designed for pipes and ladders. (A)** Traversing a row of horizontal cylinders using feedforward vertical undulation. **(B)** Traversing a tunnel using concertina modulated by force feedback control.

**(C-D)** Traversing a tunnel using geometric shape planning. **(E-F)** Traversing a ladder using geometric shape planning. **(G-H)** Traversing perch-like structures by rolling around centerline. Reproduced from (Inazawa et al., 2021; Jurestovsky et al., 2021; Rollinson and Choset, 2013; Rollinson et al., 2014; Sawabe et al., 2019; Takemori et al., 2018a; Wang and Kamegawa, 2022; Yim et al., 2006).

Some studies tried increasing the versatility of snake robots by simplifying body bending control only for steering and using actuated wheels or treads for propulsion (**Figure 8**A-D, **Figure 9**C, **Figure 10**A-E) (Arai et al., 2008; Borenstein and Hansen, 2007; Kimura and Hirose, 2002; Komura et al., 2015; Kouno et al., 2013; Nakajima et al., 2022; Pfotzer et al., 2014; Takaoka et al., 2011; Tanaka and Tanaka, 2015b; Tanaka et al., 2018). Some of the robots are able to automatically adapt to unstructured terrain by passive deformation of compliant joints under gravity (Arai et al., 2008; Borenstein and Hansen, 2007; Kimura and Hirose, 2002; Suzuki et al., 2012) or feedback control of joints (Kouno et al., 2013). However, the additional active propellers use up considerable weight, space, and energy consumption, which limits their bending ability. It is also more challenging to seal this type of robots for applications in wet or dusty environments, such as beaches and deserts. The simple follow-the-leader strategy used by many of them also often results in the entire body staying inside a vertical plane, which can be unstable laterally in 3-D environments. A wide body similar to pythons (**Figure 8**D) can mitigate the instability when using this strategy (Tanaka et al., 2018) but the width can barely match the amplitude of vertical bending without limiting the robot’s capacity of bending laterally. Some snake robots have the same risk of lateral instability when forming a rolling loop in the vertical plane (**Figure 8**F) to traverse small obstacles or with torque control to conform to larger obstacles (Jing et al., 2017; Ohashi et al., 2010; Yim et al., 2001; Yim et al., 2006).

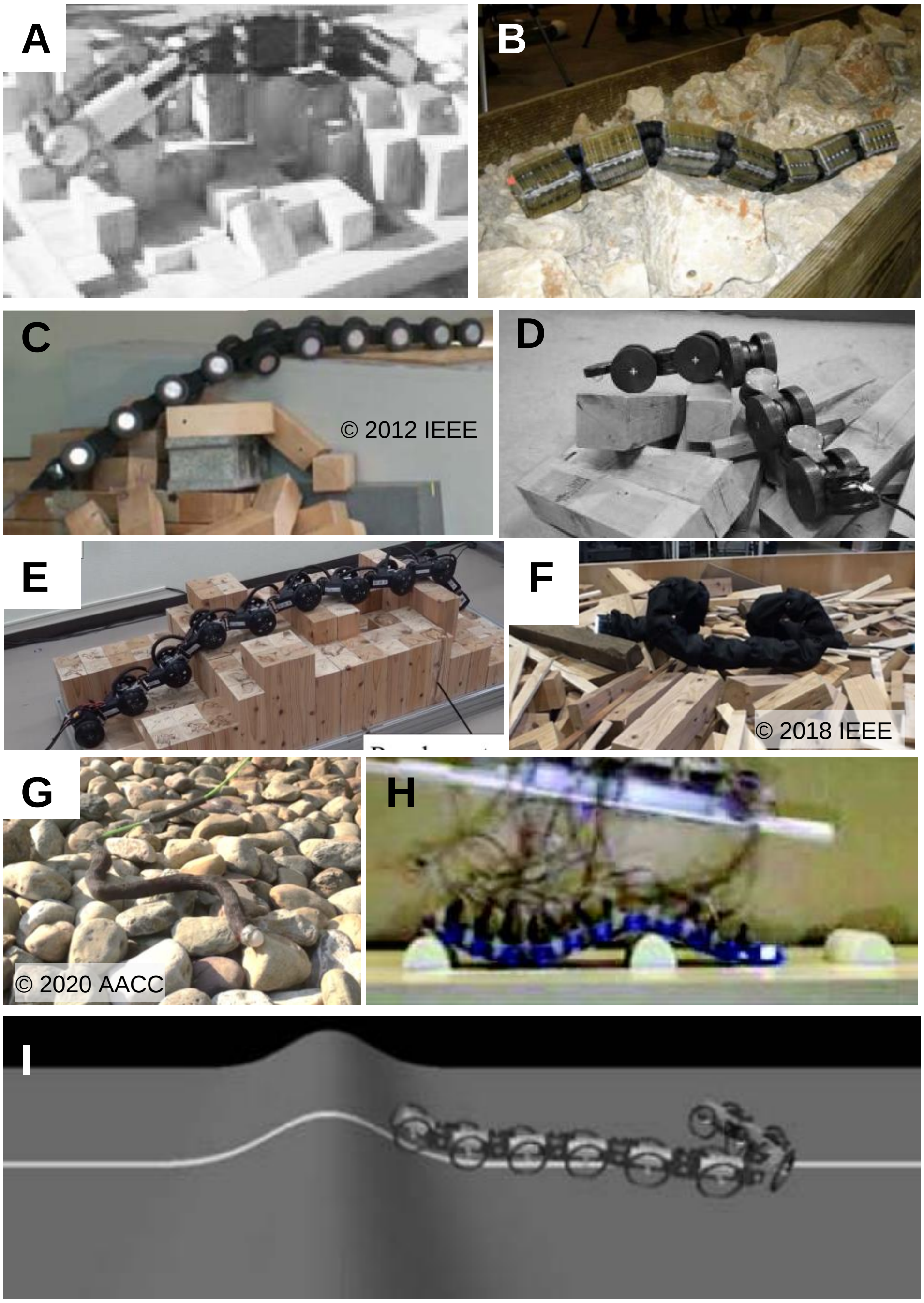

**Figure 10. Representative snake robots adapting to and traversing other 3-D terrain. (A-E)** Traversing unstructured terrain using active propellers for propulsion. Conformation to ground is realized by elastic joints (A-C), feedback control of joint torque (D), and manual control (A-E). **(F-I)** Traversing using gaits designed for flat surfaces (F-G), vertical body bending (H), or 3-D body

bending (I). Conformation to ground is realized by limit of joint torque (F), proprioception-based feedback control (G, I), or control based on a combination of contact force feedback, joint angle feedback, and manual input of head bending (H). Reproduced from (Arai et al., 2008; Borenstein and Hansen, 2007; Date and Takita, 2005; Kano et al., 2014; Kouno et al., 2013; Nakajima et al., 2022; Suzuki et al., 2012; Takemori et al., 2018b; Wang et al., 2020a).

Some robots can traverse unstructured 3-D environments using gaits designed for flat surfaces and accommodate height variations by compliant shells and joints or proprioception-based feedback control (**Figure 10**F, G) (Takemori et al., 2018b; Travers et al., 2018; Wang et al., 2020a; Yim et al., 2006). However, they suffer from severe slipping, presumably because of frequent loss of contact with the uneven terrain and increasing resistance from pushing against terrain asperities. Only a few snake robots consider using vertical body bending modulated by sensory feedback control to push against these asperities aside from passively conforming to them (**Figure 10**H, I). However, they either still rely on human input of head movement (Kano et al., 2014) or are only designed for ideally smooth surfaces without considering external forces such as friction and gravity (Date and Takita, 2005).

### 1.2.3 Knowledge gaps

Despite the rich understanding of snake locomotion on flat surfaces and applications in snake robots, snake locomotion on 3-D terrain has not been well studied (**Figure 7**). The majority of studies that investigated snake 3-D locomotion focused on arboreal environments, whereas terrestrial snake locomotion on 3-D surfaces was rarely studied. The principles of arboreal snake locomotion discovered in these studies can be significantly different from those of terrestrial locomotion, considering the difference in terrain geometry (Astley and Jayne, 2007; Higham and Jayne, 2004; Jayne, 2020). For example, arboreal environments contain cylindrical perches with variable diameters and inclines, which require snakes to frequently surmount, grip or brace against perches for stability. In some terrestrial environments that are too large, smooth and rigid, such as on large boulders or a step, snakes do not have such perch-like structures (Astley and Jayne, 2009; Jayne and Byrnes, 2015; Jayne and Riley, 2007; Lillywhite et al., 2000) to grip for stability and cannot create structures to brace against by deforming the substrate like on sandy terrain (Marvi et al., 2014). In addition, the difference in terrain geometry also results in difference in modes used by snakes to generate propulsion when traversing the simplest structures in arboreal and terrestrial environments (a single perch and a flat surface, respectively) (Jayne, 2020), presumably leading to bigger difference for more complex 3-D environments. The lack of understanding also results in inferior performance of snake robots when traversing 3-D terrain. They either lack the versatility to adapt to diverse 3-D terrain without human operation (**Figure 8**, **Figure 9**, **Figure 10**A-E, H), suffer from severe slipping (**Figure 10**F-G), or are too idealized to accommodate real-world challenges such as friction (**Figure 10**I).

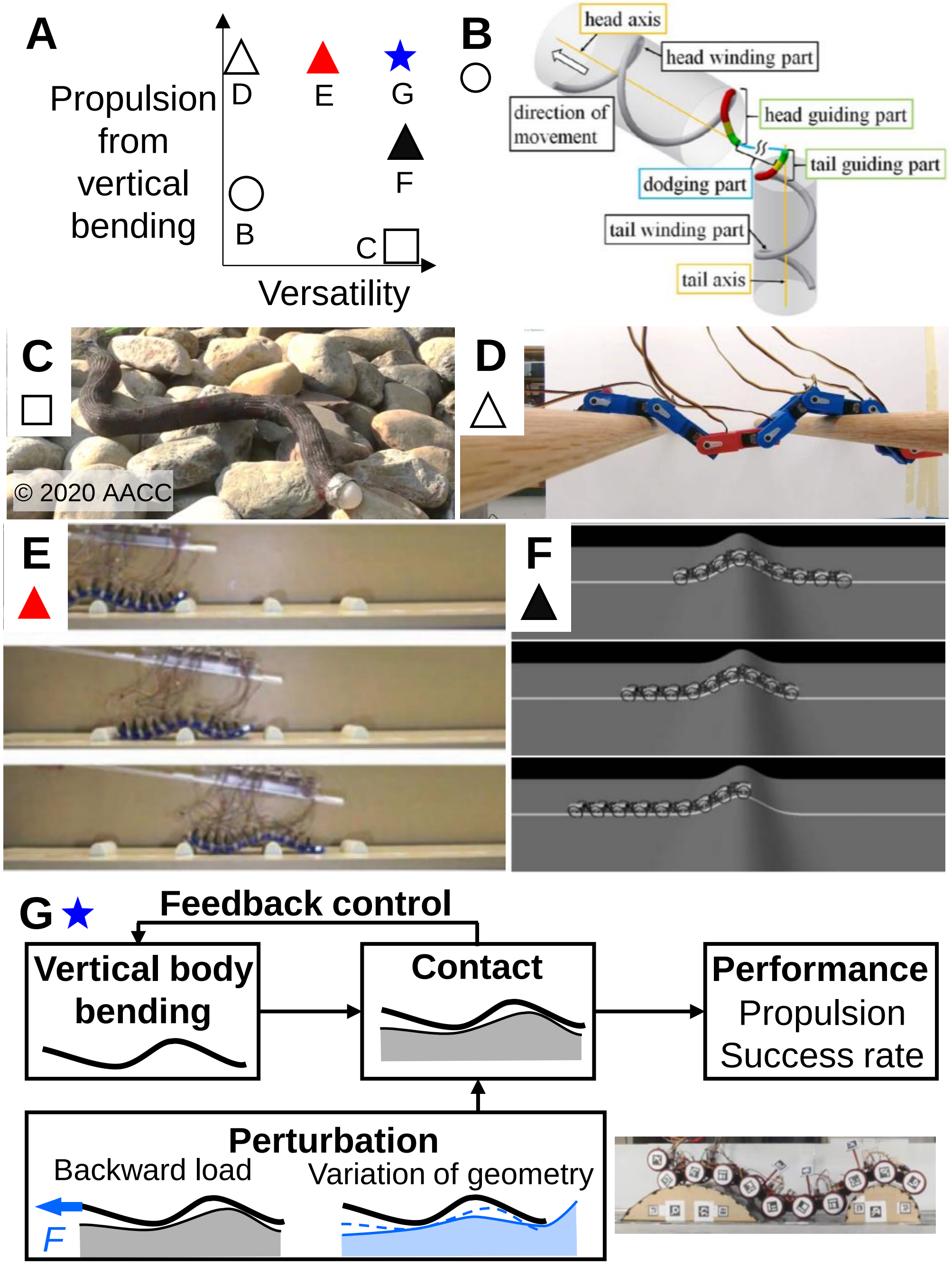

**Figure 11. Usage of vertical body bending in snake robots traversing 3-D terrain. (A)** Qualitative comparison along two dimensions: versatility to adapt to various terrain surfaces automatically and propulsion generated by vertical body bending. **(B-C)** Vertical bending for bridging height differences using feedforward planning (B) or feedback control (C). **(D-F)** Vertical

bending for propulsion generation. Robots in (D-F) are controlled using prescribed patterns matching terrain geometry (D), a combination of manual operation and feedback control (E), or only feedback control (F). Propulsion generated in (F) is questionable because of ignoring fore-aft resistance such as friction. **(G)** Our study on principles of using vertical bending for propulsion and how perturbations and feedback control affect performance by modulating body-terrain contact. Our robot does not require human input and is robust against perturbation. Reproduced from (Date and Takita, 2005; Fu and Li, 2021; Inazawa et al., 2021; Jurestovsky et al., 2021; Kano et al., 2014; Wang et al., 2020a).

What sets snake and snake robot locomotion in complex 3-D environments different from their well-studied locomotion on flat surfaces is the broader usage of vertical body bending. Vertical bending can induce stability challenges because of the elevated center of mass. While some previous studies have proposed using the base of support formed by contact points to analyze the challenge and address it on flat surfaces (Cappo and Choset, 2014; Chong et al., 2021; Marvi et al., 2014), little is known about such challenges in 3-D environments and how to address them, especially on smooth surfaces lacking gripping points like steps (**Figure 8**).

Because of the direction of gravity, the contribution of vertical bending to locomotion is presumably different from that of lateral body bending and remains to be understood. On flat surfaces, vertical body bending is used to improve efficiency by reducing frictional drag such as in sidewinding (Marvi et al., 2014) and sinus-lifting (Hirose, 1993; Hirose and Mori, 2004; Hu et al., 2009; Toyoshima and Matsuno, 2012). Previous snake and snake robot studies in 3-D environments have considered the contribution of vertical bending in bridging height differences (**Figure 11**B, C) (Gart et al., 2019; Inazawa et al., 2021; Jayne, 2020; Wang et al., 2020a). Recently, a study demonstrates that vertical bending can also be used for propulsion by pushing against uneven terrain below the body (**Figure 11**D) (Jurestovsky et al., 2021). However, the principles of producing vertical bending patterns effective in generating propulsion on random uneven terrain have not been understood (**Figure 11**A). Thus, the few snake robots that deliberately use vertical bending for propulsion either rely on manual control of bending shapes (**Figure 11**D, E) (Jurestovsky et al., 2021; Kano et al., 2014; Takanashi et al., 2022; Takemori et al., 2018a) or lack consideration of external forces (**Figure 11**F) (Date and Takita, 2005). Understanding how vertical body bending contributes to propulsion will not only better inform why snakes use 3-D body bending, but also allow snake robots to fully exploit their entire body bending in 3-D to push against a wider range of terrain asperities to improve efficiency and maneuverability.

To understand either the stability challenge or the propulsive value of vertical bending, it is important to understand how 3-D body bending affects physical interaction between the body and the terrain, and how this in turn affects the performance of the entire system. While most of the snake studies in 3-D environments have quantified their performances under different conditions, few have measured snake body bending and physical interactions in 3-D (but see (Byrnes and Jayne, 2012; Byrnes and Jayne, 2014; Gart et al., 2019; Jurestovsky et al., 2021; Sharpe et al., 2015)). The quantification is also lacking in snake robot studies in 3-D terrestrial environments, partially because only a few robots that attempted traversing such environments are equipped with contact force sensors (Kano et al., 2014).

Because of the importance of physical interaction in bridging vertical bending and system performance, it is also essential to understand how to modulate this interaction in response to perturbations such as variation of terrain geometry or external force that can cause slipping. Although evidence of sensory feedback-based modulation of lateral body bending has been discovered in animals and the principles have been well investigated in snake robots (Chapter 1.2.1.2), whether and how the modulation applies to vertical bending remain to be investigated considering the effects of gravity.

## 1.3 Research questions and objectives

To advance our understanding of 3-D terrestrial snake locomotion, we integrated biological and robotic studies in three representative 3-D environments (**Figure 12**). Observations in animal experiments provided inspirations to the development of the robotic models, while the robots as amenable physical models allowed us to systematically test hypotheses that arose from animal experiments (see Chapter 1.4). We discovered several mechanisms that affected traversal performance such as stability and propulsion, mostly via affecting body-terrain contact (**Figure 12**). The details are reported in Chapters 2 to 4, and below is a summary of the rationales and methods.

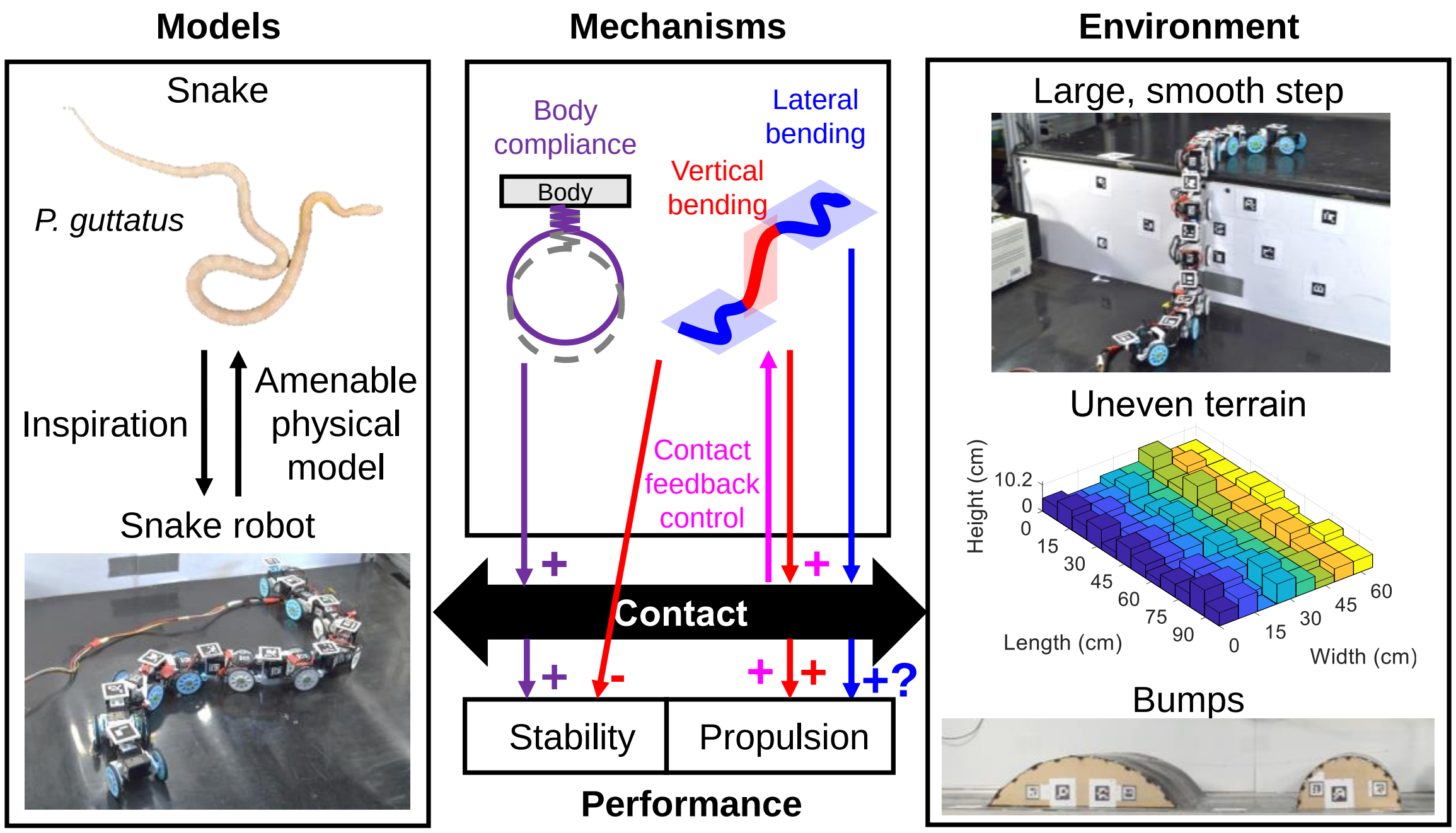

**Figure 12. Overview of this dissertation research.** By testing different models in three representative 3-D environments, we discovered several mechanisms affecting performance shown by thin arrows, mostly via affecting body-environment contact. Reproduced from (Fu and Li, 2020; Fu and Li, 2021; Fu et al., 2022).

### 1.3.1 Stability principles when traversing a large, smooth step

In Chapter 2, as an initial step to understanding the stability principles of snakes using 3-D body bending to traverse complex, 3-D environments, we started with a single terrain model, a large, smooth step. Compared to arboreal environments such as a vertical gap between branches (**Figure 7**D), this type of terrain such as boulders or fallen trees induces a bigger challenge of stability because of lacking extrusions to grip or brace against.

Our lab previously studied the kinematics of kingsnakes traversing such terrain and found that a snake partitions its body into 3 sections performing distinct patterns (**Figure 13**): the anterior and posterior body sections use lateral undulation on horizontal surfaces, the middle section cantilevers in a vertical plane to bridge the two undulating sections, and this partition travels down the body while the snake moves forward and upward the step. The snakes maintained perfect stability in all trials, likely because of the large base of support provided by the laterally undulating body sections.

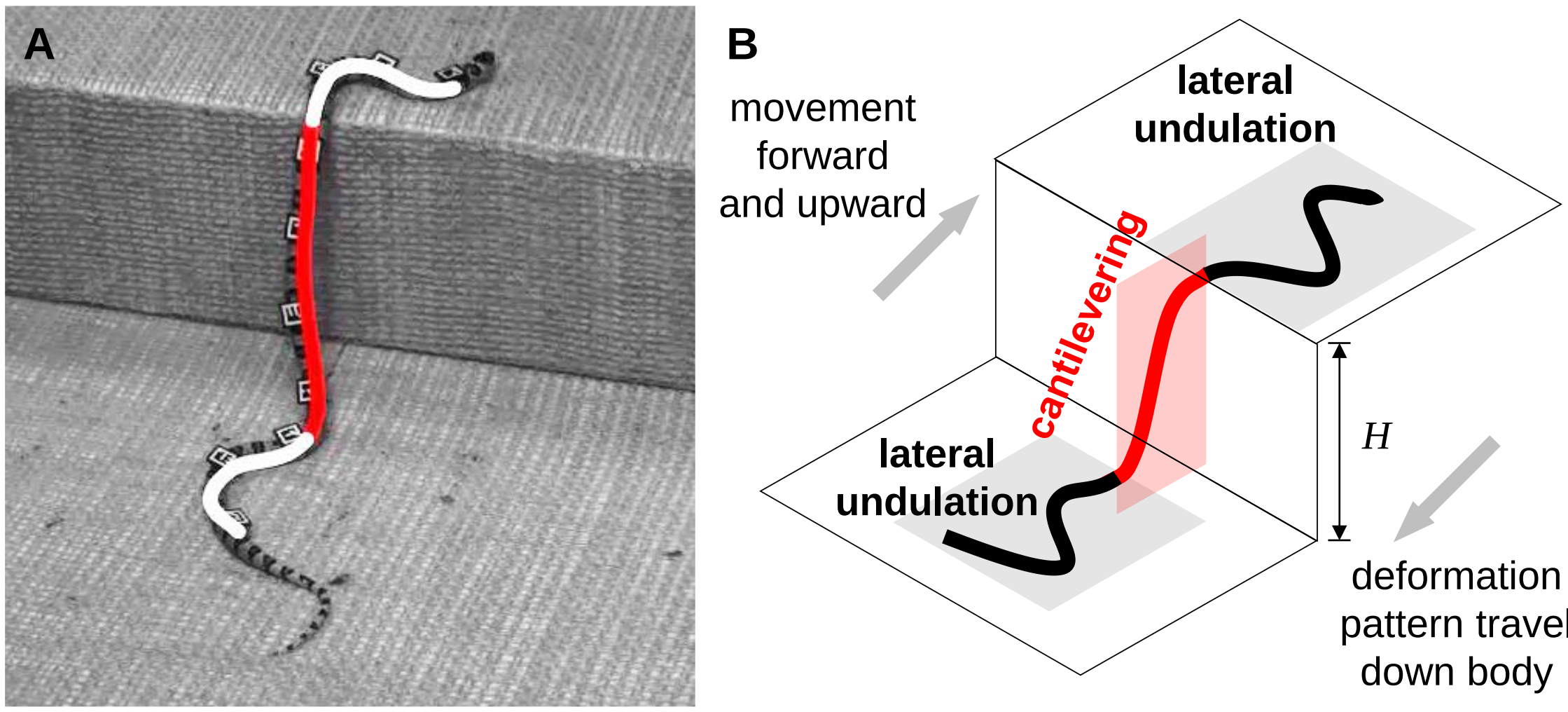

**Figure 13. Kingsnake traversing a large, smooth step by partitioning its body into three sections. (A)** Kingsnake with its tracked midline partitioned into three sections, two using lateral

undulation on horizontal surfaces (white) and one cantilevering in between (red). **(B)** Schematic of partitioned gait. Reproduced from (Gart et al., 2019).

To understand the stability principles of snakes traversing large step traversal using this 3-D partitioned gait, in Chapter 2 we developed a snake robot as an amenable physical model (**Figure 8**I). The robot has variable body compliance, can use 3-D body bending and can move forward using lateral undulation. We challenged it with a large step and measured how the body-terrain contact and stability changed with the height of the step and body compliance. We tested two hypotheses: (1) roll stability diminishes as the step becomes higher and (2) body compliance improves surface contact and reduces roll instability. This work has been published in *Royal Society Open Science* as an article entitled *Robotic modelling of snake traversing large, smooth obstacles reveals stability benefits of body compliance* authored by Qiyuan Fu and Chen Li (Fu and Li, 2020).

### 1.3.2 Kinematics and body-terrain interaction of a snake traversing more complex uneven terrain

In Chapter 3, to further elucidate how snakes use 3-D body bending to interact with and traverse more complex 3-D terrain, we tested how corn snakes traverse an uneven arena constructed by blocks of random height (**Figure 7**G, H). We measured the 3-D body bending of snakes, quantified the resulting body-terrain interaction, and analyzed how they relate to performance such as propulsion generation. This is a step towards filling the knowledge gap of lacking quantification of 3-D kinematics and physical interaction for 3-D terrestrial snake locomotion (Chapter 1.2.3; **Figure 7**). This work has been published in *Bioinspiration and Biomimetics* as an article entitled *Snakes combine vertical and lateral bending to traverse uneven terrain* authored by Qiyuan Fu, Henry Astley, and Chen Li (Fu et al., 2022).

### 1.3.3 Propulsion generation using vertical bending and benefits of contact feedback control

Chapter 4 details how we investigated the principles of propulsion generation using vertical body bending and the benefits of contact feedback control in this process. We hypothesized that vertical bending can generate propulsion substantial to accommodate friction and large additional resistance if contact is well maintained. In addition, we hypothesized that although perturbations that break body-terrain contact can result in more failure, contact feedback control can help mitigate the effects. To test these hypotheses, we equipped a snake robot with contact force sensors (Figure 11G) and used it as a physical model. We challenged the robot to traverse large bumps using vertical body bending only. We varied the control loops, challenged the robot with perturbations such as an additional backward load or variation of terrain geometry, and analyzed how body-terrain contact and performance change with control and perturbations. This work has been posted on arXiv as a preprint entitled *Snake robot traversing large obstacles using vertical bending reveals importance of contact feedback for propulsion generation* authored by Qiyuan Fu and Chen Li (Fu and Li, 2021). This work has also been submitted to *Bioinspiration and Biomimetics* and is in revision at the time of submission of this dissertation.

## 1.4 Using robots as physical models to study locomotion

In this dissertation, we constructed robots as amenable physical models (Aguilar et al., 2016) to test hypotheses that arose from snake experiments. This enabled the following benefits in this dissertation research:

(1) Help with understanding the behaviors of a complex animal in a complex environment using simplified models tested under controlled laboratory settings. For example, in Chapters 2 and 4 we used simplified snake robots traversing a step and bumps. A snake contains hundreds of vertebrae and muscles and the natural environments it resides in contain numerous detailed structures like unevenness and hybrid materials. It would be extremely complex to model every detail of them. In contrast, the robots and the model terrain we used were simplified to represent minimal features needed to approach the scientific questions of interest, such as the 3-D body bending ability of the robot on the large, smooth 3-D terrain in Chapter 2. Despite these simplifications, the studies allowed us to discover principles that can reproduce a snake's traversal of similar terrain (Gart et al., 2019; Jurestovsky et al., 2021) using the robots and may explain snake locomotion in similar natural environments like on large boulders.

(2) Enablement of discoveries of fundamental physical principles that can be generalized to a wider range of scenarios. This is challenging by directly studying a complex system interacting with complex environments but can be realized by systematically testing simplified robotic devices in controlled laboratory settings. In Chapter 2, we discovered how larger vertical body bending and body compliance affect stability. Revealed from systematic experiments of a robotic model much simpler than a snake on a simple step, the principles can be generalized to snakes or other snake robots traversing more complex 3-D environments. Similarly, in Chapter 4 this method allowed us to reveal principles of propulsion generation using vertical body bending that can be generalized to more complex scenarios.

(3) Systematic variation of mechanisms of interest, such as body compliance (Chapter 2) and feedback control loops (Chapter 4), without affecting other components in the system. Unlike studies using well-established model organisms such as *C. elegans*, such variation is especially difficult to perform on snakes, which have complex neuromechanical systems that are yet to be understood (Astley, 2020a).

(4) Repeatable experiments against the physical environment to ensure reliability of models and reveal novel mechanisms. For example, in Chapter 2 we systematically tested the stability of the robot when using a previously revealed partitioned gait model (Gart et al., 2019) and further identified the benefits of body compliance to stability.

(5) Easier measurement of physical interactions with real-world environments, such as contact conditions (Chapters 2 and 4) and contact forces (Chapter 4). This is challenging to perform on a snake with a deformable body without substantial assumptions (Chapter 3).

(6) Testing extreme performance that snakes are unwilling to attempt, such as a large step that significantly challenges stability (Chapter 2) or a large backward load applied externally (Chapter 4).

(7) Providing design principles for robust snake robot platforms for practical applications in complex environments, such as the benefits of adding compliant structures and combing both lateral and vertical body bending for stability and propulsion (Chapters 2 to 4).

Aside from these, the robophysical method has many benefits not reflected by this dissertation. The method can also be used to study extinct locomotion (Nyakatura et al., 2019) and to construct a performance landscape of a system by sweeping through the entire parameter space of a morphology (Schultz et al., 2021). All these benefits have enabled considerable advancement in diverse areas including animal locomotion, physics, and robotics using this method (Aguilar et al., 2016).

# Chapter 2

# Robotic modeling of snake traversing large, smooth obstacles reveals stability benefits of body compliance

## 2.1 Author contributions

Qiyuan Fu designed study, developed robot, performed experiments, analyzed data, and wrote the paper; Chen Li designed and oversaw study and revised the paper.

## 2.2 Summary

Snakes can move through almost any terrain. Although their locomotion on flat surfaces using planar gaits is inherently stable, when snakes deform their body out of plane to traverse complex terrain, maintaining stability becomes a challenge. On trees and desert dunes, snakes grip branches or brace against depressed sand for stability. However, how they stably surmount obstacles like boulders too large and smooth to gain such “anchor points” is less understood. Similarly, snake robots are challenged to stably traverse large, smooth obstacles for search and rescue and building inspection. Our recent study discovered that snakes combine body lateral undulation and cantilevering to stably traverse large steps. Here, we developed a snake robot with this gait and snake-like anisotropic friction and used it as a physical model to understand stability principles. The robot traversed steps as high as a third of its body length rapidly and stably. However, on higher steps, it was more likely to fail due to more frequent rolling and flipping over, which was absent in the snake with a compliant body. Adding body compliance reduced the robot’s roll instability by statistically improving surface contact, without reducing speed. Besides advancing understanding of snake locomotion, our robot achieved high traversal speed surpassing most previous snake robots and approaching snakes, while maintaining high traversal probability.

## 2.3 Introduction

Snakes are masters of locomotion across different environments (Houssaye et al., 2013). With their elongate, flexible body (Penning and Moon, 2017) of many degrees of freedom (Voris, 1975), snakes can use various planar gaits to move on flat surfaces, be it open (Gray, 1946; Jayne, 1986; Marvi and Hu, 2012), confined (Gray, 1946; Jayne, 1986; Marvi and Hu, 2012), or with small obstacles that can be circumvented (Gray and Lissmann, 1950; Jayne, 1986). Snakes can also deform their body out of plane to move across complex environments (Astley and Jayne, 2009; Byrnes and Jayne, 2014; Gart et al., 2019; Jayne and Riley, 2007; Marvi et al., 2014) (for a review, see (Gart et al., 2019) Supplementary Information). In these situations, out-of-plane body deformation can challenge stable locomotion (Gong et al., 2012; Hatton and Choset, 2010; Marvi et al., 2014), which is rarely an issue on flat surfaces with planar gaits. To maintain stability, arboreal snakes grip or brace against branches (Astley and Jayne, 2009; Jayne and Byrnes, 2015; Jayne and Riley, 2007; Lillywhite et al., 2000) and carefully distribute body weight (Jayne and Byrnes, 2015; Jayne and Herrmann, 2011); sidewinders depress portions of the body into sand and brace against it without causing avalanche, while minimizing out-of-plane deformation (Marvi et al., 2014). However, we still know relatively little about how snakes maintain stability when surmounting obstacles such as boulders that are too large and smooth to gain such "anchor points" by gripping or bracing.

With a snake-like slender, reconfigurable body, snake robots hold the promise as versatile platforms to traverse diverse environments (Hirose, 1993; Nie et al., 2013; Walker et al., 2016) for critical applications like search and rescue and building inspection (Osuka and Kitajima, 2003; Whitman et al., 2018). Similar to snakes, snake robots are inherently stable when they use planar gaits on flat surfaces (Dowling, 1999; Hirose and Mori, 2004) but face stability challenges when they deform out of plane in more complex environments (Cappo and Choset, 2014; Gong et al., 2012; Hatton and Choset, 2010; Marvi et al., 2014; Toyoshima and Matsuno, 2012; Yim et al., 2001). In branch-like terrain and confined spaces and on sandy slopes, snake robots also maintain stability by gripping or bracing against the surfaces or depressed sand (Lipkin et al., 2007; Marvi et al., 2014; Melo and Paez, 2012; Takemori et al., 2018a). Surmounting large, smooth obstacles like steps has often been achieved by using a simple, follow-the-leader gait (Birkenhofer, 2010; Jing et al., 2017; Komura et al., 2015; Kurokawa et al., 2008; Nakajima et al., 2018; Scholl et al., 2000; Takaoka et al., 2011; Takayama and Hirose, 2000; Tanaka and Tanaka, 2013; Tanaka et al., 2018; Yamada and Hirose, 2006), in which the body deforms nearly within a vertical plane with little lateral deformation and hence a narrow base of ground support. Only two previous snake robots deliberately used lateral body deformation for a wide base of support when traversing large steps (Lipkin et al., 2007; Nakajima et al., 2018; Tanaka and Tanaka, 2013). Regardless, all these previous snake robots rely on careful planning and control of motion to maintain continuous static stability and thus often traverse at low speeds. Better understanding of the stability challenges of high speed locomotion over large, smooth obstacles can help snake robots traverse more rapidly and stably.

In a recent study (Gart et al., 2019), our group studied the generalist kingsnake traversing large steps by partitioning its body into sections with distinct functions (**Figure 14**). The body sections below and above the step always undulate laterally on horizontal surfaces to propel the animal forward, while the body section in between cantilevers in the air in a vertical plane to bridge the height increase. An important insight was that lateral body undulation helps maintain stability by creating a wide base of ground support to resist lateral perturbations (**Figure 14**, red regions). Without it, when a long body section is cantilevering in the air but has not reached the upper surface (**Figure 14**B), a significant roll perturbation can tip the animal over. Thanks to body partitioning with lateral undulation, the snake traversed steps as high as 25% body length (or 30% snout-vent length) with perfect stability (Gart et al., 2019). A signature of its perfect stability was that the laterally undulating body sections never lifted off involuntarily from horizontal surfaces before and after cantilevering.

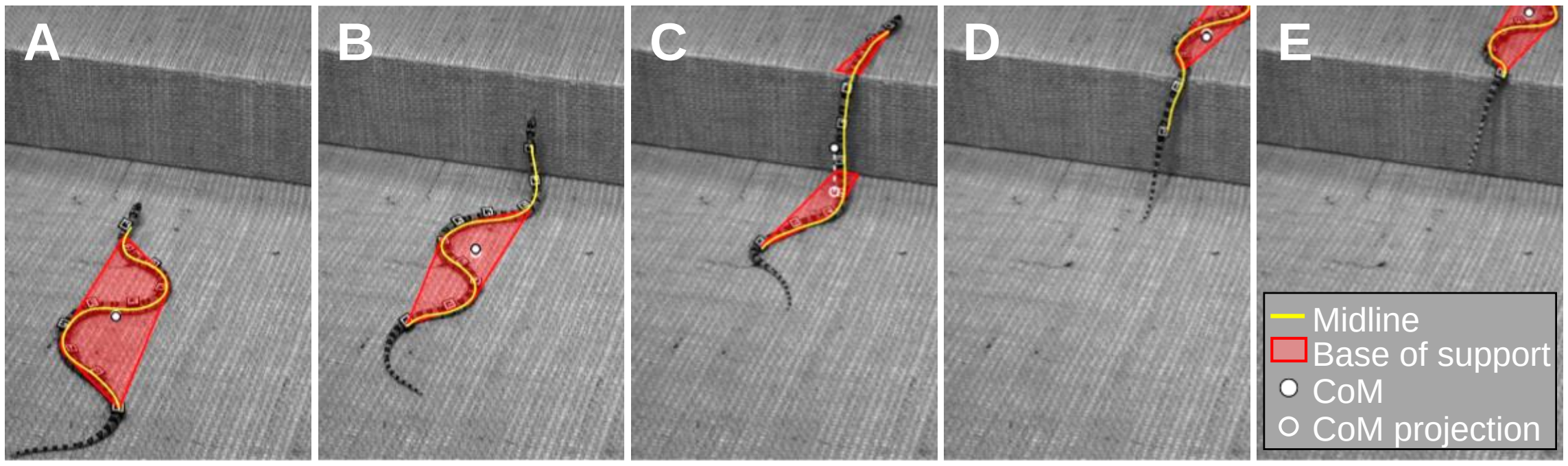

**Figure 14. A snake combines body lateral undulation and cantilevering to traverse a large step stably.** Representative snapshots of a kingsnake traversing a large step (oblique view) with base of support and center of mass overlaid. **(A)** Before cantilevering. **(B)** During cantilevering but before reaching upper surface. **(C)** After reaching upper surface. **(D)** Lifting off lower surface. **(E)** After the entire body reaches upper surface. In each snapshot, yellow curve shows body midline, red polygon shows base of support formed by body sections in contact with horizontal surfaces, white point shows center of mass, and white circle and dashed line show projection of center of mass onto a horizontal surface. Reproduced from (Gart et al., 2019).

In this study, we take the next step in understanding the stability principles of large step traversal using lateral undulation combined with cantilevering, by testing two hypotheses: (1) roll stability diminishes as step becomes higher; and (2) body compliance improves surface contact statistically and reduces roll instability. The kingsnake did not attempt to traverse steps higher than 25% body length (on which it maintains perfect stability) and their body compliance cannot be modified without affecting locomotion. Thus, to test these two hypotheses, we developed a snake robot as a physical model which we could challenge with higher steps and whose body compliance could be modified. The second hypothesis was motivated by the observation during preliminary experiments that the robot with a rigid body often rolled to the extent of involuntary lift-off from horizontal surfaces, in contrast to the snake with compliant body (Penning and Moon, 2017) that never did so (see Chapter 2.4.5 for details).

# 2.4 Physical modeling with rigid snake robot

## 2.4.1 Mechanical design

Our snake robot used the partitioned gait template (**Figure 15**A) from our recent animal observations (Gart et al., 2019). The robot was 107 cm long, 8.2 cm tall, and 6.5 cm wide and weighed 2.36 kg excluding off-board controllers and extension cables. To enable large body deformation both laterally and dorsoventrally for traversing large steps (and complex 3-D terrain in general), the robot consisted of 19 segments with 19 servo motors connected by alternating pitch (red) and yaw (yellow) joints (**Figure 15**B, Movie 1, see details in Chapter 2.8.1), similar to (Kouno et al., 2013). We refer to segments containing pitch or yaw joint servo motors as pitch or yaw segments, respectively.

An anisotropic friction profile, with smaller forward friction than backward and lateral friction, is critical to snakes' ability to move on flat surfaces using lateral undulation (Hu et al., 2009). To achieve this in the robot, we added to each pitch segment a pair of one-way wheels (48 mm diameter, with a rubber O-ring on each wheel) realized by a ratchet mechanism similar to (Chirikjian and Burdick, 1995) (**Figure 15**B, C, blue, Movie 1). The one-way wheels unlocked when rotating forward and locked when rotating backward, resulting in a small forward rolling friction and a large backward sliding friction, besides a large lateral sliding friction. We measured the kinetic friction coefficient at various body orientation (**Figure 15**D, see details in Chapter 2.8.3) and confirmed that forward friction was indeed smaller than backward and lateral friction (**Figure 15**E).

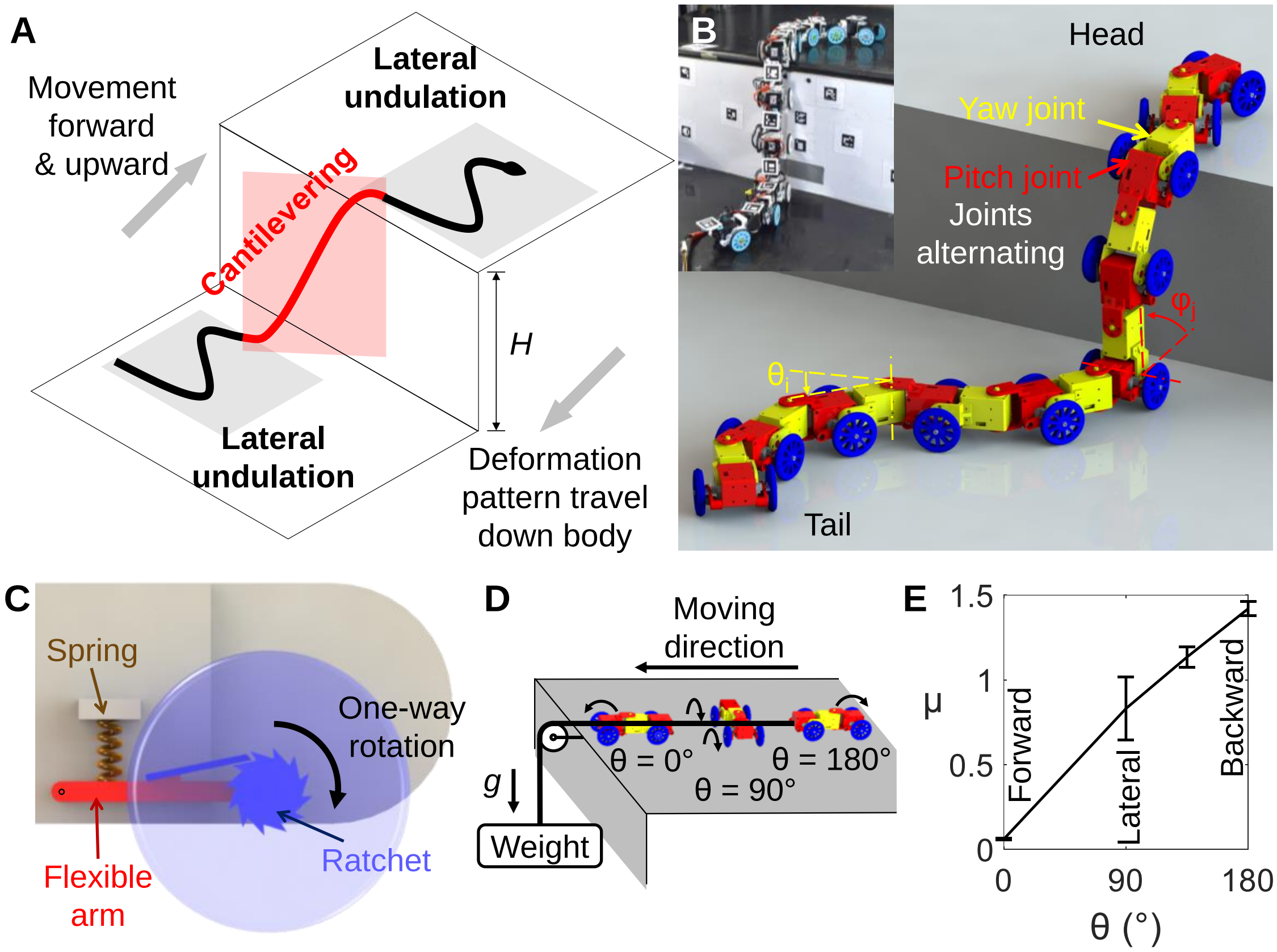

**Figure 15. Gait and mechanical design of snake robot. (A)** Partitioned gait template from kingsnakes combining lateral undulation and cantilevering to traverse a large step (Gart et al., 2019). Lateral undulation can be simply controlled and varied using a few wave parameters, such as wavelength, amplitude, and frequency. **(B)** The snake robot consists of serially connected segments with alternating pitch (red) and yaw (yellow) joints and one-way wheels (blue). **(C)** Close-up view of a one-way wheel (blue) attached to each pitch segment, which only rotates forward via a ratchet mechanism. To add mechanical compliance, wheel connects to body segment via a suspension system with a spring (brown) and a flexible arm (red). Suspension is disabled in rigid robot experiments by inserting a lightweight (0.4 g) block with the same length as natural length of spring. **(D)** Experimental setup to measure kinetic friction coefficient. Three segments are dragged by a weight using a string through a pulley with various orientation angle θ between body long axis and direction of drag force. θ = 0°, 90°, and 180° are for sliding in body's forward, lateral, and backward directions, respectively. Arrow on one-way wheel shows its direction of free rotation. **(E)** Kinetic friction coefficient as a function of body orientation θ. Error bars show ± 1 s.d. See Movie 1 for demonstration of robot mechanisms.

### 2.4.2 Control of body lateral undulation and cantilevering

To generate lateral undulation on the robot's body sections below and above the step, we applied a serpenoid traveling wave spatially on the body shape with sinusoidal curvature (Hirose, 1993) in the horizontal plane, which propagated from the head to the tail. The wave form, with a wavenumber of 1.125, was discretized onto the robot's yaw segments in these two sections, by actuating each yaw joint to follow a temporal sinusoidal wave with an amplitude of 30°, a frequency of 0.25 Hz, and a phase difference of 45° between adjacent yaw joints. The traveling wave form in the section below the step immediately followed that above, as if they formed a single wave, if the cantilevering section was not considered. We chose a serpenoid traveling wave because it is similar to that used by kingsnakes and easy to implement in snake robots (Hirose, 1993; Tanaka and Tanaka, 2015a; Transeth et al., 2007). We chose wave parameters from preliminary experiments and kept them constant in this study to study the effect of step height and body compliance.

To generate cantilevering on the section in between, for each step height tested, we used the minimal number of pitch segments required to bridge across the step. The cantilevering section was kept straight and as vertical as possible, except that the two most anterior pitch segments pitched forward for the anterior undulating body section to gain proper contact with the upper surface (**Figure 24**A, see details in Chapter 2.8.4). This shape was calculated based on the step height measured from online camera tracking before body cantilevering started and remained the same while traveling down the body.

Overall, with this partitioned gait template, control of the robot's many degrees of freedom was simplified to using only a few wave parameters.

To propagate the three partitioned sections down the robot as it moved forward and upward onto the step, we used the measured positions of the robot's segments to conduct feedback logic control (**Figure 24**C) similar to (Tanaka and Tanaka, 2013). An online camera tracked ArUco markers attached to each pitch segment and the step surfaces, and the distance of each segment relative to the step in the forward and upward direction was calculated. This distance was used to determine when each segment should transition from lateral undulation to cantilevering or conversely. Below we refer to this process as section division propagation. See more technical details of robot control in Chapter 2.8.1**.**

Apart from the experimenter starting and stopping it, the robot's motion to traverse the step was automatically controlled by a computer (**Figure 24**B). The experimenter stopped robot when it: (1) flipped over, (2) became stuck for over 10 undulation cycles, or (3) traversed the step.

### 2.4.3 Traversal probability diminishes as step becomes higher

To test our first hypothesis, we challenged the robot to traverse increasingly large, high friction step obstacles (**Figure 23**A), with step height $H$ = 33, 38, 41, and 43 cm, or 31, 36, 38, and 40% of robot length $L$ (see representative trial in **Figure 16**A, Movies 2, and Movie 3, left). Using body lateral undulation combined with cantilevering, the robot traversed a step as high as near a third of body length (31% $L$) with a high probability of 90% (**Figure 16**C, black dashed). In addition, its motion during traversal was more dynamic than previous snake robots that traverse steps using quasi-static motion (Movie 2 and Movie 3, left). However, as it attempted to traverse higher steps, the robot struggled (Movie 3, left) and its traversal probability quickly decreased ($P < 0.005$, simple logistic regression), diminishing to 20% when step height reached 40% $L$.

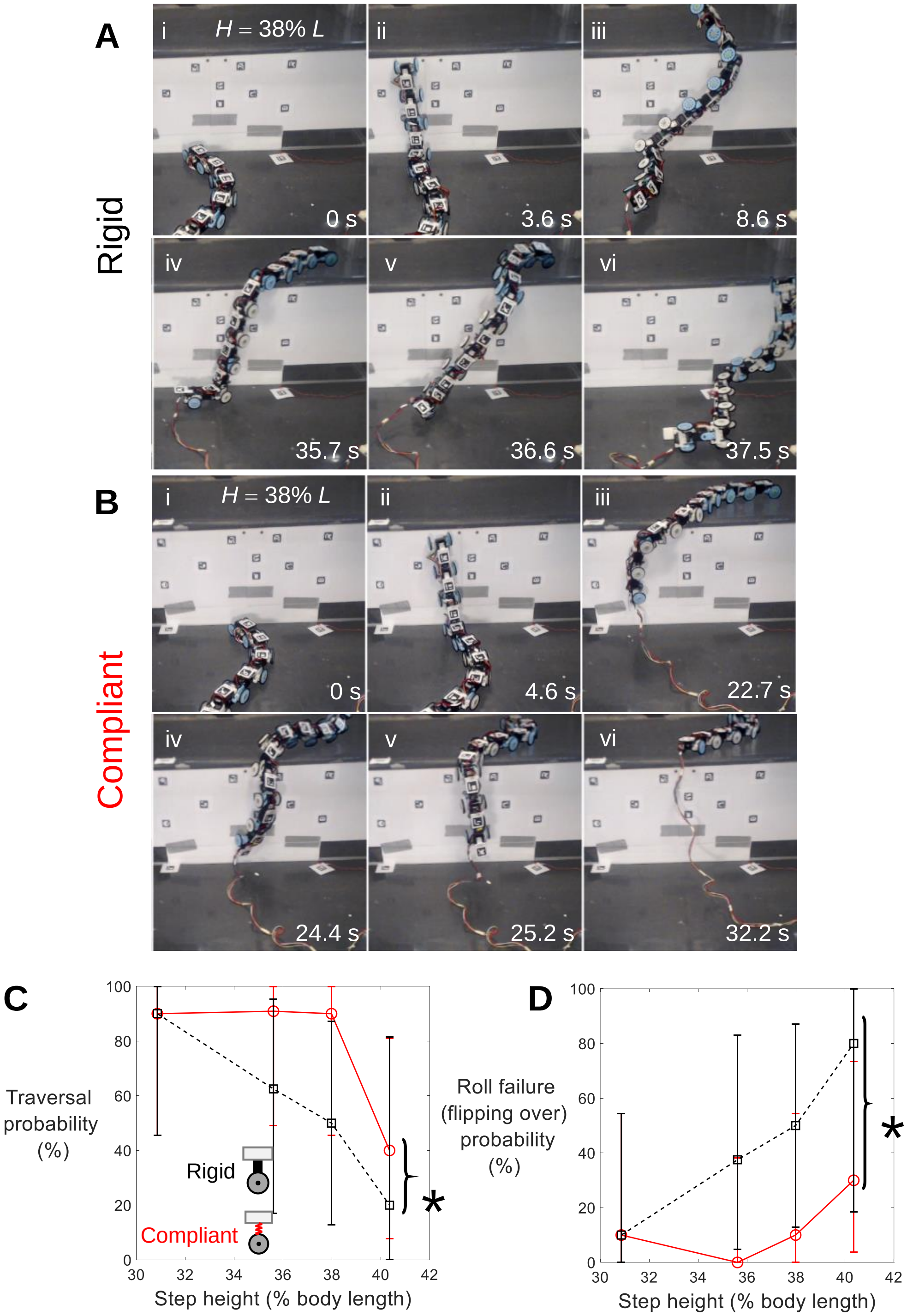

**Figure 16. Traversal performance of robot. (A, B)** Representative snapshots of robot with a rigid body (A) and a compliant body (B) traversing a step as high as 38% body length. Body rolling back and forth (wobble) is visible in (A) iii-vi and (B) iii-v. Rigid body robot failed to recover from rolling

and eventually flipped over; see (A) vi and Movie 3, left for a representative video. Rolling is less severe for compliant body robot, which often recovers or transitions to other events and rarely flips over; see (B) vi and Movie 3, right for a representative video. **(C)** Traversal probability as a function of step height. Bracket and asterisk show a significant difference between rigid and compliant body robot ($P$ < 0.05, multiple logistic regression). **(D)** Effect of body compliance on probability of roll failure (i.e., flipping over, see Chapter 2.5.3). Bracket and asterisk represent a significant difference between rigid and compliant body robot ($P$ < 0.005, multiple logistic regression). In (C) and (D), black dashed is for rigid body robot; red solid is for compliant body robot. Error bars show 95% confidence intervals.

### 2.4.4 Poorer roll stability on higher steps increases failure

To determine whether the diminishing traversal probability was caused by diminishing roll stability, we recorded high speed videos of all experimental trials (**Figure 23**B). Observation of these videos revealed that failure to traverse was a result of one or a sequence of adverse events (**Figure 17**, **Figure 18**, Movie 4).

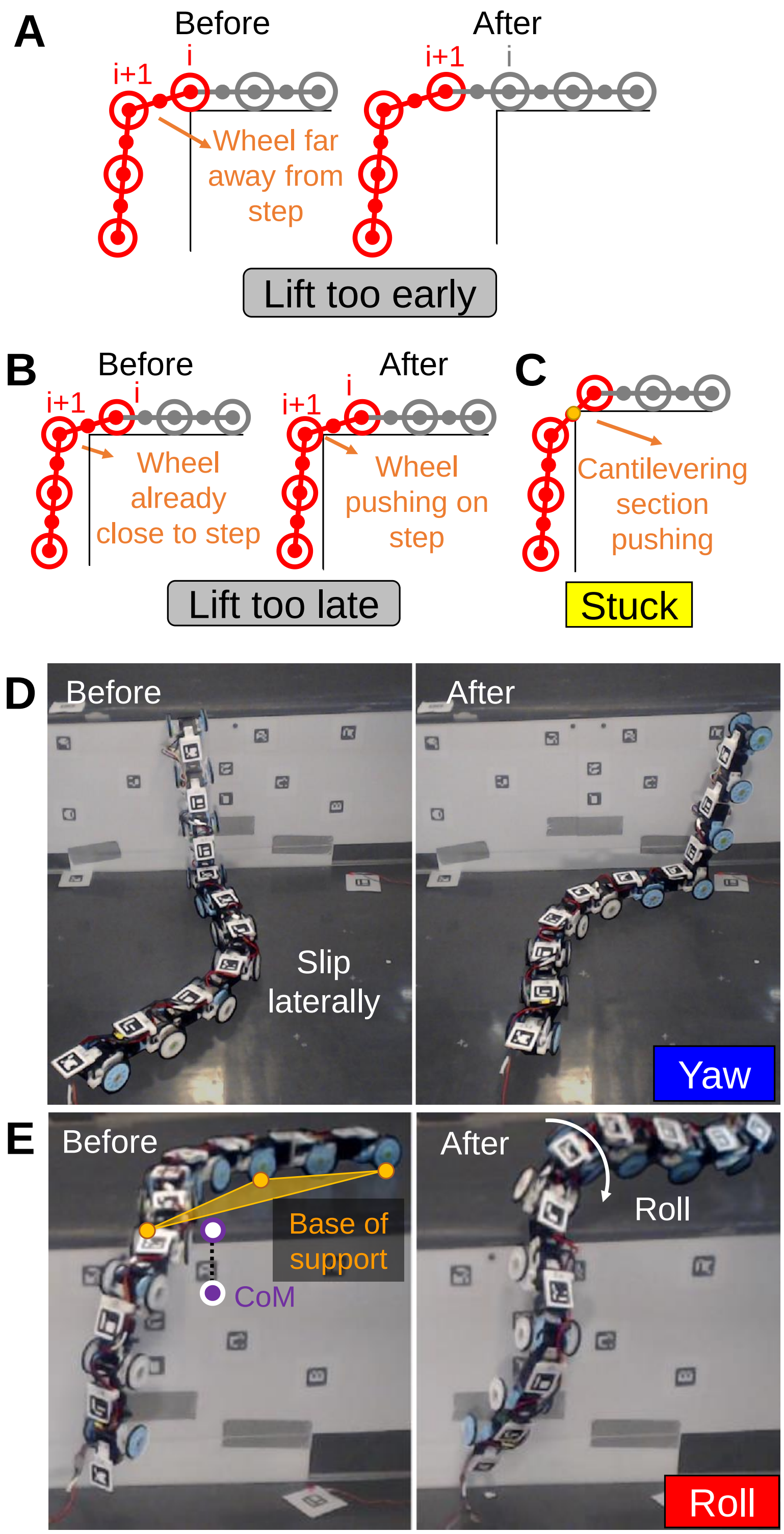

**Figure 17. Adverse events leading to failure. (A)** Lift early: robot lifts up a segment too early before preceding segment moves onto the upper surface. **(B)** Lift late: robot lifts up a segment too

late and a cantilevering segment closest to top edge of step pushes against it. In (A, B), indices above wheels are from head to tail. **(C)** Stuck: robot becomes stuck when cantilevering section pushes against step, with no overall direction or position change. This happens when robot belly instead of wheels contacts top edge of step. In (A-C), red and gray sections are cantilevering and undulating sections. **(D)** Yaw: robot yaws due to lateral slipping of body sections below and/or above step. **(E)** Roll: robot rolls with loss of contact below and/or above step, purple and white points are the center of mass (CoM) and its projection onto the upper horizontal surface. In (C) and (E), orange points are contact points, and orange shade is base of support. See Movie 4 for examples of (C-E).

(1) Imperfect lift timing (**Figure 17**A-B; **Figure 18**, gray). This includes lifting too early (**Figure 17**A) or late (**Figure 17**B) due to inaccurate estimation of body forward position relative to the step. Noise in the system, both mechanical (e.g., variation in robot segment and surface friction) and in feedback control (e.g., camera noise, controller delay), resulted in trial-to-trial variation of robot motion and interaction with the step, leading to inaccurate position estimation. With underestimation, a segment still far away from the step was lifted too early (**Figure 17**A, segment between $i$ and $i + 1$). With overestimation, a segment close to the step was lifted too late and pushed against the step (**Figure 17**B, segment between $i$ and $i + 1$). These control imperfections often triggered other adverse events stochastically, as described below.

(2) Stuck (**Figure 17**C; **Figure 18**, yellow). The robot was occasionally stuck when its cantilevering section pushed against the step with no substantial body yawing or rolling. After becoming stuck, the robot always eventually recovered and succeeded in traversing within 10 undulation periods (**Figure 18**, no purple arrows).

(3) Yawing (**Figure 17**D; **Figure 18**, blue). The robot often yawed substantially when the sections below and/or above step slipped laterally. This was always triggered by imperfect lift timing (**Figure 18**, arrows from gray to blue box) which led the cantilevering section to push against the vertical surface (even with lifting too early, the extra weight suspended in the air sometimes resulted in sagging of the cantilevering section, which in turn pushed against the vertical surface). The push resulted in a yawing torque too large to be overcome by the frictional force on the undulating sections from the horizontal surfaces. Because of this, yawing was often accompanied by small lift-off and slip of the undulating segments, which could lead to rolling (**Figure 18**, blue arrows), described below. Yawing could also compromise segment position estimation and sometimes led to further imperfect lift timing (**Figure 18**, arrows from blue to gray box).

(4) Rolling (**Figure 17**E; **Figure 18**, red). The robot rolled about the fore-aft axis substantially as the center of mass (**Figure 17**E, purple point) projection (white point) moved out of the base of support (orange shade). Sometimes, this instability was induced by a sudden shift of center of mass position during segment lifting. At other times, this instability was induced by sudden shrinking of base of support due to loss of surface contact, which resulted from either small lift-off and/or slip of segments due to yawing or the last segment lifting off the lower surface. When rolling occurred, the robot suddenly lost many contacts from the horizontal surfaces and often lost thrust and stability. Sometimes the robot could roll back by its own bending motion (**Figure 16**B iv-v). If not, it would flip over (and sometimes fell off the step) (**Figure 16**A, v-vi; Movie 4), resulting in failure to traverse (**Figure 18**, red arrows). Hereafter, we refer to the robot flipping over due to rolling as roll failure.

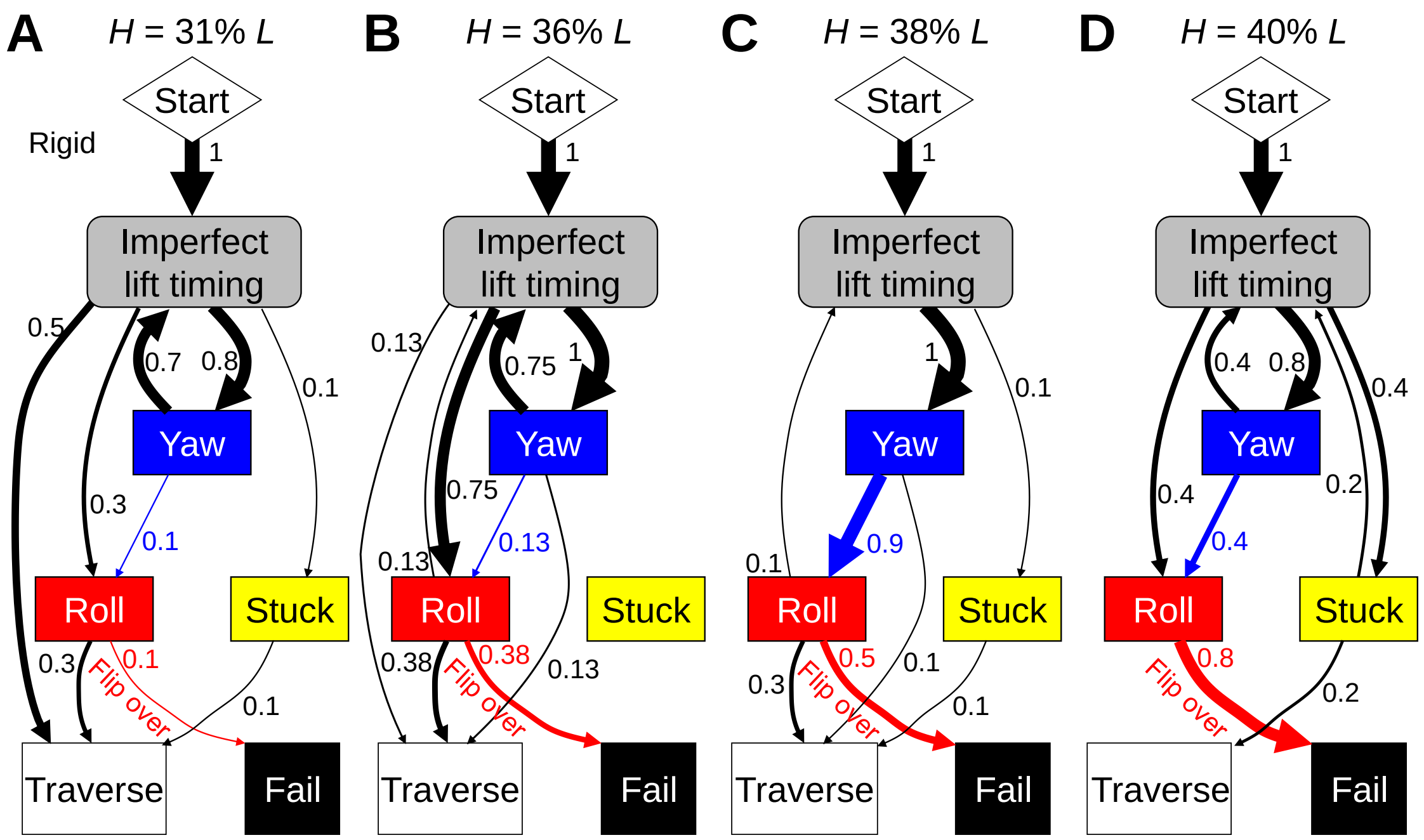

**Figure 18. Transition pathways of rigid body robot among adverse events to traverse or fail. (A)** Step height *H* = 31% robot length *L*. **(B)** *H* = 36% *L*. **(C)** *H* = 38% *L*. **(D)** *H* = 40% *L*. Each arrow is a transition between nodes, with arrow thickness proportional to its probability of occurrence, shown by number next to it. Probability of occurrence here is the ratio of the number of trials in which a transition occurs to the total number of trials; it is different from transition probability in Markov chains. If a transition occurs multiple times in a trial, it is only counted once.

For all step heights tested, we observed a diversity of pathways stochastically transitioning among these adverse events (**Figure 18**). Given the diverse, stochastic transitions, statistical trends emerged in their pathways. First, failure was always directly resulting from rolling to the extent of flipping over, i.e. roll failure (**Figure 18**, red arrows). In addition, as step height increased from 31% *L* to 40% *L*, roll failure (flipping over) became more likely, from 10% to 80% probability (**Figure 18**A-D, red arrows; **Figure 16**D, black; $P < 0.05$, simple logistic regression), which resulted in decreasing traversal probability from 90% to 20% (**Figure 16**C, black dashed). This confirmed our first hypothesis that increasing step height diminishes roll stability. This diminish was a direct result of the shorter undulating body sections for lateral support as the cantilevering body section lengthened as step height increased.

### 2.4.5 Comparison with snakes

Comparison of robot with animal observations (Gart et al., 2019) revealed and elucidated the snake's better ability to maintain stability over the robot. First, the robot always suffered imperfect body lift timing (**Figure 18**, arrows from start to gray box), which was rarely observed in the snake (Gart et al., 2019). Second, the robot's laterally undulating body sections often suffered large yawing (≥ 80% probability, **Figure 18**, blue) and rolling (≥ 40%, **Figure 18**, red). In contrast, the snake's undulating body sections rarely rolled on the horizontal surfaces, even when step friction was low and the animal slipped substantially (Gart et al., 2019). Third, when step friction was high, the robot sometimes became stuck (**Figure 18**, yellow), whereas the snake always smoothly traversed (Gart et al., 2019). These indicate that the snake is better at accommodating noise and perturbations in control, design, and the environment (e.g., improper timing, unexpected forces and slippage, variation in step surface height and friction) to maintain effective body-terrain interaction.

Besides the animal's better ability to use sensory feedback to control movement in complex environments (Dickinson et al., 2000), two morphological features of the snake body likely contributed to its better ability to maintain stability—being more continuous (over 200 vertebrae (Voris, 1975) vs. the robot's 19 segments) and more compliant (Penning and Moon, 2017). The latter is particularly plausible considering that the introduction of mechanical compliance to end effectors has proven crucial in robotic tasks where contact with the environment is essential, e.g., grasping (Mason et al., 2012;

Shimoga and Goldenberg, 1996; Truby et al., 2019), polishing (Furukawa et al., 1996), and climbing (Asbeck, 2010; Ruotolo et al., 2019) robots (for a review, see (Hawkes and Cutkosky, 2018)). Although many snake robots have used compliance in control to adapt overall body shape to obstacles (Gong et al., 2014; Kano and Ishiguro, 2013; Takemori et al., 2018b; Travers et al., 2018), the use of mechanical compliance to better conform to surfaces locally was less considered (Takemori et al., 2018b; Togawa et al., 2000), especially for improving stability. These considerations inspired us to test our second hypothesis that body compliance improves surface contact statistically and reduces roll instability.

## 2.5 Physical modeling with compliant snake robot

### 2.5.1 Suspension to add mechanical compliance

To test our second hypothesis, we added mechanical compliance to the robot by inserting between each one-way wheel and its body segment a suspension system inspired by (Togawa et al., 2000) (**Figure 15**C). The suspension of each wheel (even the left and right on the same segment) could passively compress independently to conform to surface variation (by up to 10 mm displacement of each wheel). From our second hypothesis that body compliance improves surface contact statistically and reduces roll instability, we predicted that this passive conformation would increase traversal probability and reduce roll failure (flipping over) probability, especially for larger steps. The suspension system was present but disengaged in the rigid robot experiments for direct comparison.

### 2.5.2 Body compliance increases traversal probability

The compliant body robot maintained high traversal probability over a larger range of step height than the rigid robot, consistently succeeding 90% of the time as step height increased from 31% *L* to 38% *L* (**Figure 16**C, red solid). Like the rigid body robot, the compliant body robot's motion during traversal was also more dynamic than previous robots using quasi-static gaits (Movie 3, right). Traversal probability only decreased for step height beyond 38% *L* ($P < 0.05$, pairwise Chi-square test). For the large step heights tested, adding body compliance increased traversal probability ($P < 0.05$, multiple logistic regression). These observations were in accord with our prediction from the second hypothesis.

### 2.5.3 Body compliance reduces roll failure probability

To test our second hypothesis, specifically that body compliance reduces roll instability, we compared the compliant body robot's transition pathways among adverse events (**Figure 19**) to those of the rigid body robot (**Figure 18**). The compliant body robot still stochastically transitioned among adverse events. However, two improvements were observed in the statistical trends of the transition pathways.

First, the compliant body robot suffered roll failure (flipping over) less frequently (**Figure 18** vs. **Figure 19**, red arrows; **Figure 16**D; $P < 0.005$, multiple logistic regression), especially after yawing occurred (**Figure 18** vs. **Figure 19**, blue arrows). It also experienced less frequent back and forth rolling (wobbling) (**Figure 16**B; Movie 3, right) than the rigid body robot (**Figure 16**A; Movie 3, left). This confirmed our hypothesis that body compliance reduces roll instability. However, body compliance did not eliminate rolling, and the compliant robot still slipped frequently. These observations were in accord with our prediction from the second hypothesis.

Second, the compliant body robot suffered imperfect lift timing less frequently than the rigid body robot (**Figure 18** vs. **Figure 19**, green arrows), which eventually resulted in more frequent traversal for all step heights above 31% *L* (**Figure 18** vs. **Figure 19**, sum of all black arrows into white box). This was because, even when lifting was too early or too late, the compliant body robot could often afford to push against the vertical surface without triggering catastrophic failure from yawing or rolling before section division propagation resumed.

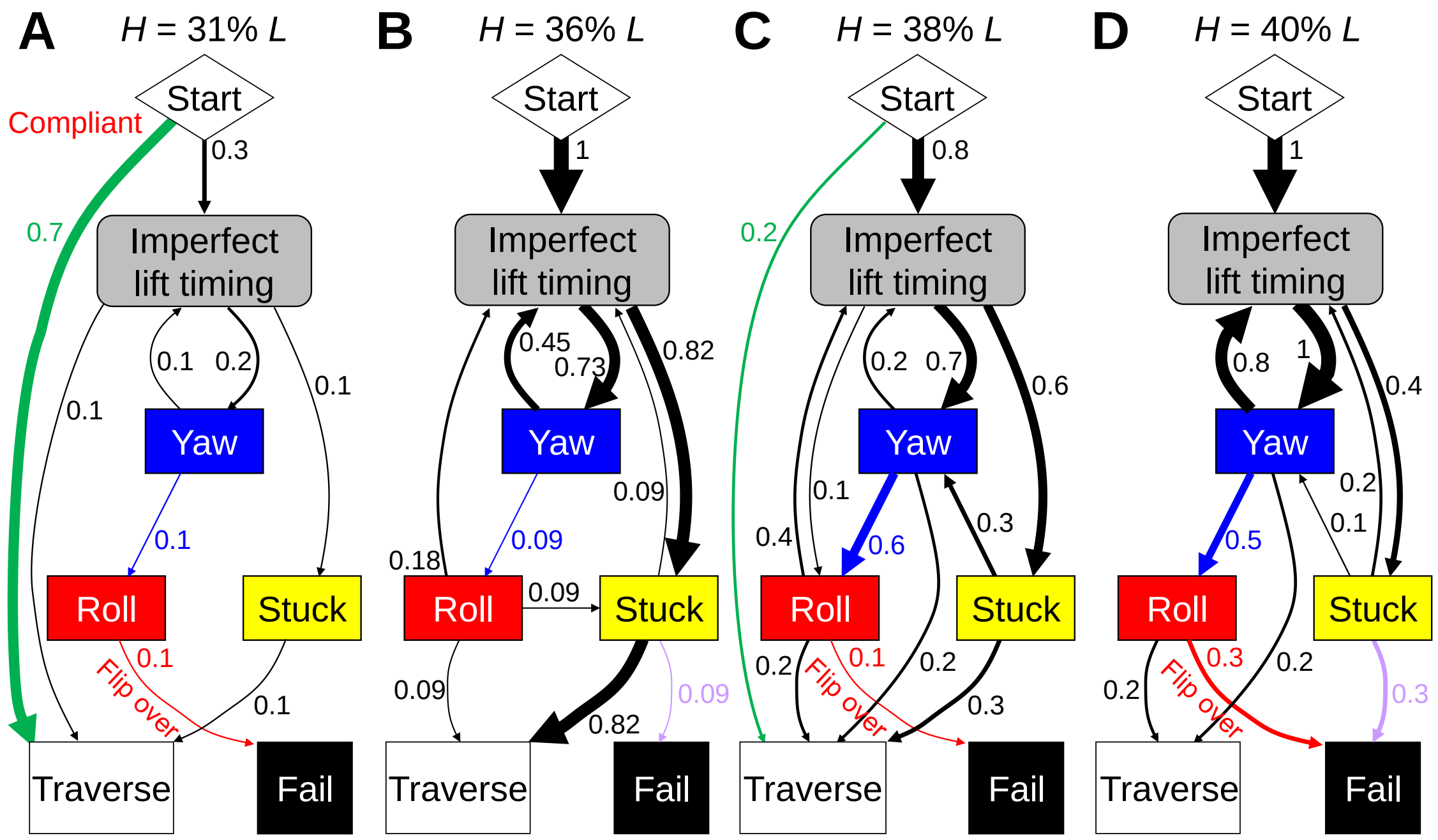

**Figure 19. Transition pathways of compliant body robot among adverse events to traverse or fail. (A)** Step height *H* = 31% robot length *L*. **(B)** *H* = 36% *L*. **(C)** *H* = 38% *L*. **(D)** *H* = 40% *L*. See **Figure 18** for definition of transition diagrams.

However, the compliant body robot became stuck more frequently (**Figure 18** vs. **Figure 19**, arrow from gray to yellow box) and failed more frequently as a result (**Figure 18** vs. **Figure 19**, purple arrow). This was because compression of the suspension lowered ground clearance of the segments. This stuck failure mode was always directly triggered by lifting too early. This is a limitation of the robot's discrete, few degree-of-freedom body.

## 2.5.4 Body compliance improves contact statistically

To further test our second hypothesis, specifically that body compliance improves surface contact statistically, we compared contact probability between the rigid and compliant body robot. Contact probability was defined as the ratio of the number of wheels contacting horizontal surfaces in the laterally undulating body sections to the total number of wheels in these two sections (**Figure 20**A). In addition, we calculated body deformation as how much each wheel suspension was compressed for these two sections (**Figure 20**B). Both wheel contact and body deformation were determined by examining whether any part of each wheel penetrated the step surface assuming no suspension compression, based on 3-D reconstruction of the robot from high speed videos. Both were averaged spatiotemporally over the traversal process across all pitch segments in these two sections combined for each trial. See details in Chapter 2.8.10.

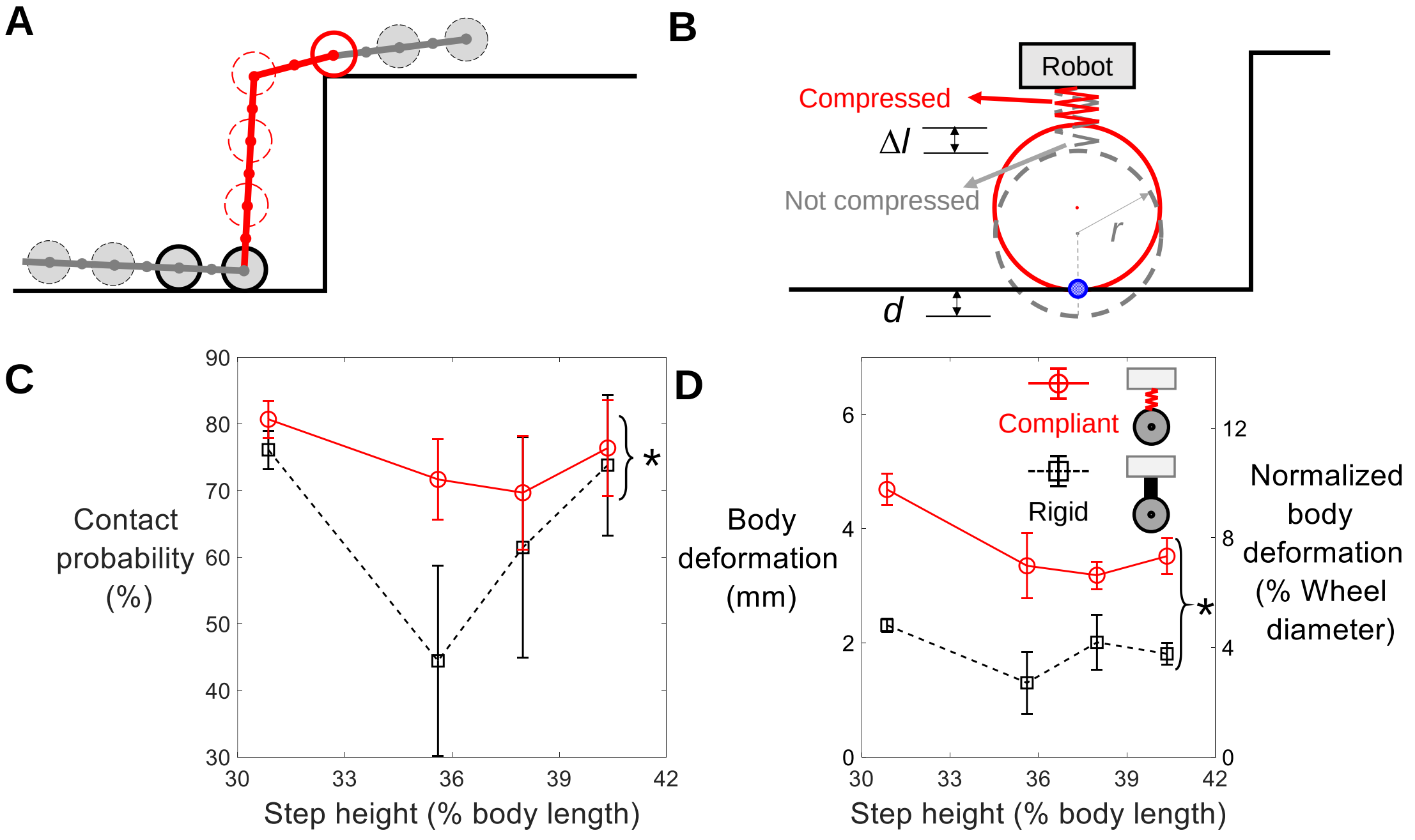

**Figure 20. Effect of body compliance on contact probability and body deformation. (A)** Example side view schematic to define contact probability. Gray are laterally undulating body

sections and red is cantilevering body section. In this example, three wheels of undulating section are in contact with surface (solid) and four are not (dashed), and contact probability = 3/(3 + 4) = 43%. **(B)** Definition of body deformation. Red solid schematic shows actual wheel position with suspension compressed, and gray dashed one shows wheel position assuming no suspension compression. Blue circle shows point of contact between wheel and surface. Body deformation $\Delta l$ is approximated by virtual wheel penetration $d$ into surface. **(C)** Contact probability as a function of step height. **(D)** Body deformation as a function of step height. In (C) and (D), black dashed is for rigid body robot; red solid is for compliant body robot. Error bars show ± 1 s.d. Brackets and asterisks represent a significant difference between rigid and compliant body robot ($P < 0.001$, ANCOVA).

For all step heights tested, the compliant body robot had a higher contact probability than the rigid body robot (**Figure 20**C; $P < 0.001$, ANCOVA). This improvement was a direct result of larger body deformation (**Figure 20**D; $P < 0.0001$, ANCOVA): the rigid body robot only deformed around 2 mm or 4% wheel diameter (which occurred in the rubber on both the wheels and step surfaces); by contrast, the compliant robot's suspension deformed around 4 mm or 8% wheel diameter, a 100% increase.

### 2.5.5 Body compliance reduces severity of body rolling

Body rolling would result in lateral asymmetry in how the robot conforms with the surface between the left and right sides of the body. Thus, to quantify the severity of body rolling, we calculated the difference (absolute value) in surface conformation between left and right wheels (**Figure 21**) excluding the cantilevering section. Surface conformation was defined as the virtual penetration of a wheel in contact (positive distance) or the minimal distance from the surface of a wheel lifted off (negative distance) (**Figure 21**A inset, right or left wheel). A larger difference (absolute value) in surface conformation between left and right sides means more severe rolling. Surface conformation difference was averaged spatiotemporally over the traversal process across all pitch segments separately for the body sections below and above the step for each trial. See details in Chapter 2.8.10.

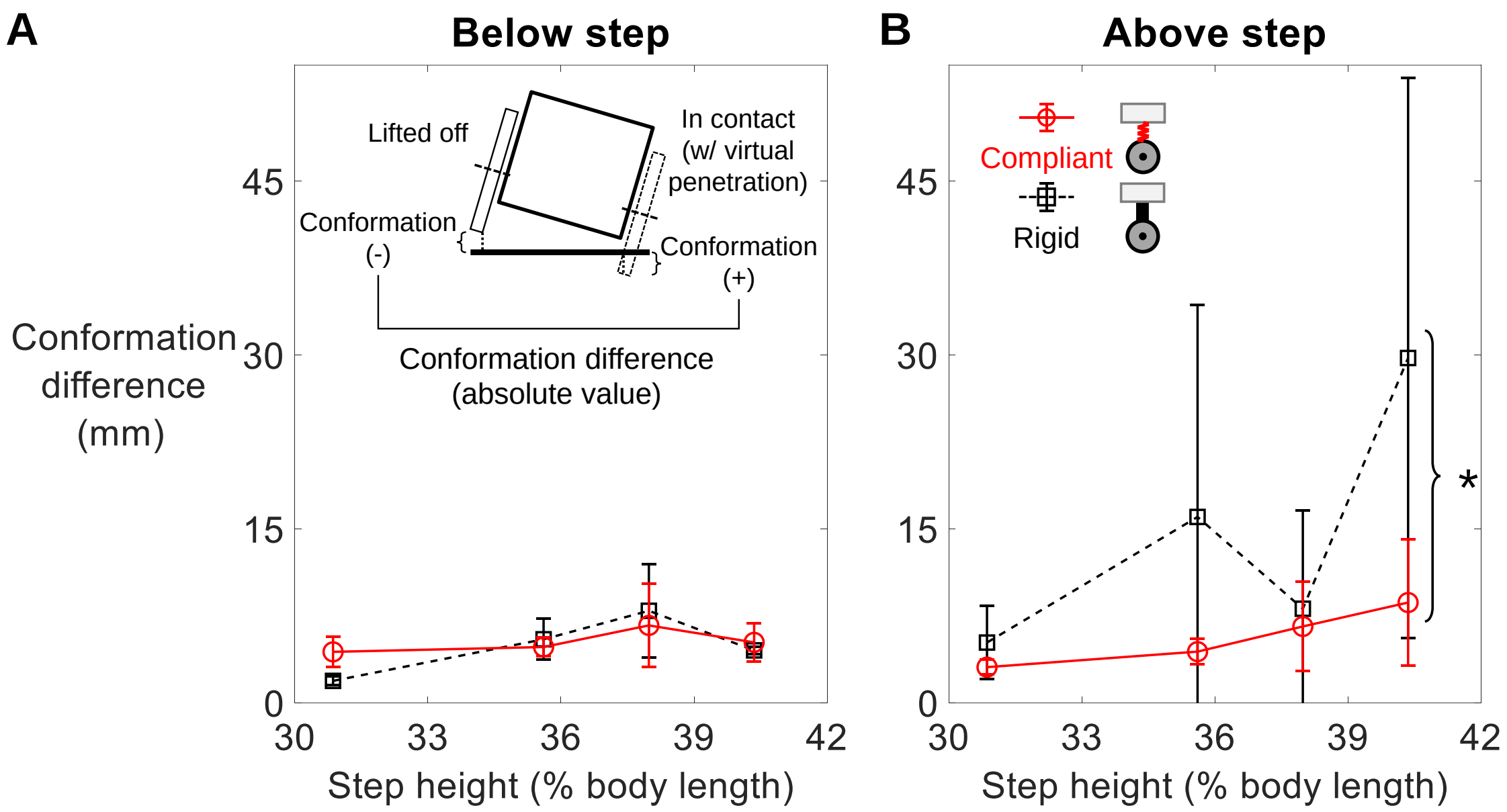

**Figure 21. Effect of body compliance on surface conformation difference. (A, B)** Surface conformation difference between left and right wheels as a function of step height for body sections below (A) and above (B) step. Black dashed is for rigid body robot; red solid is for compliant body robot. Error bars show ± 1 s.d. Brackets and asterisks represent a significant difference between rigid and compliant body robot ($P < 0.005$, ANCOVA). Inset in (A) shows front view schematic to define surface conformation difference (see text for detail).

For all step heights tested, body compliance reduced lateral asymmetry in surface conformation for the body section above the step (**Figure 21**B; $P < 0.005$, ANCOVA), although not for the section below (**Figure 21**A; $P > 0.05$, ANCOVA). This means that the compliant body robot had less severe body rolling above the step (**Figure 16**B; Movie 3, right) and was more stable during traversal. Such better surface conformation likely allowed the compliant body robot to generate ground reaction forces more evenly along the body (Kim et al., 2007) to better propel itself forward and upward, which the rigid robot with poorer surface conformation struggled to do. The compliant body robot still wobbled and slipped more substantially than the snakes (Gart et al., 2019).

All these observations from the compliant body robot confirmed our second hypothesis that body compliance improves surface contact statistically and reduces roll instability.

### 2.5.6 Body compliance increases energetic cost

Not surprisingly, these benefits came with a price. The electrical power consumed by the robot increased when body compliance was added (**Figure 25**; $P < 0.0001$, ANCOVA; see details in Chapter 2.8.11). We speculate that this was due to an increase in energy dissipation from larger friction dissipation against the surfaces due to higher contact probability, viscoelastic response (Hawkes and Cutkosky, 2018) of the suspension, and more motor stalling and wheel sliding due to more frequently getting stuck. Electrical power consumption decreased with step height (**Figure 25**; $P < 0.0001$, ANCOVA), which may result from the decrease in the number of laterally undulating segments that dissipated energy during sliding against the surfaces. We noted that the electrical energy consumed (power integrated over time) during traversal was two orders of magnitude larger than the mechanical work needed to lift the robot onto the step; the majority of the energy was not used to do useful work.

## 2.6 Contribution to robotics

Our study advanced the performance of snake robots traversing large steps. When normalized to body length, our robot (both rigid and compliant body, **Figure 22**, red and black) achieved step traversal speed higher than most previous snake robots (gray) and approaching that of kingsnakes (blue). In addition, our compliant robot maintained high traversal probability (90%) even when the step was as high as 38% body length (**Figure 16**C, red solid), without loss of traversal speed compared to the rigid body robot ($P > 0.05$, pairwise two-sample *t*-test). These improvements were attributed to the inherent roll stability from body lateral undulation and an improved ability to maintain surface contact via body compliance, which alleviated the need for precise control for static stability and enabled dynamic traversal. The only other snake robot that achieved comparable step traversal speed ($T^2$ Snake-3 (Tanaka et al., 2018)) used active wheels to drive the body while only deforming in a vertical plane and thus had low roll stability. It also pushed against and propelled its wheels up the vertical surface, which reduced the load of pitching segments and increased the maximal cantilevering length. We note that our robot still has the potential to achieve even higher speeds with high traversal probability, because in our experiments the motors were actuated at only 50% full speed to protect the robot from breaking after drastically flipping over as well as motor overload due to inertial forces and we have yet to systematically test and identify optimal serpenoid wave parameters (Saito et al., 2002).

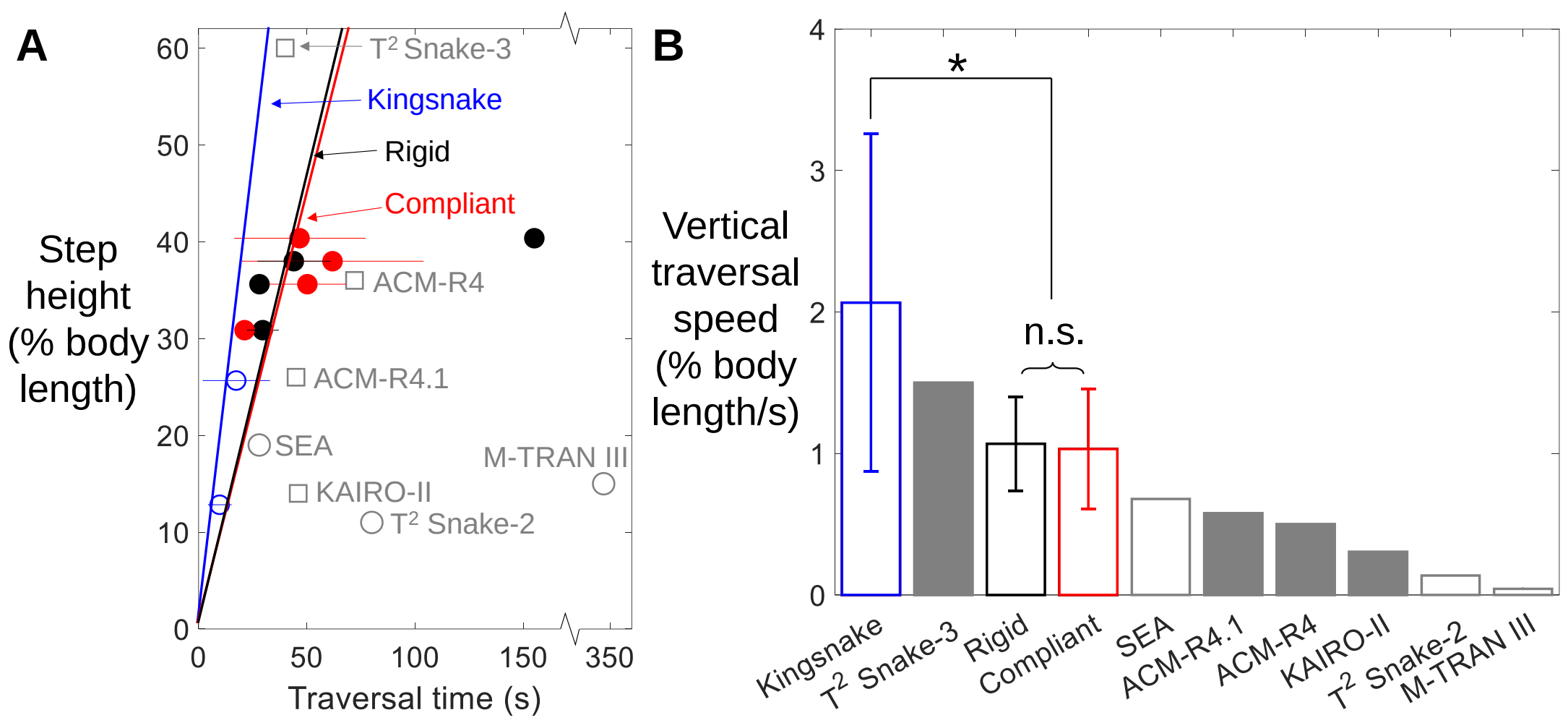

**Figure 22. Comparison of traversal performance of our robot with previous snake robots and the kingsnake. (A)** Maximal traversable step height (normalized to body length) as a function of traversal time for kingsnake (blue), our robot with rigid (black) and (red) compliant body, and previous snake robots with data available (gray squares: with active propellers; gray circles: no active propellers). Several previous robots with no traversal time reported are not included (Borenstein and Hansen, 2007; Kouno et al., 2013; Osuka and Kitajima, 2003; Scholl et al., 2000; Suzuki et al., 2012; Takayama and Hirose, 2000). **(B)** Vertical traversal speed normalized to body length. Vertical traversal speed, i.e., normalized step height divided by traversal time, is the slope of lines connecting each data point to the origin in (A). Thus, a higher slope indicates a larger vertical traversal speed. Speeds of previous robots are the fastest reported values from (Birkenhofer, 2010; Kurokawa et al., 2008; Lipkin et al., 2007; Takaoka et al., 2011; Tanaka and Tanaka, 2013; Tanaka et al., 2018; Yamada and Hirose, 2006) or accompanying videos. See Chapter 2.8.10 for details of speed calculation. Bracket and asterisk represent a significant difference in vertical traversal speed ($P < 0.005$, pairwise two-sample $t$-test). n.s. represents no significant difference ($P > 0.05$, pairwise two-sample $t$-test). In (A, B), error bars show ± 1 s.d.

## 2.7 Summary and future work

Inspired by our recent observations in snakes, we developed a snake robot as a physical model and performed systematic experiments to study stability principles of large step traversal using a partitioned gait that combines lateral body undulation and cantilevering. Our experiments confirmed two hypotheses: (1) roll stability diminishes as step becomes higher; and (2) body compliance improves surface contact statistically and reduces roll instability. In addition, thanks to the integration of lateral body undulation to resist roll instability with anisotropic friction for thrust, our snake robot traversed large step obstacles more dynamically than previous robots with higher traversal speeds (normalized to body length), approaching animal performance. Moreover, by further adding body compliance to improve surface contact, our snake robot better maintained high traversal probability on high steps without loss in traversal speed. Although our discoveries were made on a simple large step with only vertical body compliance, the use of body lateral undulation and compliance to achieve a large base of support with reliable contact for roll stability is broadly useful for snakes and snake robots traversing other large, smooth obstacles in terrain like non-parallel steps (Nakajima et al., 2018; Tanaka et al., 2018), stairs (Komura et al., 2015; Lipkin et al., 2007; Tanaka et al., 2018), boulders, and rubble (Takemori et al., 2018b; Whitman et al., 2018).

Given these advances, the snake's locomotion over large obstacles is still superior, without any visible wobble or slip on high friction steps (Gart et al., 2019). This is likely attributed to the animal's more continuous body, additional body compliance in other directions (e.g., rolling, lateral), and ability to actively adjust its body (Jayne, 1988) using sensory feedback (Proske, 1969) to conform to the terrain beyond achievable by passive body compliance. Future studies should elucidate how snakes, and how snake robot should, use tactile sensory feedback control (Kano and Ishiguro, 2013; Kano and Ishiguro, 2020; Kim et al., 2007; Liljebäck et al., 2014b; Yim et al., 2006) and combine control compliance (Kano and Ishiguro, 2013; Takaoka et al., 2011) with mechanical compliance (Hawkes and Cutkosky, 2018; Kim et al., 2007; Roberts and Koditschek, 2019) along multiple directions (Ruotolo et al., 2019) to stably traverse large, smooth obstacles.

Finally, as our study begins to demonstrate, locomotion in 3-D terrain with many large obstacles often involves stochastic transitions, which are statistically affected by locomotor-terrain physical interaction (Byl and Tedrake, 2009; Gart and Li, 2018; Han et al., 2021; Li et al., 2015) (e.g., step height and body compliance here). A new statistical physics view of locomotor transitions (Othayoth et al., 2020) will help accelerate the understanding of how animals use or mitigate such statistical dependence and its application in robotic obstacle traversal using physical interaction in the stochastic world.

## 2.8 Materials and Methods

### 2.8.1 Robot parts

The robot was actuated with 19 Dynamixel XM430-W350-R servo motors operating at 14 V, powered by an external DC power supply (TekPower, CA, USA). The rubber O-rings wrapping each wheel were oil-resistant soft buna-n O-rings with an outer diameter of 48.1 mm and a width of 5.3 mm (McMaster-Carr, Elmhurst, IL, USA). The springs used in the suspension were compression springs with a length of 9.5 mm and an outer diameter of 3.1 mm (McMaster-Carr, Elmhurst, IL, USA). The maximal compression of each spring was 5 mm, which, when amplified by the lever arm (**Figure 15**, red), limited the suspension deformation of each wheel to within 10 mm.

### 2.8.2 Large step obstacle track

We constructed a 180 cm long, 120 cm wide obstacle track using extruded T-slotted aluminum and acrylic sheets (McMaster-Carr, Elmhurst, IL, USA) (**Figure 23**A). The step spanned the entire width of the track. To reduce slipping of the robot, we covered the horizontal surfaces of the step with a high friction rubber sheet (EPDM 60A 1.6 mm thick rubber sheet, Rubber-Cal, CA, USA).

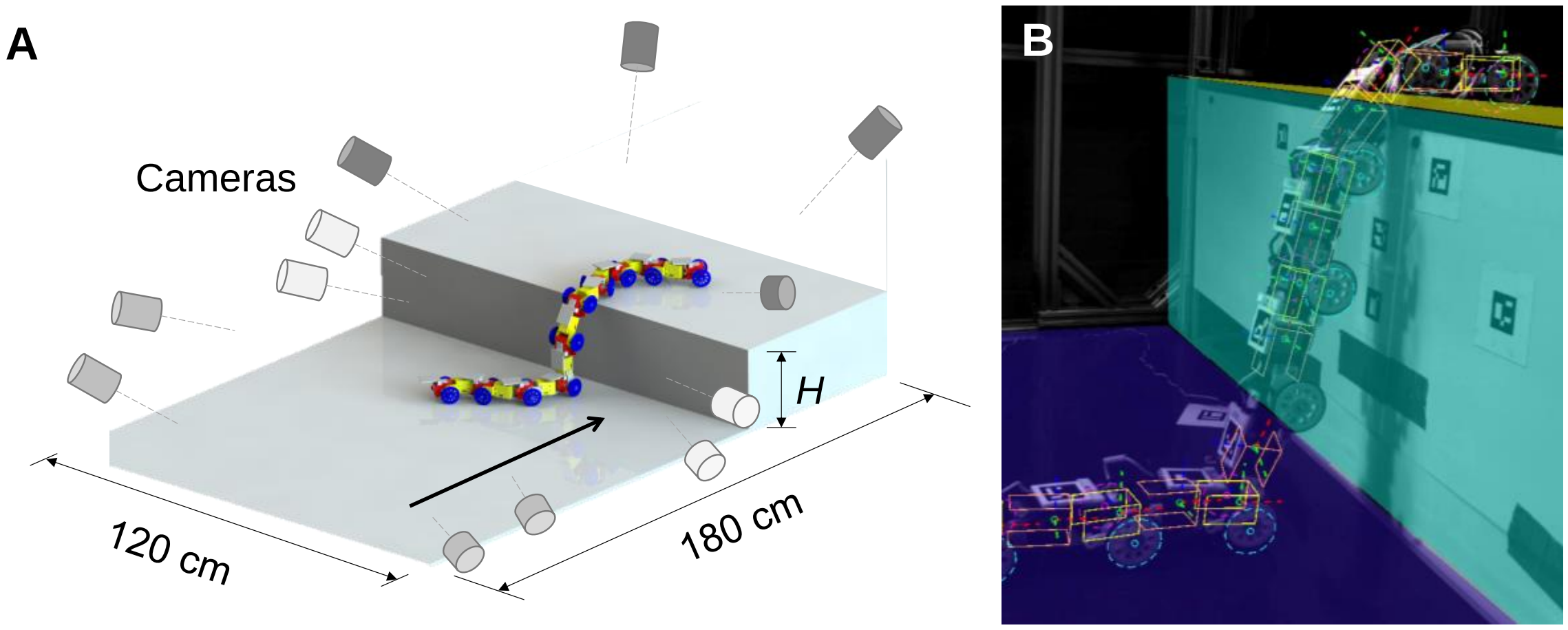

**Figure 23. Experimental setup and 3-D kinematics reconstruction. (A)** Schematic of experimental setup. Twelve high-speed cameras are used for 3-D kinematics reconstruction, divided into groups of four (different shades) focusing on three step surfaces. **(B)** High-speed video snapshot of robot traversing step, with projection of reconstructed body segments, wheels, and step surfaces. Yellow and orange boxes are reconstructed yaw and pitch servo motors. Dashed magenta and cyan circles are reconstructed left and right wheels assuming no suspension compression. Violet, cyan, and gold surfaces are reconstructed lower horizontal, vertical, and upper horizontal surfaces.

### 2.8.3 Friction measurement

In friction experiments, we measured the position as a function of time of three body segments being dragged by a weight, by tracking ArUco tags in videos captured by Logitech C920 webcam at 30 frames/s. Then, by fitting a quadratic function of displacement as a function of time to estimate acceleration, we calculated kinetic friction coefficient as:

$$\mu = \frac{m_2 g - (m_1 + m_2)a}{m_1 g}$$

where $m_1$ is the mass of the weight, $m_2$ is the total mass of the segments, $a$ is the fitted acceleration, $g$ is the local gravitational acceleration (9.81514 m/s$^2$).

### 2.8.4 Motor actuation to achieve partitioned gait

The actuation profile of the yaw joints in the laterally undulating body sections, defined as the angular displacement from the straight body pose (**Figure 15**B, yellow angle) as a function of time and segment index, followed the serpenoid gait (Hirose, 1993):

$$\theta_i = \begin{cases} A\sin(\omega t + \phi + (i-1)\Delta\phi), i = 1,2, \dots, k_1 \\ A\sin(\omega t + \phi + (i + k_1 - k_2)\Delta\phi), i = k_2, \dots, 9 \end{cases}$$

where $i$ is for the $i$th yaw joint from the robot head, $A = \pi/6$ and $\omega = \pi/2$ are the amplitude and angular velocity of each yaw joint angle waveform, $\phi = 0$ is the initial phase (at time zero) of the first yaw joint, and $\Delta\phi = \pi/4$ is the phase difference between adjacent yaw joints. $\Delta\phi$ determines the wavenumber of the entire serpenoid wave in the robot, $k = 9|\Delta\phi|/2\pi$. The $k_1$th yaw joint is the last yaw joint in the undulating section above the step, and the $k_2$th yaw joint is the first yaw joint in the undulating section below the step, $k_2 - k_1$ is the number of pitch segments in the cantilevering section. The pitch angles of all pitch segments in these two undulating sections were set to zero (**Figure 24**A, gray) to maintain contact with the horizontal surfaces.

The actuation profile of the joints of the cantilevering section (**Figure 24**A, red) was designed to bridge across the large step with the minimal number of segments necessary. The yaw angles of all yaw segments in this section were set to zero. The pitch angle of the most anterior pitch joint in the undulating section below (**Figure 24**A, joint c) was set to its maximal possible value $\phi_{max}$ so that the cantilevering section was as vertical as possible to minimize cantilevering length. The two most anterior pitch joints in the cantilevering section (**Figure 24**A, joints a and b) were set to keep the section above in contact with the upper horizontal surface. Their pitch angles were calculated as follows: $\phi_a = \phi_b - \phi_{max}$, $\phi_b = -\sin^{-1}[(H - nh\sin\phi_{max})/L]$, $n = \text{floor}[H/(h\sin\phi_{max})]$, where $H$ is step height, $h$ is the distance between two adjacent pitch axes when the robot is straight, $n$ is the maximum number of pitch and yaw segments that can be kept straight in the cantilevering section.

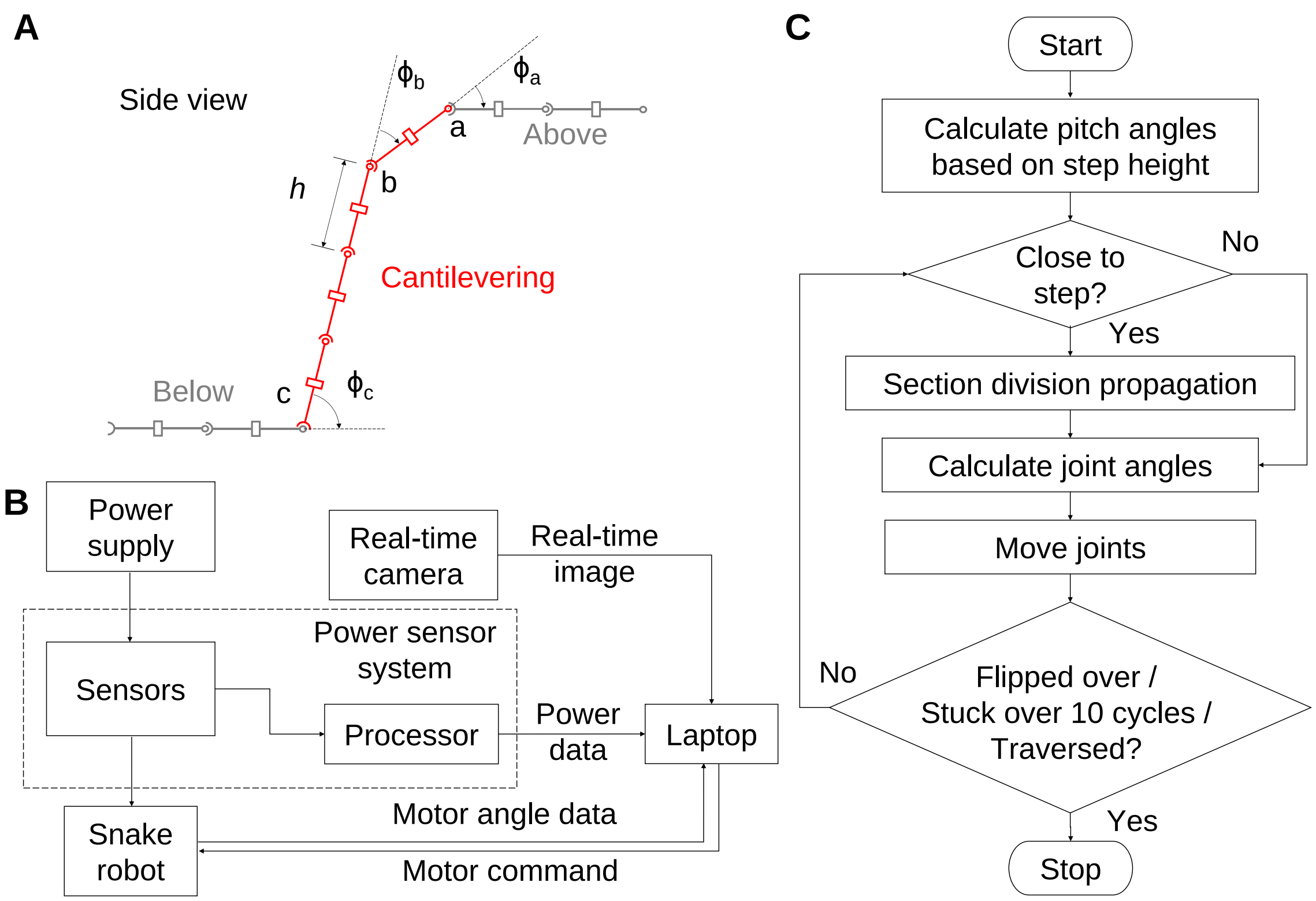

**Figure 24. Controller design. (A)** Side view schematic of partitioned gait design to show control of cantilevering section (red). Three pitch angles are calculated based on measured step height, including: $\phi_a$ and $\phi_b$ of the two most anterior pitch joint of the cantilevering section and $\phi_c$ of the most anterior pitch joint of the undulating section below the step. **(B)** Data acquisition system. **(C)** Flow chart of robot control. For (B) and (C), see Chapters 2.4.2, 2.8.5, and 2.8.6 for detailed description.

## 2.8.5 Marker-based feedback logic control

For feedback logic control (Tanaka and Tanaka, 2013) of the robot, a 3 × 3 cm ArUco marker (Garrido-Jurado et al., 2014) was fixed to the top of each pitch segment and on both the upper and lower horizontal surfaces near the top and bottom edge of the step (**Figure 23**B). Their positions captured by a camera were tracked before each trial to measure the step height for adjusting the robot gait and then tracked online to locate the position of each pitch segment relative to the step. We used a webcam (C920, Logitech, Lausanne, Switzerland) with 1920 × 1080 resolution for experiments with step height $H \leq 38\% L$. We used another camera (Flea3, FLIR, OR, USA) with 1280 × 1024 resolution and a 12.5 mm lens (Fujinon CF12.5HA-1, Fujifilm, Minato, Japan) for experiments with $H > 38\% L$ because the webcam could not capture the entire setup with its limited focus length and angle of view.

The snake robot was controlled by a custom Robot Operating System (ROS) package running on an Ubuntu laptop connected with the online camera and a power sensor system to measure electrical power consumption (see below) (**Figure 24**B). The laptop sent joint position commands to the servo motors and received motor angle readings at around 20 Hz. The online camera sent images to the laptop for online tracking of the ArUco markers at 20 Hz.

The feedback logic control algorithm is shown in flow chart (**Figure 24**C). Before entering the main loop of online servo motor control at 25 Hz (in ROS time), the actuation profile of pitch segments was first calculated based on the step height acquired. In each control loop, the controller determined whether section division needed to be propagated down the body by checking: (1) whether the middle point of the motor axle line segment of the most posterior pitch segment in the cantilevering section had crossed a vertical plane 4 cm before and parallel to the vertical surface of the step but was no higher than 10 cm above the upper horizontal surface; or (2) whether the middle point of the motor axle line segment of the most anterior pitch segment in the undulating section below the step had crossed a vertical plane 12 cm before and parallel to the vertical surface of the step. If either was true, the controller calculated the updated joint angles and sent angle commands to the servo motors. The controller continued this loop until a termination signal sent by the experimenter was received.

### 2.8.6 Electrical power measurement

We used two current sensors (Adafruit, NY, USA) between the servo motors and the power supply to record both voltage and current and measure electrical power of the robot (**Figure 25**) at 100-135 Hz. The two sensors were installed on the power cord near the power supply in parallel to accommodate the large current drawn. The DC current and voltage data were sent to the laptop for recording with timestamps via an Arduino-based Single Chip Processor (SCP) communicating with the laptop.

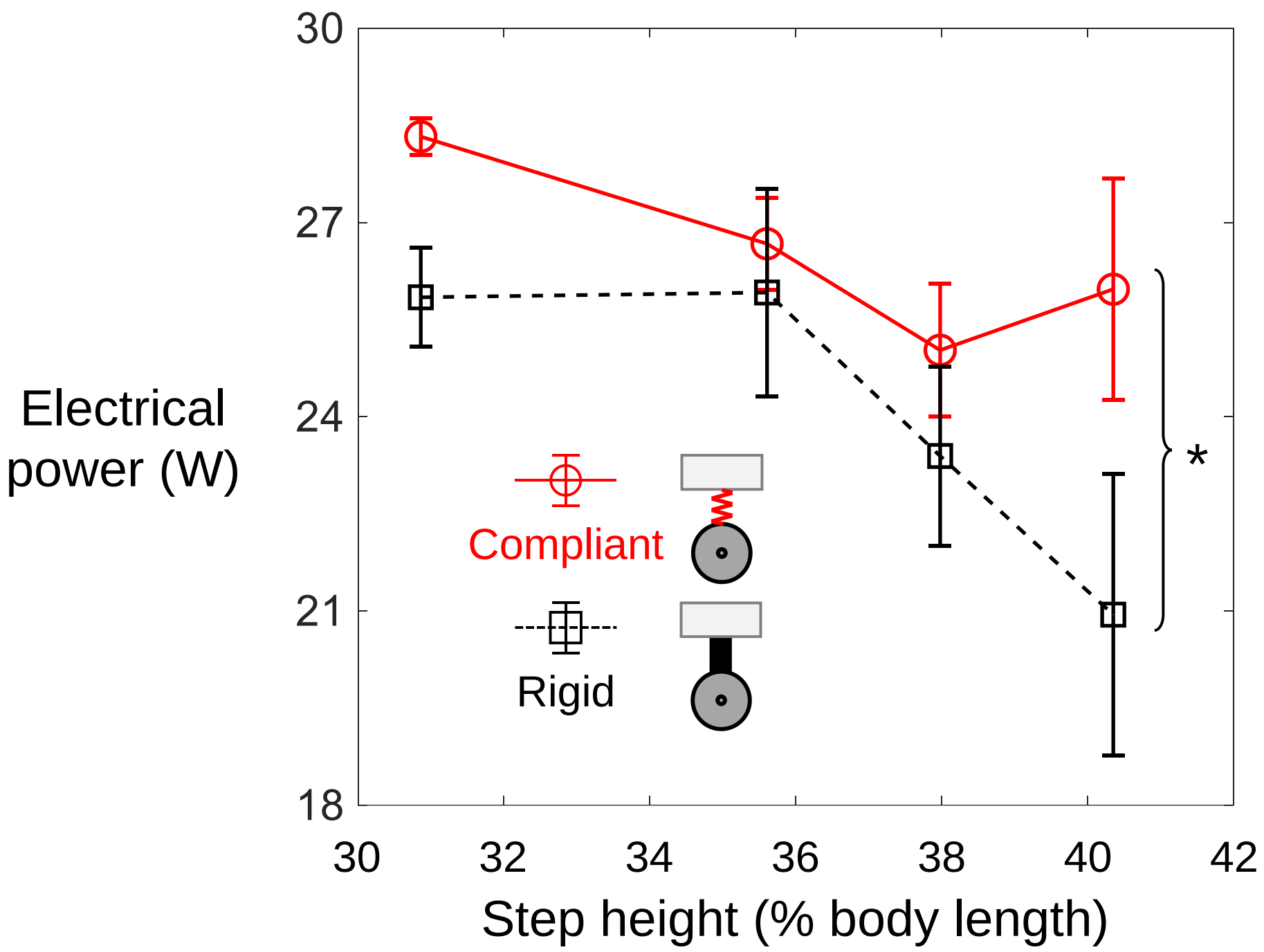

**Figure 25. Effect of body compliance on electrical power.** Electrical power of robot as a function of step height. Black dashed is for rigid body robot; red solid is for compliant body robot. Error bars show ± 1 s.d. Bracket and asterisk represent a significant difference between rigid and compliant body robot ($P < 0.0001$, ANCOVA).

### 2.8.7 Data synchronization

To synchronize motor angle data and electrical power data recorded by the laptop with the high-speed camera videos recorded on a desktop server, the power measurement circuit included a switch to turn on/off an LED bulb placed in the field of view of the high-speed cameras. When the LED was switched on/off, the SCP detected the voltage increase/drop and began/stopped recording power data. By aligning the initial and final power data points with the LED on/off frames in the videos and interpolating the motor position and electrical power data to the same sampling frequency as high-speed video frame rate (100 Hz), these data were synchronized.

### 2.8.8 Experiment protocol

At the beginning of each trial, we placed the robot on the surface below the step at the same initial position and orientation. The robot was set straight with its body longitudinal axis perpendicular to the vertical surface of the step. Its distance was set to be 16.5 cm from the wheel axle of the first segment to the vertical surface. This distance was selected so that the forward direction of most anterior segment in the undulating section below the step was perpendicular to the vertical surface before it began to cantilever. We then started high-speed video recording and switched on the LED in the SCP circuit. Next, we started the robot motion and monitored traversal progress until a termination condition was met. After the robot motion was terminated, the LED was first switched off, then the high-speed camera recording was stopped, and the setup was reset for the next trial while high speed videos were saved.

### 2.8.9 3-D kinematics reconstruction

To reconstruct 3-D kinematics of the entire robot traversing the large step obstacle, we recorded the experiments using twelve high-speed cameras (Adimec, Eindhoven, Netherlands) with a resolution of 2592 × 2048 pixels at 100 frames $s^{-1}$ (**Figure 23**A). The experiment arena was illuminated by four 500 W halogen lamps and four LED lights placed from the top and side.

To calibrate the cameras over the entire working space for 3-D reconstruction, we built a three section, step-like calibration object using T-slotted aluminum and Lego Duplo bricks (The Lego Group, Denmark). The calibration object consisted of 23 landmarks with 83 BEEtags (Crall et al., 2015) facing different directions for automatic tracking. We then used the tracked 2-D coordinates of the BEEtag center points for 3-D calibration using Direct Linear Transformation (DLT) (Hedrick, 2008). To obtain 3-D kinematics of the robot relative to the step, we used the 10 ArUco markers attached to the robot (one on each pitch segment), the two attached near the top and bottom edge of the step, and 13 additional ones temporarily placed on the three step surfaces before the first trial of each step height treatment. After all the experiments, we used a custom C++ script to track the 2-D coordinates of the corner points of each ArUco marker in each camera view. We checked and rejected ArUco tracking data whose four marker corners did not form a square shape with a small tolerance (10% side length).

Using the tracked 2-D coordinates from multiple camera views, we obtained 3-D coordinates of each tracked marker via DLT using a custom MATLAB script. We rejected marker data where there was an unrealistic large acceleration ($> 10$ m/$s^2$), resulting from a marker suddenly disappearing in one camera view while appearing in another in the same frame. We then obtained 3-D position and orientation of each pitch segment by offsetting its marker 3-D position and orientation using the 3-D transformation matrices from the marker to the segment, which was measured from the CAD model of the robot. We also measured the step geometry by fitting a plane to the markers on each of its three surfaces and generated a point cloud using the fit equation and the dimension of the three surfaces.

For yaw segments without markers and the pitch segments whose markers were not tracked due to occlusions or large rotation, we inferred their 3-D positions and orientations using kinematic constraints. We first tried inputting motor angles recorded by these segment motors into the robot forward kinematics to solve for their transformation matrices from other reconstructed segments. If their motor angles were not properly recorded, we tried inferring their positions and orientations from the two adjacent segments (as long as they were reconstructed). To do so, we first obtained all servo motor angles in this missing section by solving an inverse kinematics problem, then derived the transformation matrices of the missing segments from the forward kinematics. Finally, if both methods failed, we interpolated temporally from adjacent frames to fill in the missing transformation matrices. The interpolation was linearly applied on the twists of transformation matrices. We compared joint angles from the reconstructed segments to motor position data and rejected those with an error larger than 10°. To reduce high frequency tracking noise, we applied a window average filter temporally (*smooth2a*, averaging over 11 frames) to the 3-D positions of each segment after reconstructing all segments.

We verified the fidelity of 3-D kinematics reconstruction by projecting reconstruction back onto the high-speed videos and visually examined the match (**Figure**

**23**B). The thresholds used in this process were selected by trial and error, with the aim of removing substantial visible projection errors while rejecting as few data as possible.

### 2.8.10 Data analysis

To quantify traversal performance, we measured traversal probability defined as the ratio of the number of trials in which the entire robot reached the surface above the step to the total number of trials for each step height. To quantify roll instability, we measured roll failure (flipping over) probability, defined as the ratio of the number of failed trials in which the robot flipped over due to rolling to the total number of trials for each step height. To determine whether a wheel contacted a surface, we examined whether any point in the wheel point cloud (**Figure 20**B, grey dashed circle) penetrated the surface assuming no suspension compression. Unrealistic body deformation values from tracking errors larger than the 10 mm limit from the mechanical structure were set to 10 mm.

To compare electrical power during traversal across step height and body compliance treatments, we analyzed electrical power over the traversal process, defined as from when the first pitch segment lifted to cantilever, to when the last pitch segment crossed the top edge of the step for successful trials, or to when the robot flipped over (roll failure) or the trial was terminated due to robot getting stuck (stuck failure) for failed trials.

To compare traversal performance of our robot with previous snake robots and the kingsnake, we calculated vertical traversal speed for each robot and the animal. For our snake robot and the kingsnake with multiple trials, we first calculated vertical traversal speed of each trial by dividing step height normalized to body length by traversal time and then pooled speed data of all trials from all step heights for each body compliance treatments (for the robot) to obtain average speed. The slopes shown in **Figure 22** are average vertical traversal speed for each robot and the animal.

During experiments, we rejected trials in which the robot moved out of the obstacle track before successfully traversing the step or failing to traverse due to occasional crash of the control program. We collected around 10 trials for each combination of step height and suspension setting (rigid and compliant). For the rigid body, 40% *L* step treatment only 5 trials were collected, because the 3-D printed segment connectors were often damaged by ground collisions during roll failure (flipping over) and had to be replaced. Detailed sample size is shown in **Table 1**.

**Table 1. Sample size.**

| | *H* = 31% *L* | *H* = 36% *L* | *H* = 38% *L* | *H* = 40% *L* |
|---|---|---|---|---|
| Rigid | 10 | 8 | 10 | 5 |
| Compliant | 10 | 11 | 10 | 10 |

Records of traversal success and roll failure (flipping over) were binomial values (1 for success and 0 for failure) for each trial and averaged across trials to obtain their probabilities for each step height and body compliance treatment. For each trial, contact probability, body deformation, and surface conformation difference were averaged spatiotemporally over time and across all pitch segments in the undulating sections above and below the step combined. Electrical power was averaged over time for each trial. Finally, these trial averages were further averaged across trials for each step height and body compliance treatment to obtain treatment means and standard deviations (s.d.) or confidence intervals, which are reported in figures.

### 2.8.11 Statistics

To test whether traversal probability and roll failure (flipping over) probability depended on step height, for the rigid or compliant body robot, we used a simple logistic regression separately for each of these measurements, with step height as a continuous independent factor and records of traversal success or roll failure (flipping over) as a nominal dependent factor.

To test whether traversal probability and roll failure (flipping over) probability further depended on body compliance while taking into account the effect of step height, we used a multiple logistic regression for each of these measurements with data from rigid and compliant body robot combined, with body compliance as a nominal independent factor and step height as a continuous independent factor and records of traversal success or roll failure (flipping over) as a nominal dependent factor.

To test whether traversal probability differed between each adjacent pair of step heights for the rigid or compliant body robot, we used a pairwise chi-square test for each pair of step heights, with step height as a nominal independent factor and traversal success record as a nominal dependent factor.

To test whether contact probability, body deformation, surface conformation difference, and electrical power differed between rigid and compliant body robot, we used an ANCOVA for each of these measurements. We first set body compliance, step height, and their interaction term as independent factors and each of these measurements as a nominal/continuous dependent factor. If the $P$ value of the interaction term was less than 0.05, we then re-ran the same test excluding the interaction term.

To test whether vertical traversal speed differed between rigid and compliant body robot and the kingsnake, we used a two-sample *t*-test for each pair of the three subjects, with the subject as a nominal independent factor and vertical traversal speed as a continuous dependent factor.

All the statistical tests followed the SAS examples in (McDonald, 2014) and were performed using JMP Pro 13 (SAS Institute, Cary, NC, USA).

# Chapter 3

# Snakes combine vertical and lateral bending to traverse uneven terrain

This chapter was previously published in *Bioinspiration and Biomimetics* as an article entitled *Snakes combine vertical and lateral bending to traverse uneven terrain* authored by Qiyuan Fu, Henry Astley, and Chen Li (Fu et al., 2022). We re-used the article in this chapter with the permission from the publisher and all the authors. We revised Chapter 3.4.4 to better introduce the interpolation method and added **Figure 31**D and relevant texts to discuss the potential braking effects of body-terrain contact.

## 3.1 Author contributions

Qiyuan Fu designed study, performed experiments, validated results, analyzed data, and wrote the paper; Henry Astley designed study and revised the paper; Chen Li designed and oversaw study and revised the paper.

## 3.2 Summary

Terrestrial locomotion requires generating appropriate ground reaction forces which depend on substrate geometry and physical properties. The richness of positions and orientations of terrain features in the 3-D world gives limbless animals like snakes that can bend their body versatility to generate forces from different contact areas for propulsion. Despite many previous studies of how snakes use lateral body bending for propulsion on relatively flat surfaces with lateral contact points, little is known about whether and how much snakes use vertical body bending in combination with lateral bending in 3-D terrain. This lack had contributed to snake robots being inferior to animals in stability, efficiency, and versatility when traversing complex 3-D environments. Here, to begin to elucidate this, we studied how the generalist corn snake traversed an uneven arena of blocks of random height variation 5 times its body height. The animal traversed the uneven terrain with perfect stability by propagating 3-D bending down its body with little transverse motion (11° slip angle). Although the animal preferred moving through valleys with higher neighboring blocks, it did not prefer lateral bending. Among body-terrain contact regions that potentially provide propulsion, 52% were formed by vertical body bending and 48% by lateral bending. The combination of vertical and lateral bending may dramatically expand the sources of propulsive forces available to limbless locomotors by utilizing various asperities available in 3-D terrain. Direct measurements of contact forces are necessary to further understand how snakes coordinate 3-D bending along the entire body via sensory feedback to propel through 3-D terrain. These studies will open a path to new propulsive mechanisms for snake robots, potentially increasing the performance and versatility in 3-D terrain.

## 3.3 Introduction

Unlike limbed animals, which typically generate support and propulsive forces at a few points in the environment with distinct anatomical structures (feet), elongate, limbless animals such as snakes can use their entire body to create a large number of contact points with the surrounding environment to move through (Gray and Lissmann, 1950). This enables the body of the snake to interact with the substrate at a wide range of local positions and orientations, and then modulate these interactions by altering force distribution among them. Lateral slithering motion has been the focus of much of the literature, focusing on either frictional interactions with smooth, rigid surfaces (Hirose, 1993; Hu et al., 2009), interactions with granular media (Schiebel et al., 2020a) and artificial turf (Gerald and Wass, 2019; Jayne and Bennett, 1989; Jayne and Bennett, 1990; Walton et al., 1990), or, most often, interactions with arrays of vertical structures replicating natural terrain objects such as plants and rocks on flat surfaces (Gray and Lissmann, 1950; Jayne, 1986; Jayne and Byrnes, 2015; Jayne et al., 2013; Kano et al., 2012; Schiebel et al., 2020b). These have inspired many snake robots to traverse similar environments using lateral bending (Gong et al., 2016; Hirose, 1993; Kano and Ishiguro, 2013; Sanfilippo et al., 2017; Wang et al., 2020a). Lateral body bending in these scenarios relies on objects with special anisotropic properties or positions and orientations to press against. However, slithering is still commonly observed in a variety of 3-D terrains that lack such objects (Jayne and Byrnes, 2015; Jayne and Herrmann, 2011). This indicates that slithering snakes are able to generate propulsion by interacting with a wider range of terrain asperities using body deformation in all three dimensions.

Vertical bending during terrestrial snake locomotion is rarely studied. Previous work has focused on the use of vertical lifting to improve efficiency either by reducing frictional drag, such as in sidewinding (Marvi et al., 2014) and in sinus-lifting (Hirose, 1993), or by raising the body to reach higher surfaces (Gart et al., 2019). Recent studies have revealed that vertical body bending can be utilized by snakes to interact with terrains with significant height variations and generate propulsive forces to traverse them. For example, when the corn snake traverses a horizontal ladder lacking lateral contact points, it can generate substantial propulsive force and propulsive impulse by posteriorly propagating vertical waves with minimal lateral motion (Jurestovsky et al., 2021). The propulsive value of pure vertical bending was further confirmed by the success to traverse a similar terrain of a robophysical model replicating only the vertical bending (Jurestovsky et al., 2021). These recent observations in simplified environments with no lateral contact points suggested that vertical bending is promising for expanding the source of propulsion in natural 3-D environments by pressing against suitably oriented terrain asperities below the body, similar to lateral bending pushing against lateral contact points.

Understanding of these basic principles will have a major impact on snake robot locomotion in complex 3-D environments. Some previous snake robots traversed 3-D complex terrains using geometric gait designs that only apply to limited scenarios (Fu and Li, 2020; Jurestovsky et al., 2021; Lipkin et al., 2007; Takemori et al., 2018a; Tanaka and Tanaka, 2013). Some robots adapted simple cyclic gaits originally used on flat surfaces and passively conformed to vertical height variation of terrain by mechanical or control compliance (Takemori et al., 2018b; Travers et al., 2018; Wang et al., 2020a). However, there is still a significant gap in snake robots' stability, efficiency, and versatility compared to animals in complex environments, in large part due to a lack of principled understanding of how to use vertical body bending to generate propulsion. In previous snake robots, vertical bending was used either to improve efficiency by reducing frictional drag (Marvi et al., 2014; Toyoshima and Matsuno, 2012) or to reach different terrain surfaces (Fu and Li, 2020; Lipkin et al., 2007; Takemori et al., 2018a; Takemori et al., 2018b; Tanaka and Tanaka, 2013; Wang et al., 2020a). Only one snake robot used vertical bending to traverse a single cylindrical obstacle (Date and Takita, 2005), but that study assumed that there is no longitudinal friction, no lateral slipping, and no gravity. However, recent demonstration of a simple snake robot traversing a horizontal ladder in a similar fashion as snakes suggested that vertical bending can be used to generate propulsion over more complex 3-D terrain (Jurestovsky et al., 2021).

Inspired by these recent insights, here we take the next step in studying how snakes use 3-D body bending to move through the 3-D world. We hypothesize that, when both vertical and horizontal contact points are available, generalist snakes can use vertical body bending as frequently as lateral body bending to interact with and traverse 3-D terrain. As an initial step to test the hypothesis, we measured the kinematics of generalist corn snakes during their traversal of uneven terrain and analyzed their contact between their body and terrain surfaces. The uneven terrain allowed the animal to use both lateral and vertical bending for contact due to the variation of geometry in both vertical and horizontal directions. We evaluated performance by analyzing longitudinal vs. transverse motions and static stability. To estimate the relative contribution of lateral and vertical bending, we compared: (1) the number of body bends of each type contacting the terrain and (2) the number of horizontal and vertical bends which would potentially allow propulsive force generation.

## 3.4 Methods

### 3.4.1 Animals

We used three captive-bred juvenile corn snakes [*Pantherophis guttatus* (Utiger et al., 2002)] purchased from an online vendor (Reptiles by Alex, Wichita, KS, USA). These species is a locomotor generalist commonly used in snake locomotion studies, and its natural habitats include substantial fine-scale terrain height variations such as forests and rocky hillsides (Burbrink, 2002; Conant and Collins, 1998). We housed snakes individually in 45.7 × 19.1 cm or 50.8 × 38.1 cm containers on a 12 h:12 h light:dark cycle at a temperature of 30 °C on the warm end and 25 °C on the cool end. Snakes were fed water and pinky mice. The snakes' full body length measured 82.2 ± 5.7 cm, and they weighed 165.0 ± 16.2 g. To measure length, we digitized dorsal view photos of each snake by tracing the body midline and scaling its length from pixels to centimeters (Astley et al., 2017). To quantify body tapering, we measured the cross-sectional height of the snakes at different locations by digitizing lateral view photos and interpolated values in between by fitting a quadratic polynomial to measured heights (**Figure 32**A). All animal experiments were approved by and in compliance with The Johns Hopkins University Animal Care and Use Committee (protocol RE19A165).

### 3.4.2 Uneven terrain arena

We constructed a 65 cm wide, 97.5 cm long uneven terrain arena using 96 blocks in an 8 × 12 array (**Figure 26**A-B). Each block has a horizontal footprint of 8.2 × 8.2 cm. Heights of all blocks follow a normal distribution (**Figure 26**A) with a mean of 5.1 cm and a standard deviation of 1.7 cm. Positions and heights of all blocks were kept unchanged during all trials. Each block consisted of a rectangular PVC pipe (McMaster-Carr, Elmhurst, IL, USA) cut to desired height as side walls and a piece of acrylic sheet (McMaster-Carr, Elmhurst, IL, USA) laser cut to the same shape as the cross section of the PVC pipe as the top surface (**Figure 26**C). The PVC pipe and the acrylic sheet were hot glued from the inside to keep the outer surface clean. Adjacent blocks were rigidly connected by 3-D printed clamps from bottom. To measure the 3-D positions of terrain blocks, we attached one 3.8 cm × 3.8 cm BEEtag marker (Crall et al., 2015) to each top surface of blocks and covered the marker with packaging tape (3M, Maplewood, MN, USA) to reduce friction. Three 61.5 cm tall wooden sheets were used as sidewalls to prevent the snakes from escaping and clamped together with the blocks using 3-D printed clamps from bottom.

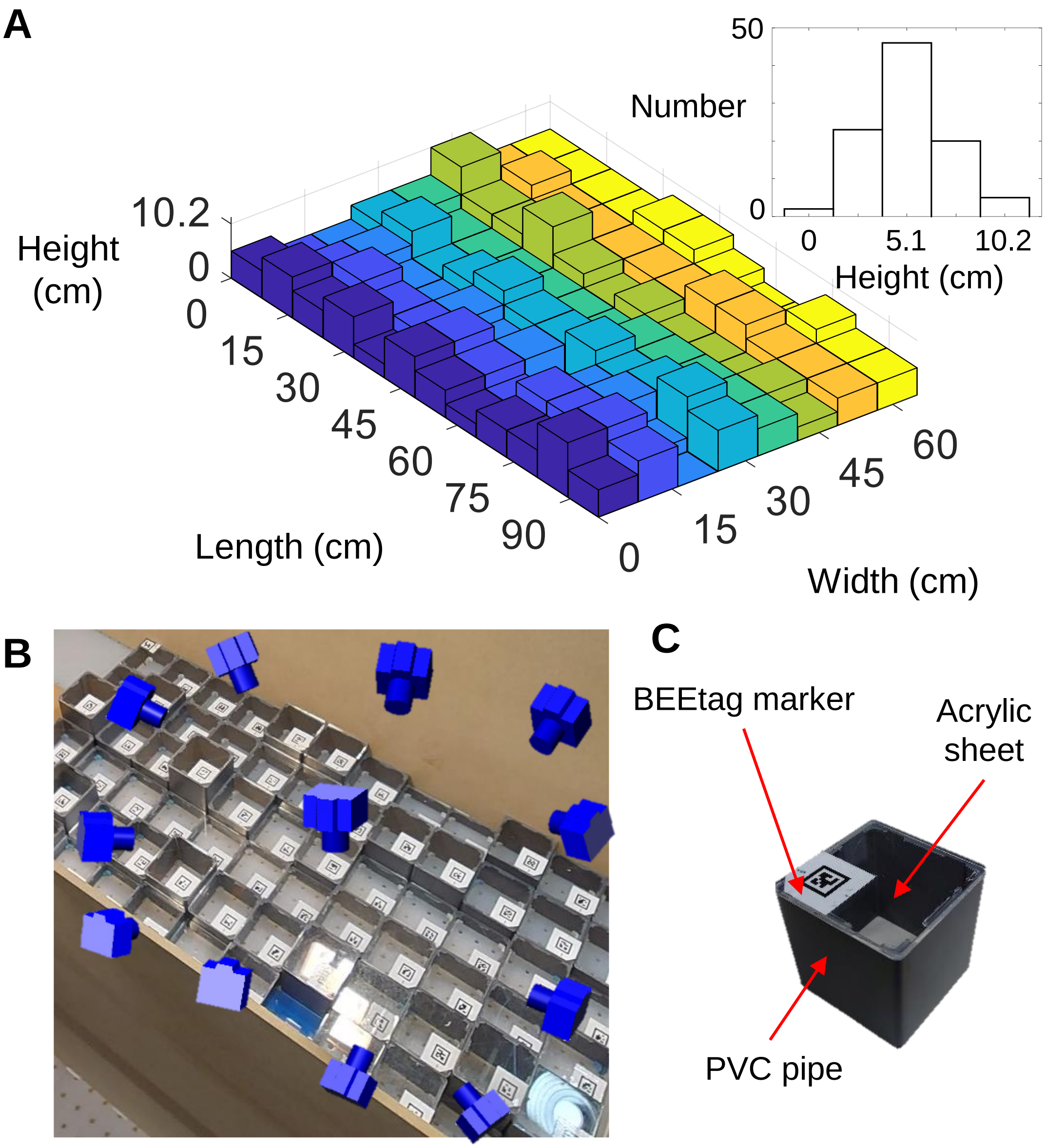

**Figure 26. Experimental setup. (A)** Height distribution of terrain blocks. Inset shows histogram of block heights. **(B)** Photo of experimental setup. Blue shows twelve high-speed cameras used for 3-D reconstruction of markers. **(C)** Components of a terrain block.

We measured the kinetic friction coefficient between the snake body and the terrain blocks using a 3-axis force/torque sensor (**Figure 33**A; ATI mini 40, Apex, NC, USA). The sensor measured normal force and friction while a snake was sliding against a plate rigidly connected to the sensor. We fit a line through origin to the normal force and friction collected during the middle 50% of time in each slide and calculated the friction coefficient from the slope. Each material that made up terrain blocks was used as the top surface of the plate for 5 measurements along each direction (forward, backward, left, and right) for each of the 3 animals. The kinetic friction coefficients between the snake body and acrylic, PVC, and packaging tape (covering the BEEtag markers), were $\mu = 0.32 \pm 0.05$, $0.28 \pm 0.07$, and $0.19 \pm 0.02$ (mean ± s.d.), respectively.

### 3.4.3 Locomotion experiment protocol

Snake locomotion was recorded using 12 high-speed cameras (**Figure 26**B; Adimec, Eindhoven, The Netherlands) at 50 frames $s^{-1}$ with a resolution of 2592 × 2048 pixels. To illuminate the arena, two 500 W halogen lamps and two LED lamps were placed dorsally above the arena. The surface of the test area was heated to 32 °C during experiments. To calibrate the cameras for 3-D reconstruction, we made a 61 × 66 cm calibration grid out of DUPLO bricks (The Lego Group, Bilund, Denmark) and attached BEEtag markers (Crall et al., 2015) on it. We placed the calibration grid in the arena before experiments and recorded snapshots using the 12 cameras. To track the 3-D movements of the snake, we attached 10 to 12 lightweight (0.3 g) BEEtag markers along the dorsal side of the snake equally spaced (≈ 6.6 cm) between neck and vent (**Figure 32**B) using lightly adhesive tape (0.4 × 1.2 cm).

The snake was kept in a hide near the test area at a temperature of 30 °C prior to experiments. We placed the snake on random locations inside the arena and encouraged it to traverse blocks by light tapping on the tail and a shaded shelter near the test area. A trial was ended if any part of the snake moved out of the test area or the snake stopped moving for more than 15 seconds. After each trial, the snake was removed from the test area, placed in the hide, and allowed to rest for 1 to 2 minutes.

After experiments, we tracked 2-D coordinates of the markers attached to the calibration grid and obtained intrinsic and extrinsic camera parameters using direct linear transformation (DLTdv5) (Hedrick, 2008). BEEtag markers attached to snakes and blocks were tracked in 2-D camera views and reconstructed for 3-D positions and orientations using custom MATLAB scripts (Crall et al., 2015; Hedrick, 2008). The geometry of terrain blocks was then reconstructed using measured dimensions and tracked positions and orientations.

### 3.4.4 Continuous body 3-D kinematics interpolation

To obtain continuous 3-D kinematics of snake body for contact analyses, we interpolated the midline of each section of body (both position and orientation) between adjacent markers using a custom method (**Figure 27**A) (Fu et al., 2021). Most previous studies that quantified the movement of a limbless animal in 3-D used discrete points along their elongate bodies (Byrnes and Jayne, 2012; Kwon et al., 2013; Marvi et al., 2014; Socha et al., 2005) or a linear combination of basis curves described by simple analytical functions (Fontaine et al., 2008; Padmanabhan et al., 2012; Schiebel et al., 2019; Sharpe et al., 2015; Yeaton et al., 2020). However, these methods cannot accurately quantify body orientation (especially rolling around the centerline), which can largely affect the interaction forces such as anisotropic friction and granular resistive force (Hu et al., 2009; Li et al., 2013).

In our custom method, we first tracked the 3-D positions and orientations of the BEEtag markers (Crall et al., 2015) attached to a snake's body as described in Chapter 3.4.3. To interpolate a continuous body section between two adjacent markers, we approximated it as an elastic rod (Cao and Tucker, 2008; Dill, 1992; Zhang et al., 2019) subject to end constraints imposed by the tracked markers. We used a rod with a constant radius throughout the body length for simplicity. By applying an optimization method (Chirikjian, 2015), we determined one possible solution that minimizes the elastic energy of the rod. Despite the difference between the elastic rod and the real musculoskeletal structure of a snake, the method results in a 50% reduction in errors when interpolating both position and orientation than commonly used B-spline methods. The low position error (17% of body diameter on average (Fu et al., 2021)) of the interpolated midline enabled us to use it to reconstruct the surface of the snake body for contact analyses (see Chapter 3.4.6). When reconstructing the body surface (see below), we used measured body tapering instead of a constant radius for higher accuracy. Unless otherwise noted, we used the entire reconstructed midline in our analyses.

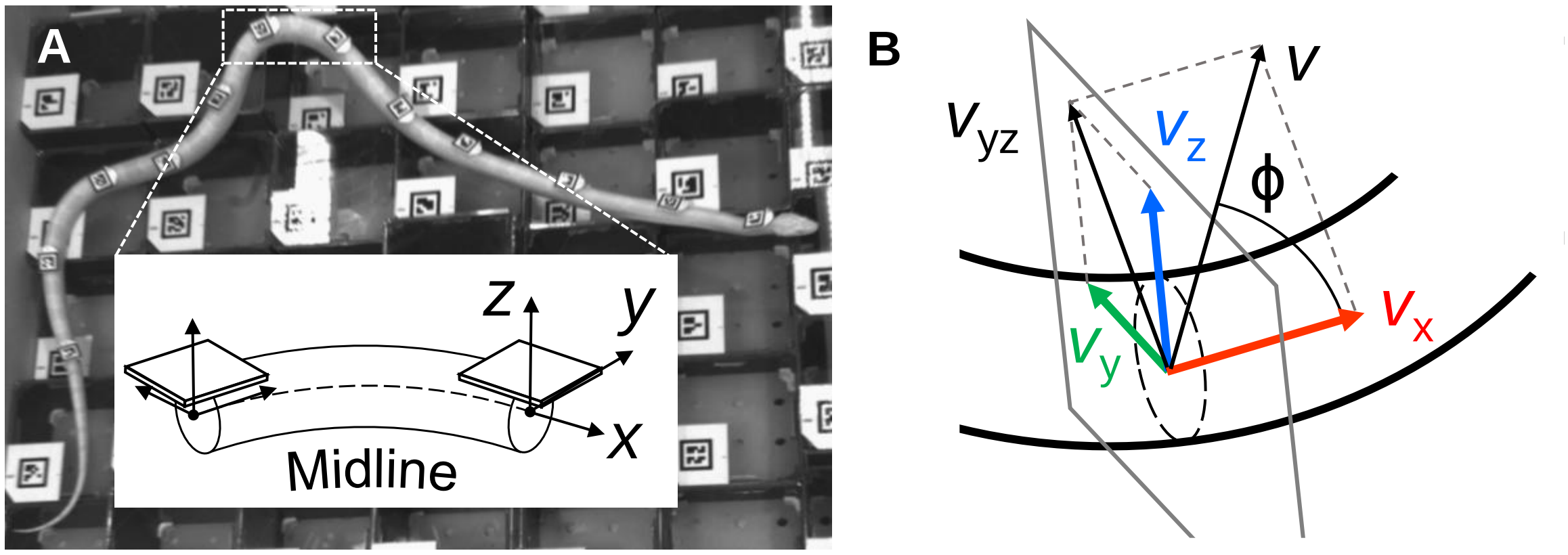

**Figure 27. 3-D snake body reconstruction and definition of metrics to evaluate transverse motion. (A)** A reconstructed snake body segment between two adjacent markers and its midline (dashed). **(B)** Definition of local body velocity $v$, longitudinal velocity $v_x$, lateral velocity $v_y$, dorsoventral velocity $v_z$, transverse velocity $v_{yz}$, and slip angle $\phi$. Dashed circle shows cross section and gray parallelogram shows local normal plane.

### 3.4.5 Performance analysis

To assess traversal performance, we calculated the accumulated distance traveled by the mid-body position (the mid point of the reconstructed midline) along its trajectory and the duration of travel using all video frames in each trial.

Occasionally, the snake stopped during traversal and most of the body remained static. In contrast, terrestrial slithering motion such as lateral undulation has forward longitudinal velocities throughout the body as the bending wave propagated down the body (Jayne, 2020). Thus, in all the following analyses, to avoid artificial bias from stopped body postures, we excluded video frames in which the average longitudinal velocity along the entire reconstructed body was small ($< 0.125$ cm $s^{-1}$) (**Figure 34**A; 6.4% of all the video frames).

To assess how effectively the corn snake used substrate irregularities in the uneven terrain arena to generate propulsion, we calculated longitudinal and transverse velocities and slip angle. During slithering locomotion, posteriorly propagating bends of the body generate reaction forces against the substrate to propel the snake forwards (Gray and Lissmann, 1950; Mosauer, 1932). On flat, rigid planes (rarely found in nature), frictional anisotropy of the scales allows generation of forward forces, but with high slipping and low speeds (Alben, 2013; Hu et al., 2009; Schiebel et al., 2020a). However, with the presence of geometric asperities against which body bends can push without constant yielding (e.g., rocks, plants, sufficiently large piles of sand), snakes will show minimal slip as if moving in a virtual tube with greatly increased speed (Gray, 1946; Gray and Lissmann, 1950; Jayne, 1986; Kelley et al., 1997; Mosauer, 1932; Schiebel et al., 2020a; Schiebel et al., 2020b).

For each infinitesimal body segment of the reconstructed midline, we calculated longitudinal velocity as the velocity component parallel to the local body segment, $v_x = \vec{v} \bullet \vec{T}$, and transverse velocity as the velocity component perpendicular to it, $v_{yz} = |\vec{v} - \vec{v} \bullet \vec{T}|$ (**Figure 27**B). We calculated slip angle as the angle between local body velocity and local body tangent in 3-D ($\phi = \cos^{-1}(\vec{v} \bullet \vec{T} / |\vec{v}|)$, which ranges from 0º to 180º, **Figure 27**B) (Sharpe et al., 2015), which measures how well the body stays within a virtual tube as it progresses. Perfectly progressing forward in a virtual tube results in a slip angle of 0º, whereas no progress or backward progress in it results in a slip angle of 90º or 90-180º, respectively.

We calculated longitudinal and transverse velocities and slip angles for all body segments regardless of contact conditions (next section) to quantify how well the overall kinematics matches ideal slithering motion within a virtual tube. We also calculated these three metrics only for the body segments in contact with the terrain, which can alter contact forces and affect propulsion performance more than that of body segments suspended in the air.

To test whether the anterior end of the snake moved transversely more than the other part of the body, we divided the snake body into two parts: the 10% of body segments closest to the nose and the other 90%, defined as the anterior region and the main body region, respectively (**Figure 32**A).

To assess whether the snake tended to move on lower blocks (as if moving in a valley), we compared average height of blocks directly below the snake body ($h_{below}$) and that of neighboring blocks ($h_{neighbor}$) (see Chapter 3.4.8). Neighboring blocks were defined as blocks that were adjacent to blocks directly below the snake body. A neighboring block may contact the lateral sides of the snake body under this definition. We also calculated the percentage of video frames in each trial when the average height of all blocks directly below the body was smaller than that of neighboring blocks.

### 3.4.6 Contact analysis

To classify contact types of different body parts (**Figure 28**A), we first determined contact between the snake body and the terrain surfaces. We sampled 200 locations evenly along the reconstructed midline and 24 points on the circumference of the cross-sectional outline (assumed to be circular) of each sampled body segment, resulting in a total of 4800 points on the reconstructed snake body surface. Each cross-sectional outline was radially expanded outward from the reconstructed midline by the fitted local body radius to account for effects of body tapering. Collision detection between these sample points and each reconstructed terrain block was performed to locate contact points using the GJK algorithm (Sheen, 2021), a common algorithm to determine collision between convex objects. Only blocks directly below and neighboring blocks were included for collision detection to save computation time.

To identify terrain surfaces that a body segment was contacting, we checked the distances between each point on the sampled outline where the outline started to penetrate blocks (**Figure 28**E inset, top yellow point) and each face of the block that this point was penetrating (**Figure 28**E, blue solid line). Faces obstructed by other blocks were not considered (**Figure 28**E). Vertical surfaces of the block directly below the body segment were not considered because otherwise the body segment sitting on a horizontal surface along an edge was falsely classified as contacting the vertical surface directly below it (**Figure 28**F).

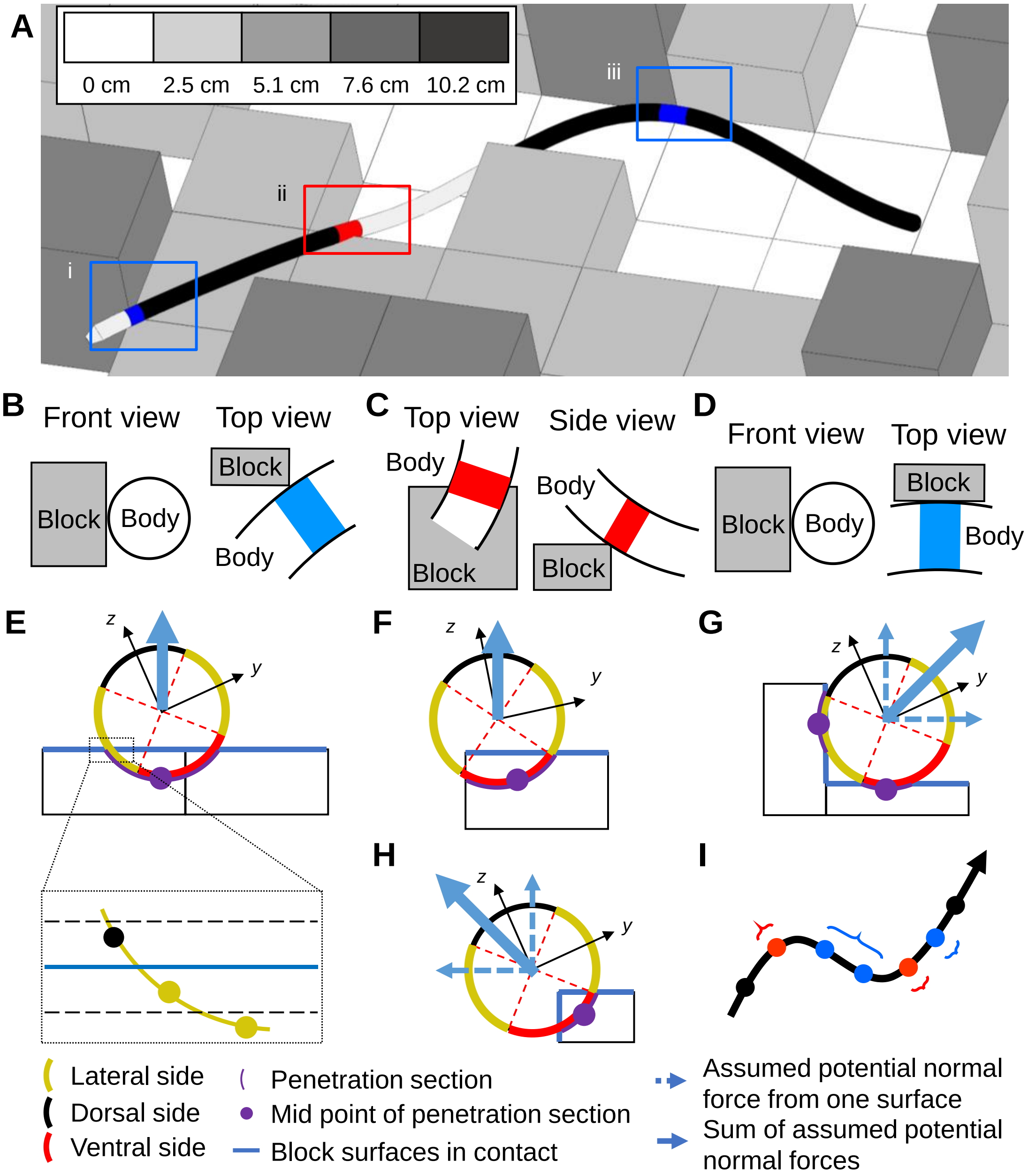

**Figure 28. Examples to determine contact types and contact regions. (A)** An example of contact types of different body regions. Blue (lateral contacting), red (vertical contacting), black (supported), and white (suspended) indicate different body-terrain contact types. Three insets show representative cases of body-terrain contact: (i) Body is laterally contacting a vertical edge. (ii) Body is vertically contacting a horizontal edge. (iii) Body is laterally contacting a vertical wall. **(B-D)**

Different views of case (i-iii) in (A). **(E-H)** Examples to show how contact types are determined by where cross-sectional outline (circle) contacts terrain blocks (boxes). (E) A supported body segment that only contacts horizontal surfaces. Inset shows identification of surfaces a body segment is contacting. Black dashed lines show range in which a sample point is considered in contact with terrain surface. Yellow and black points show sample points penetrating and outside the terrain block. (F) A special case of supported body segment that is considered to only contact a horizontal surface and vertical surface of the block below the segment is ignored because otherwise body segments sitting on a horizontal edge are falsely classified as vertical contact segments. (G) A lateral contact body segment that contacts one vertical surface on its lateral side. (H) A vertical contact body segment that contacts one vertical surface only with its ventral side. In (E-H), assumed kinetic friction force is perpendicular to sum of assumed potential normal forces and opposite to local body velocity. **(I)** An example to determine contact regions. Black line shows body midline. Points show body segments and brackets show contact regions.

To check whether a body segment contacted terrain on the lateral sides or on the ventral side, we divided the outline of each body segment into four sections of equal length (one ventral, one dorsal, and two laterals; **Figure 28**E-H, red, black, and yellow arcs, respectively) and checked into which section the midpoint (**Figure 28**E-H, purple point) of each penetration section (**Figure 28**E-H, purple arc) fell.

By checking which terrain surfaces the body segment contacted and on which sides of the body segment the contact happened, we classified sampled infinitesimal body segments into 4 types: (1) Suspended (**Figure 28**A, white): the segment was not contacting the terrain. (2) Supported (**Figure 28**A, black; **Figure 28**E): the segment was contacting horizontal terrain surfaces only. However, if the segment was also contacting vertical surfaces, it was classified as either (3) or (4) below. (3) Lateral contact (**Figure 28**A, blue; **Figure 28**B, D, G): the segment was contacting vertical walls with a lateral side.

This includes laterally contacting a vertical wall (**Figure 28**A, right blue) or a vertical edge connecting two vertical walls (**Figure 28**A, left blue). (4) Vertical contact (**Figure 28**A, red; **Figure 28**C, H): the segment was contacting vertical walls only with its ventral side. This includes contacting a horizontal edge (**Figure 28**A, red) or vertices connecting multiple horizontal edges with its ventral side. Contact types of the infinitesimal body segments not sampled for collision detection were interpolated using the values of the nearest sampled body segment. Results of this classification were visually examined by color-coding the reconstructed midline accordingly in camera videos (**Figure 29**A) and flattened sagittal views (**Figure 29**B).

To quantify vertical and lateral bending used by animals for contact, we counted the number of body sections in contact with the terrain (referred to as contact regions hereafter) formed by continuous body segments with lateral or vertical contact. A contact region was counted for each continuous section of the reconstructed midline that was made up only by body segments of one contact type (**Figure 28**I). Each contact region may have variable length and shape. We counted the number of contact regions instead of the number of body segments because a vertical contact region contacting an edge (**Figure 29**A, red) often appeared with fewer body segments than a lateral contact region contacting a surface (**Figure 29**A, blue).

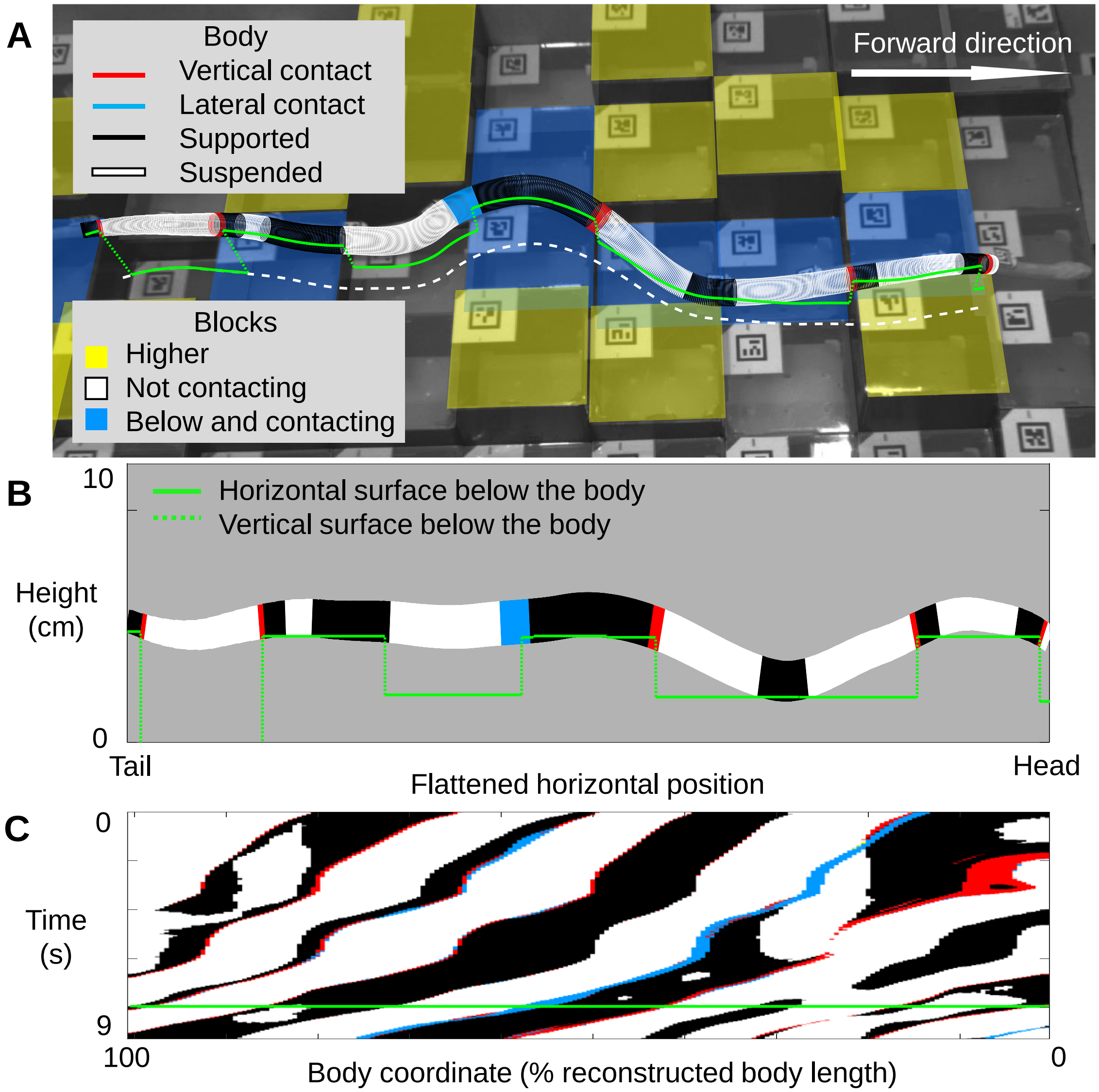

**Figure 29. Contact types of snake body at a representative moment. (A)** A representative snapshot of snake traversing uneven terrain in which the snake is moving from left to right. Colors of reconstructed snake body surface show body-terrain contact types. Colors of reconstructed top surfaces of terrain blocks show block heights with respect to nearby snake body. Green solid, green dotted, and white dashed curves indicate intersections between the curved body sagittal plane and terrain top surfaces, terrain side surfaces, and ground horizontal plane, respectively. Thus, the white dashed line is a top-down view, albeit viewed from an oblique perspective. **(B)** Flattened sagittal view of reconstructed snake body and its intersections with the different surfaces in (A).

Flattened horizontal position is the accumulated horizontal distance to the head in the curved body sagittal plane. **(C)** A representative spatiotemporal profile of contact types as a function of body coordinate and time. Body coordinate is defined as ratio of cumulative length along the body from head to reconstructed body length. Green line corresponds to snapshot in (A). This representative trial does not contain phases in which the snake stops (i.e., video frames in which the average longitudinal velocity along the entire reconstructed body is small (< 0.125 cm $s^{-1}$)). Other trials may contain such phases but these video frames are excluded from most analyses (see Chapter 3.4.5). In (B, C), body coordinate starts from 0% at the most anterior marker and ends at 100% at the most posterior marker. See Movie 5 for a representative video.

Because we could not measure forces directly, we used simple assumptions to infer the likely terrain reaction force directions considering the rectangular geometry of the arena. Regardless of force magnitude, the reaction force against any surface must be the sum of the outward force normal to the surface and the frictional force along the surface opposite to local body velocity, which is proportional to normal force by the coefficient of friction. We only considered kinetic friction because of a lack of force measurements required to determine the direction of static friction. Thus, we did not consider friction from body segments that had a small total velocity (< 0.3 cm $s^{-1}$) (**Figure 34**C; 3.3% of all the body segments in al the video frames pooled from all trials of all animals). If a body segment was contacting multiple surfaces (e.g., two vertical surfaces that met at a vertical edge), we assumed that the potential normal force generated from it (**Figure 28**G, H, blue solid arrows) was along the sum of the normal vectors of all these surfaces (**Figure 28**G, H, blue dashed arrows). We assumed a body segment to be generating propulsion if the sum of potential normal force and potential kinetic friction force generated from it has a positive projection in the forward motion of overall body movement (i.e., if the total potential force direction and the instantaneous center of mass velocity formed an angle smaller than 90°) in the top view. We assumed a body segment to be generating braking force if the projection was negative.

We considered a contact region to be potentially propulsive if any of the body segment within it was assessed to be propulsive, assuming that the animal can redistribute force to use any potentially propulsive body segment within a contact region. Otherwise, we consider a contact region to be potentially braking if all the body segments within it was assessed to be braking. Given the lack of direct force measurement, we could not assume that contact surface area is positively correlated with the force exerted by each body region, as small contact points may generate high forces and large ones may generate lower forces. This has never been assessed in limbless locomotion. Instead, we assumed that each contact region had the same potential to propel the body and counted the number of potentially propulsive contact regions. We did not consider video frames (**Figure 34**B; 7% of all frames) with slow instantaneous center of mass velocities (< 0.3 cm $s^{-1}$) when evaluating likely propulsive body segments to mitigate tracking noise.

To test whether the snake was stable, we first estimated center of mass position by averaging positions of all interpolated body segments (96% of full body volume as estimated from tapering data) weighted by cross-sectional area in each video frame. Then, we checked whether center of mass projection onto the horizontal plane (**Figure 35**, red circle) fell into the support polygon (**Figure 35**, purple polygon), a 2-D convex hull enclosing the projection of body segments in contact with horizontal surfaces (including their edges) of the terrain (Gart et al., 2019). We calculated static stability performance for each trial by dividing the number of video frames in which the snake was statically stable with the total number of video frames.

### 3.4.7 Sample size

We performed experiments using three snakes (*N* = 3) with 18 trials for each animal. After rejecting trials with large reconstruction errors because of loss of tracking of occluded markers for a long time, 8, 9, and 13 trials remained for the three individuals, respectively, resulting in a total of *n* = 30 accepted trials. Video frames in which part of the body was not reconstructed because markers were occluded by blocks (0.08% of all video frames) were excluded from statistical tests.

### 3.4.8 Data averaging

#### 3.4.8.1 Metrics directly calculated for each body segment or each terrain block

For each trial, we first averaged these metrics spatially in each video frame and then averaged them temporally across all video frames (excluding frames in which the snake stopped). These metrics include: (1) average longitudinal and transverse velocities and slip angle of the entire reconstructed body; (2) average longitudinal and transverse velocities and slip angle of the body sections in contact with the terrain; (3) slip angle of the anterior region; (4) slip angle of the main body region.

To obtain height difference between the blocks directly below snake body and the neighboring blocks, we first calculated the average height of all blocks below the body and that of all neighboring blocks, because the number of these two types of blocks were often unequal. We then calculated their difference in each video frame and averaged them temporally across all video frames (excluding frames in which the snake stopped).

### 3.4.8.2 Metrics directly calculated for each video frame

For each trial, we averaged these metrics temporally across all video frames. These measurements include: (1) accumulated distance traveled by the mid-body position along its trajectory; (2) duration of travel; (3) number of vertical and lateral contact regions; (4) number of potentially propulsive vertical and lateral contact regions. Video frames in which the snake stopped (< 0.125 cm $s^{-1}$) were excluded for (3) and (4).

### 3.4.8.3 Metrics directly calculated for each trial

No averaging was necessary for these metrics. These metrics include: (1) static stability performance; (2) percentage of video frames in which the average height of the blocks under the animal was smaller than that of the neighboring blocks; (3) percentage of video frames in which all the potentially propulsive regions were purely vertical contact region or purely lateral contact regions.

By performing a random-effect ANOVA with individual as a random factor, we verified that the intra-subject variance was larger than the between-subject variance (Leger and Didrichsons, 1994). Thus, we pooled all trials of all individuals. Finally, for metrics in categories 1 and 2, we calculated means and standard deviations of all trials. For metrics in category 3, we calculated median and interquartile range of all trials or plotted histogram of all trials in the case where the data had a non-normal distribution.

## 3.4.9 Statistical tests

To test whether two paired measurements within a trial differed consistently, we performed paired *t*-tests pooling all trials from all animals. These paired measurements include heights of blocks directly below the body versus neighboring blocks, transverse versus longitudinal velocity, slip angle of the anterior part versus the rest of the body, the number of lateral versus vertical contact regions, the number of potentially propulsive lateral versus vertical contact regions, and the number of potentially braking lateral versus vertical contact regions.

To test whether kinetic friction coefficient between the snake body and a surface material differed along different directions, for each individual and each surface material, we performed an ANOVA followed by a Tukey's honestly significant difference (HSD) test with sliding direction as an independent variable and kinetic friction coefficient as a dependent variable.

All the statistical tests followed (McDonald, 2014) and were performed using JMP Pro 15 (SAS Institute, Cary, NC, USA).

# 3.5 Results

## 3.5.1 Traversal behavior

The animal traversed the uneven terrain by propagating 3-D bending down the body with little transverse motion out of the virtual tube (**Figure 30**A, Movie 6), similar to prior studies of snakes moving in heterogeneous terrain such as artificial turf and surfaces with arrays of vertical structures (Jayne, 1986; Kano et al., 2012; Schiebel et al., 2020b). However, the snake often did not form clear periodic wave forms during traversal. The animal's mid-body position (midway from the head and from the tail) traveled along its trajectory an accumulated distance of 52.8 ± 23.2 cm (0.64 ± 0.28 body length) within 15.5 ± 6.4 s. For the entire reconstructed body, the average longitudinal velocity was 3.9 ± 1.4 cm $s^{-1}$ (4.8 ± 1.7% body length $s^{-1}$) and the highest average longitudinal velocity in all video frames from all trials was 18.0 cm $s^{-1}$ (22% body length $s^{-1}$). The highest velocity occurred after the snake was tapped lightly on its tail. Slip angle of the entire reconstructed body (which is in 3-D) was small (12° ± 3°), around 41% of that when corn snakes move on a rigid, smooth substrate (28° (Schiebel et al., 2020b). Transverse velocity was only 16% of longitudinal velocity (0.6 vs. 3.9 cm $s^{-1}$; **Figure 30**B, C; $t(29) = 15.52$, $P < 0.0001$, paired *t*-test). For the body sections in contact with the terrain, slip angle was 10° ± 3°, and transverse velocity was only 14% of longitudinal velocity (0.5 vs. 3.8 cm $s^{-1}$; **Figure 30**C; $t(29) = 15.23$, $P < 0.0001$, paired *t*-test).

The anterior region moved out of the virtual tube more than the main body region (**Figure 30**D; $t(29) = 9.35$, $P < 0.0001$, paired *t*-test), with a twice larger slip angle (anterior: 19.70° ± 6.4° vs. main: 9.4° ± 2.3°, respectively). Video observation indicated that this may result from the exploration behavior of the head which occurred in all trials (see Movie 5 for an example). The anterior region frequently moved laterally or dorsoventrally as if exploring and selecting a path, while the main body region mostly followed the path of the anterior points.

### 3.5.2 Body-terrain contact

The animal tended to move through lower "valleys" surrounded by higher neighboring blocks. Average height of blocks directly below snake body in all trials was 1.2 cm (147 % body height) smaller than that of neighboring blocks on average (**Figure 31**A; $t(29) = 6.66$, $P < 0.0001$, paired *t*-test). The average height of the blocks under the animal was smaller than that of the neighboring blocks in 100% (median) of the video frames across all trials of all individuals (**Figure 36**A).

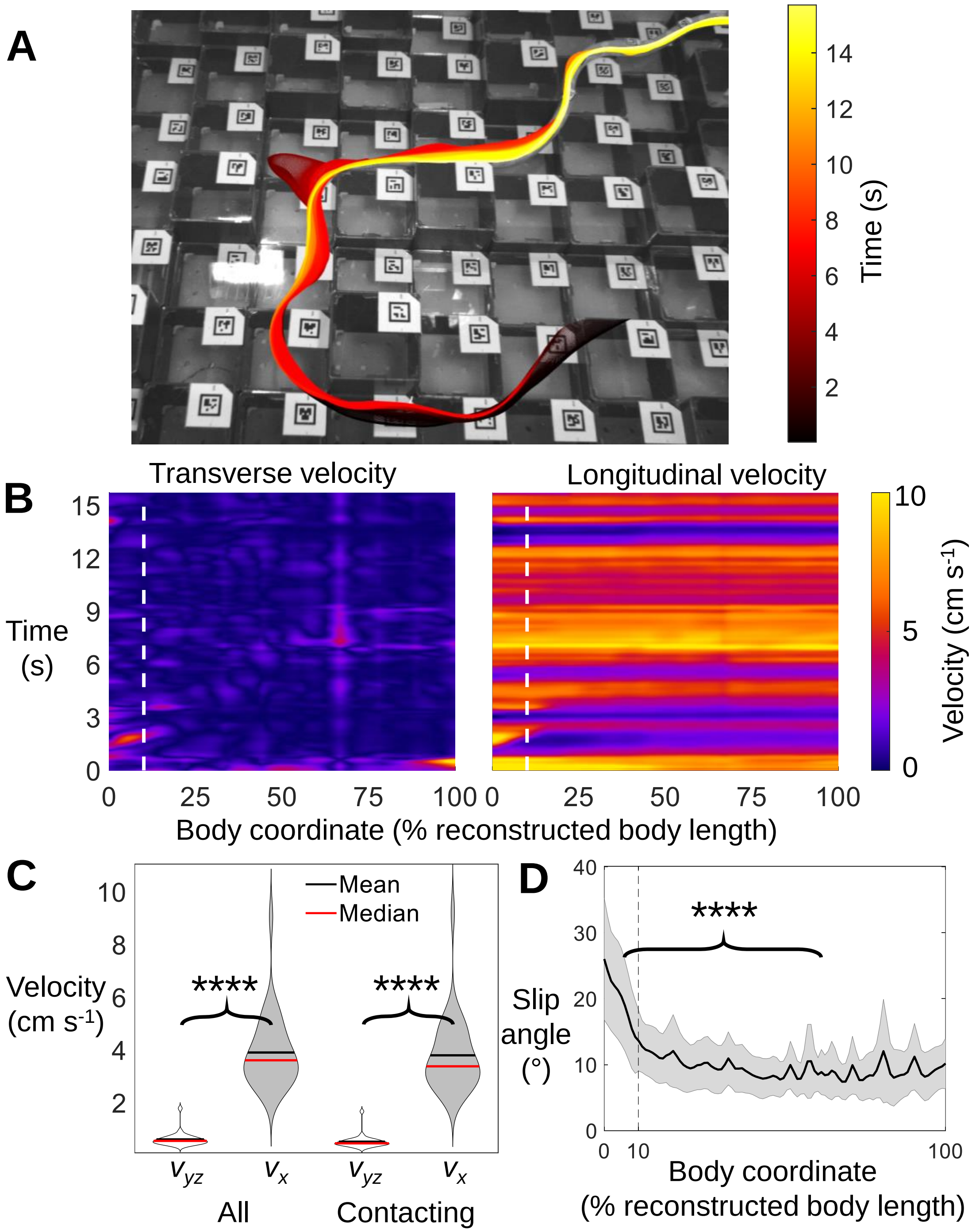

**Figure 30. Representative snapshot showing little transverse motion. (A)** Representative snapshot of snake with reconstructed midline overlaid at different time instances during traversal of uneven terrain. Midline color changes from black to light yellow with elapse of time from start to end of traversal. See Movie 6 for a representative video. **(B)** Spatiotemporal profiles of transverse velocity (left) and longitudinal velocity (right) as functions of body coordinate and time in the same trial in (A). Vertical white dashed lines indicate division between the anterior region and main body region. This representative trial does not contain phases in which the snake stops (i.e., video frames in which the average longitudinal velocity along the entire reconstructed body is small ($< 0.125$ cm $s^{-1}$)). Other trials may contain such phases but these video frames are excluded from most analyses (see Chapter 3.4.5). Body coordinate starts from 0% at the most anterior marker and ends at 100% at the most posterior marker. **(C)** Transverse ($v_{yz}$, white) and longitudinal ($v_x$, gray) velocity of all the reconstructed body sections (left) and of the body sections in contact with the terrain (right). Data is shown using violin plots. Black and red lines show mean and median. Local width of graph is proportional to smoothed probability density of data along the y-axis (Hoffmann, 2021). **(D)** Slip angle along the body. Black curve and shaded area show mean ± s.d. Gray dashed lines indicate division between the anterior region and the main body region. Brackets and asterisks represent a significant difference (****$P < 0.0001$, paired $t$-test). $N = 3$ individuals, $n = 30$ trials.

Despite this tendency, lateral contact with higher blocks was not utilized by the animal more frequently than vertical contact during traversal (**Figure 31**B). The number of vertical contact regions in all trials was statistically larger than the number of lateral contact regions (3.5 vs. 2.8; **Figure 31**B; $t(29) = 2.66$, $P < 0.05$, paired $t$-test). In 3 out of all 30 trials, the number of vertical contact regions along the body during traversal was more than 4 times that of lateral contact regions (see Movie 5 for an example). However, the maximum ratio of the number of lateral contact regions with respect to that of vertical contact regions in all trials was only 2.2.

The animal used lateral and vertical bending similarly often to form potentially propulsive contact regions, with 1.6 ± 0.7 potentially propulsive lateral and 1.7 ± 0.7 vertical contact regions across all trials, respectively (**Figure 31**C; $t(29) = 0.52$, $P = 0.61$, paired *t*-test). In 8% (median; with interquartile range of 1% to 23%) or 7% (median; with interquartile range of 1% to 20%) of video frames in each trial, all the potentially propulsive regions were purely vertical contact region or purely lateral contact regions, respectively. The animal can also form potentially braking contact regions with lateral bending (1.2 ± 0.7) and more with vertical bending (1.8 ± 0.7) bending across all trials (**Figure 31**D; $t(29) = 4.57$, $P < 0.0001$, paired *t*-test).

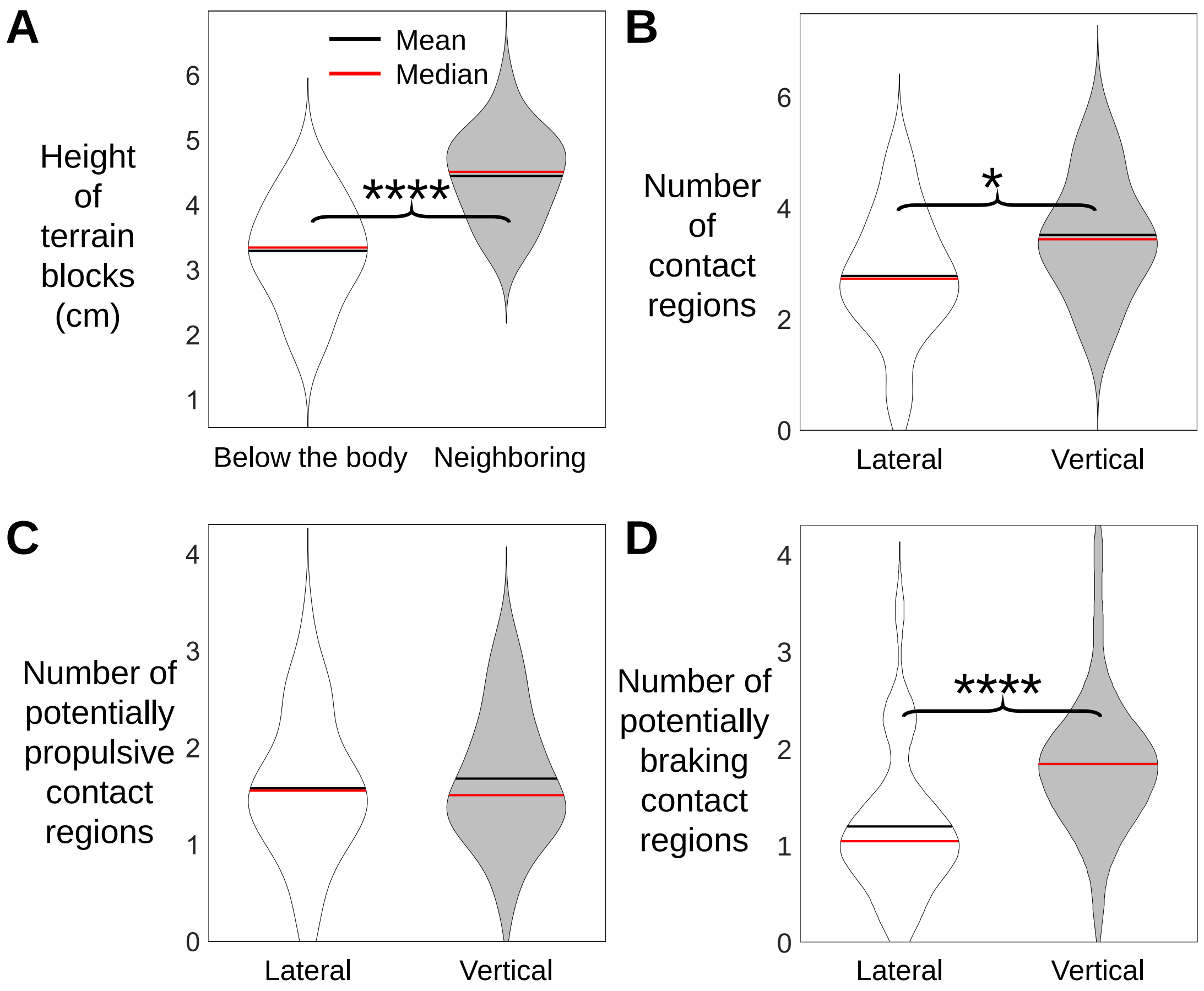

**Figure 31. Quantification of movement in a valley and comparison of contact types. (A)** Comparison of height of terrain blocks directly below snake body and height of neighboring blocks. **(B)** Comparison of number of lateral and vertical contact regions. **(C-D)** Comparison of number of lateral and vertical contact regions that are likely propulsive (C) or braking (D). Data are shown using violin plots. Black and red lines show mean and median, respectively. Local width of graph is proportional to the probability density of data along the y-axis (Hoffmann, 2021). Brackets and asterisks represent a significant difference (*****P* < 0.0001, **P* < 0.05, paired *t*-test). $N = 3$ individuals, $n = 30$ trials.

The animal traversed the uneven terrain with perfect stability, with the center of mass falling within the support polygon formed by body segments in contact with horizontal surfaces all the time (**Figure 36**B; median across all trials from all individuals: 100%). Video observation showed that the few video frames estimated to be unstable resulted from the underestimation of support polygon. This is because we could not interpolate the body shape beyond the most anterior and the most posterior markers.

# 3.6 Discussion

## 3.6.1 Contribution and implications

Our observations and quantification of types of body-terrain contact supported the hypothesis that vertical bending is used by generalist snakes to push terrain as frequently as lateral body bending during traversal of uneven terrain. We observed that supported contact regions sometimes (0.9 ± 0.8 regions across all trials) experienced kinetic friction pointing toward the center of mass velocity. These supported regions occurred when the snake was bending in a U shape in the horizontal plane. When the center of mass velocity pointed towards the head, the rear part of the U-shaped body had local velocities that were opposite to center of mass velocity and generated such friction. However, it remains to be studied whether the supported regions were actively controlled to generate propulsion using static or kinetic friction.

The combination of lateral and vertical bending in 3-D may drastically expand the range of natural surfaces available for force generation, either propulsive or braking, in all but the smoothest environments (Gart et al., 2019; Jurestovsky et al., 2021), by allowing each part of the entire body to adaptively push against its nearby terrain surfaces. This expanded range would allow snakes to better maintain propulsive forces to overcome frictional resistance continuously. This is important because, unlike legged locomotion which is affected significantly by inertial forces (except for tiny animals like ants (Clifton et al., 2020; Hooper, 2012) and mites (Weihmann et al., 2015)) and allows momentary loss of propulsive forces during continuous movement, terrestrial limbless slithering is mostly dominated by frictional forces and stops immediately after losing propulsion (Chong et al., 2021; Hu et al., 2009). The expanded range may also give snakes more redundancy to adjust distribution of contact forces to improve propulsion, stability, maneuverability, and efficiency, contributing to snake's locomotor versatility.

One potential advantage of vertical bending over lateral bending in providing propulsion is that obtaining vertical contact points is relatively easier in certain environments that have a small density of asperities large enough for lateral contact but substantial height variation over the entire body length, such as when snakes move over horizontal branches (Jurestovsky et al., 2021), travel down large boulders or move inside vertically bent tunnels. In such environments, the slender, elongated body has a high probability to ventrally contact terrain structures with height differences that are available for propulsion using vertical bending. Gravity pulls part of the body down to overcome frictional resistance, and continuous bending propagation allows such a process to continue at posterior body sections as long as height differences exist. The lower the belly friction is, the smaller slope angle is needed for the gravity to overcome frictional resistance, and thus a greater fraction of environmental surfaces can be utilized by using vertical bending to generate propulsion. However, to contact large asperities using lateral bending, it may need to reach laterally for a long distance before contacting such structures. Another potential advantage of vertical bending for propulsion is that force components along undesired directions can be easier to balance for stability by gravitational force. By contrast, lateral bending to contact vertical structures may be difficult to perform continuously without large yawing or lateral slipping unless there is a sufficient density of suitable asperities on both sides of the body (Gans, 1962).

Our results showed that snakes can seamlessly combine vertical and lateral body bending to generate propulsion in a three-dimensional complex environment. Combined with a recent study showing that snakes can generate propulsive force from vertical bending (Jurestovsky et al., 2021) much like lateral bending (Gray and Lissmann, 1950), this suggests that lateral undulation (Jayne, 2020) and vertical undulation are merely special cases induced by vertically and laterally homogenous environments, respectively, of an inherently three-dimensional behavior. This raises the question of whether such slithering using 3-D body bending propagation should be classified as a general mode of limbless locomotion (Jayne, 2020).

### 3.6.2 Limitation and future work

Our study is only an initial step towards understanding how snakes should combine vertical and lateral body bending to push against and move through the 3-D world. To further confirm our hypothesis, we must further measure 3-D contact forces between the body and terrain. This is challenging because high-fidelity commercial 3-D force sensors are expensive (Han et al., 2021; Jurestovsky et al., 2021) whereas low cost, customizable force sensors are typically 2-D and have low fidelity (Fu and Li, 2021; Kalantari et al., 2012;

Liljebäck et al., 2012b; Shimojo et al., 2007; Sundaram et al., 2019). We are developing a proof-of-concept custom 3-D force sensor achieving high fidelity with a relatively low cost. We still need to create a complex 3-D terrain platform with these force sensors embedded and controlled by data acquisition systems to ensure a sufficient sampling frequency.

To further understand how animals control 3-D body bending, measurements of muscle activity and neural signals are needed to answer the following questions: Does the animal also actively control scale (Marvi and Hu, 2012) and skin (Newman and Jayne, 2018) movement during this process? Is propulsion generation using vertical body bending actively controlled throughout the body by the propagation of epaxial muscular activation (Jayne, 1988; Moon and Gans, 1998), or can snakes use gravity (Jorgensen and Jayne, 2017) to facilitate this process? Although the 3-D shape did not change much as it was passed down the body, the animal often did not form clear alternating left and right bends (a hallmark of lateral undulation (Jayne, 2020)) or alternating up and down bends. Does bending in 3-D require using muscles differently from that during lateral undulation in terrestrial (Jayne, 1988; Moon and Gans, 1998) and arboreal (Astley and Jayne, 2009) environments?

In addition, future studies should test how generalist snakes modify their 3-D body bending to adapt to terrain properties change. Does their preference of using lateral or vertical bending depend on their habitat terrain properties, such as the spatial density of lateral and vertical push points available (Jayne and Herrmann, 2011; Schiebel et al., 2020b; Sponberg and Full, 2008) or friction (Alben, 2013; Gray, 1946; Marvi and Hu, 2012; Zhang et al., 2021)?

More broadly, it remains to be discovered how generally the combination of lateral and vertical bending is utilized by other snake species and other limbless clades in various terrains for propulsion. Aside from the locomotor generalist corn snakes studied here, other snakes including boas, pythons, sunbeam snakes, and many other colubrids have been observed to use similar movements (H. C. Astley, personal observation). Other limbless clades such as worms (Dorgan, 2015; Kwon et al., 2013) and fish (Ekeberg et al., 1995; Gidmark et al., 2011; Tatom-Naecker and Westneat, 2018) can also bend the body in three dimensions, but previous studies had focused on homogeneous environments like agar, gelatin, or sand until very recently (Pierce et al., 2021). Future studies will test this and reveal how the effectiveness of this strategy depends on the specie's specific neuromechanics, such as body bending capacity (Jurestovsky et al., 2020; Kelley et al., 1997), mechanical (Donatelli et al., 2017) and controlled (Marvi and Hu, 2012; Newman and Jayne, 2018) local compliance, muscular torque capability in each direction (Astley, 2020b; Long Jr, 1998), and sensing and neural control capacity (Sulston et al., 1983). This strategy's effectiveness is also likely affected by habitat terrain properties, such as push point density (Majmudar et al., 2012), friction (Dorgan et al., 2013), deformability (Gu et al., 2017), and heterogeneity (Mitchell and Soga, 2005).

These comparative studies will provide insight into their habitat use and the links between habitat, morphologies, biomechanics, and performance within and between species. For instance, unlike limbed animals that generate propulsion by stepping on surfaces with slope grades shallower than the coefficient of friction (i.e., operate within the friction cone (Klein and Kittivatcharapong, 1990)) to avoid slipping, limbless animals may prefer utilizing surfaces with slope grades steeper than the coefficient of friction (i.e., operate outside the friction cone) in order to slither through. This would allow limbless animals to shelter in complex, confined environments cluttered with heterogeneous structures that are challenging for limbed animals, which may explain dozens of independent evolutionary convergences of limbless species (Gans, 1986).

The combination of lateral and vertical bending in 3-D should also be used by snake robots to fully exploit environmental surfaces with various positions and orientations for propulsion and stability. The wide range of contact points available may offer snake robots robustness against unexpected perturbations such as sudden slipping, collisions from other objects, and loss of existing contact. Meanwhile, contact forces at multiple contact points must be coordinated to generate propulsion along desired directions and balanced to maintain stability. To achieve this, terrain contact force sensing and force feedback controllers (Fu and Li, 2021; Kano and Ishiguro, 2020; Liljebäck et al., 2014a; Ramesh et al., 2022) are needed to sense and adaptively control body bending to maintain contact with the terrain. Snake robots with terrain force sensors and feedback controllers can also be used as robophysical models (Aguilar et al., 2016) to study the principles of using exteroceptive feedback in control. For example, force measurement collected while systematically varying bending strategies can help understand how shape changes are related to contact changes (Fu and Li, 2021; Ramesh et al., 2022). A combination of centralized and decentralized controllers can be tested to study whether and how animal may use similar control mechanisms in the spinal cord to generate complex, robust locomotion patterns (Thandiackal et al., 2021).

## 3.7 Appendix

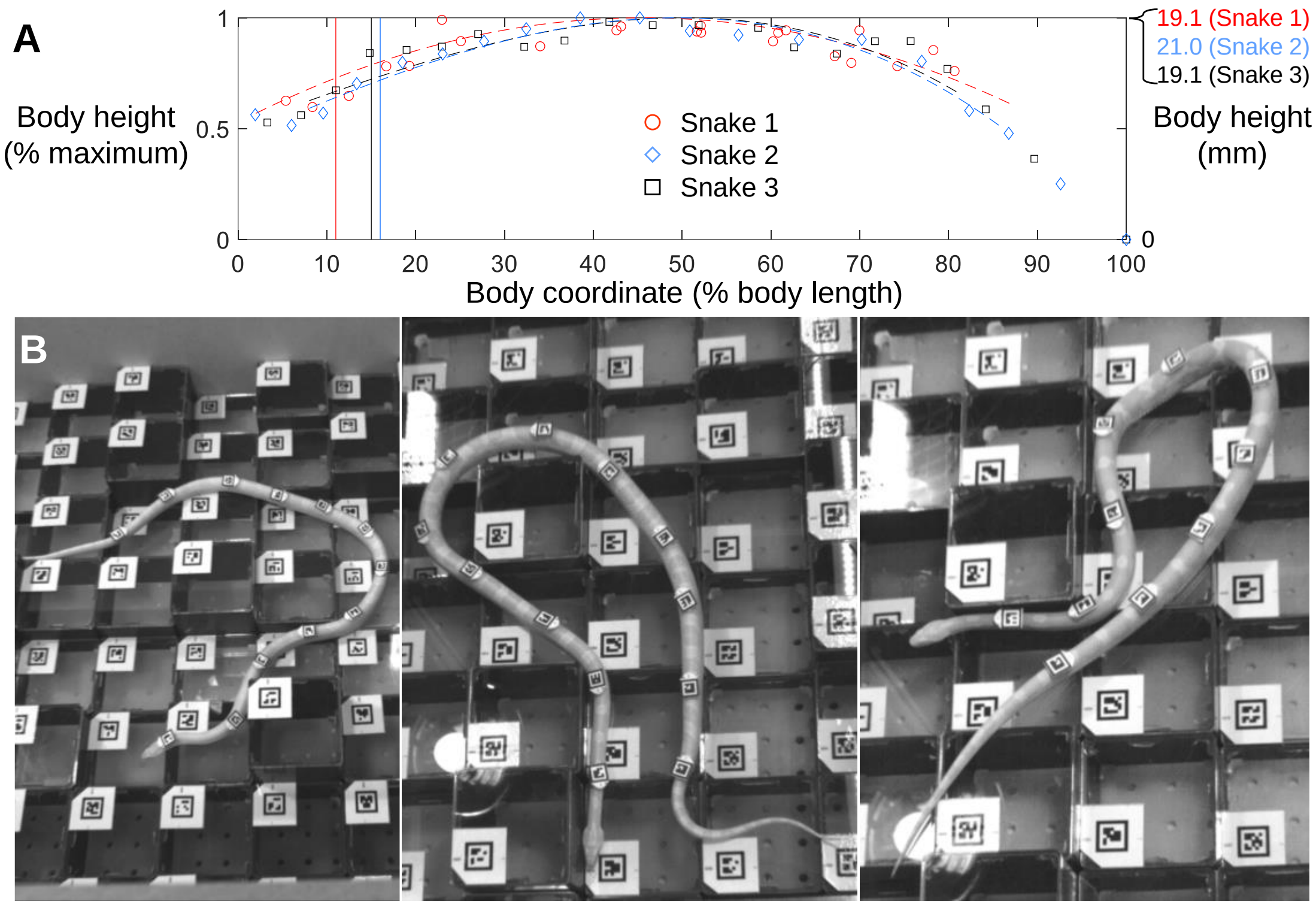

**Figure 32. Body height and marker distribution along the body. (A)** Measured and fitted height distribution along the body for each individual. Markers show measured height, dashed lines show fitted quadratic polynomials for each individual with corresponding color (trimmed to start from the most anterior marker to the most posterior marker on the body), solid lines show division between the anterior region and the main body region. **(B)** Photos of animals with BEEtag markers attached to the body. Between 10 and 12 BEEtag markers are attached along the dorsal side of the snake equally spaced between neck and vent, covering an average of 79% of full body length (96% of full body volume) as estimated from tapering data in (A).

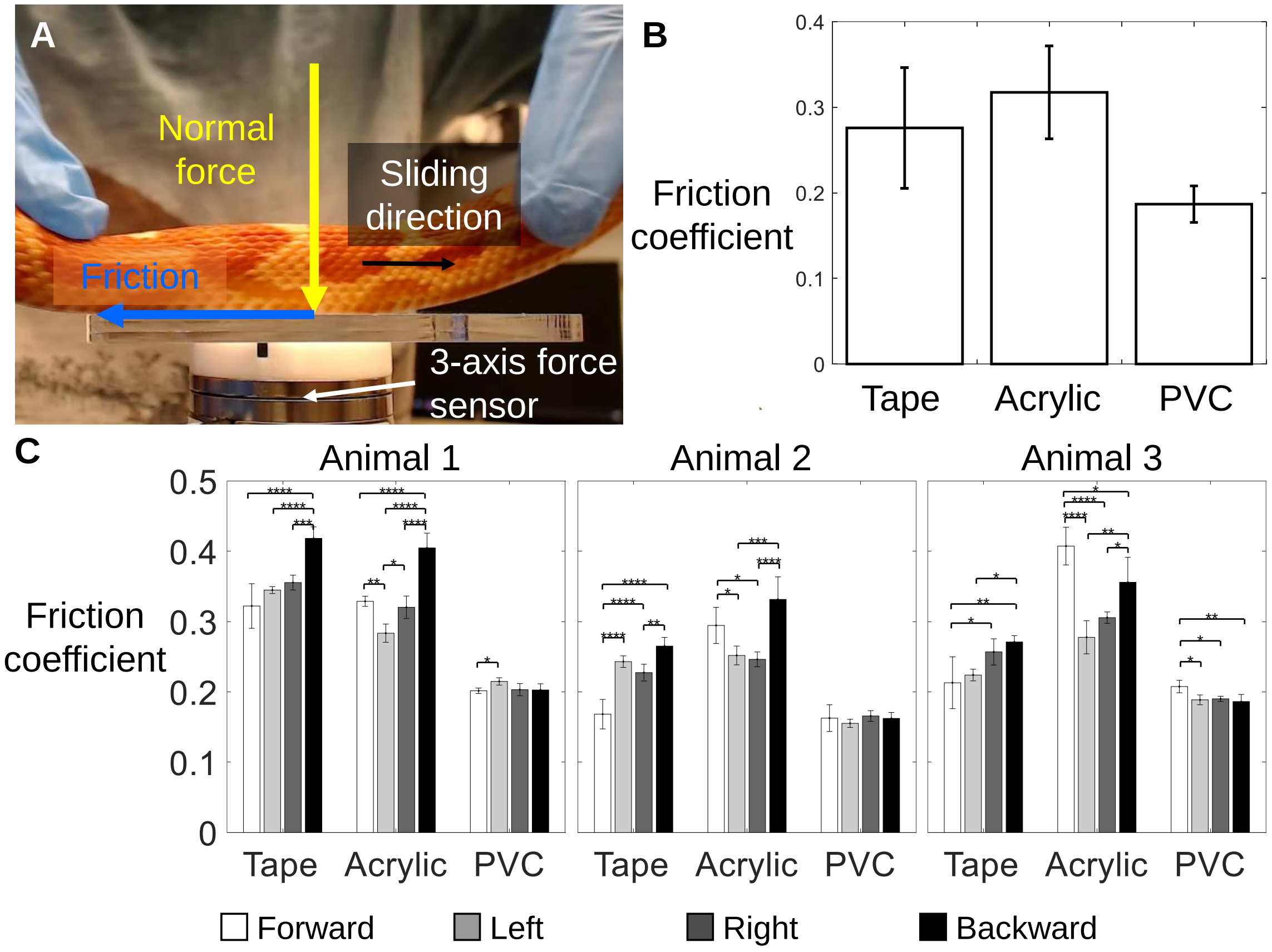

**Figure 33. Measurement of kinetic friction coefficients. (A)** Setup to measure kinetic friction coefficients between snake body and terrain surfaces. A 3-axis force sensor measures normal force and friction applied to plate while a snake is sliding against plate. Top surface of the plate is covered by acrylic, packaging tape, or PVC in different measurements. **(B-C)** Measured kinetic friction coefficient between snake body and tape, acrylic, and PVC, plotted after pooling all individuals and directions (B) and separately for each individual and each direction (C). Error bars show ± 1 s.d. Brackets and asterisks represent statistically significant differences between two directions (*$P$ < 0.05; **$P$ < 0.005; ***$P$ < 0.0005; ****$P$ < 0.0001, ANOVA, Tukey HSD).

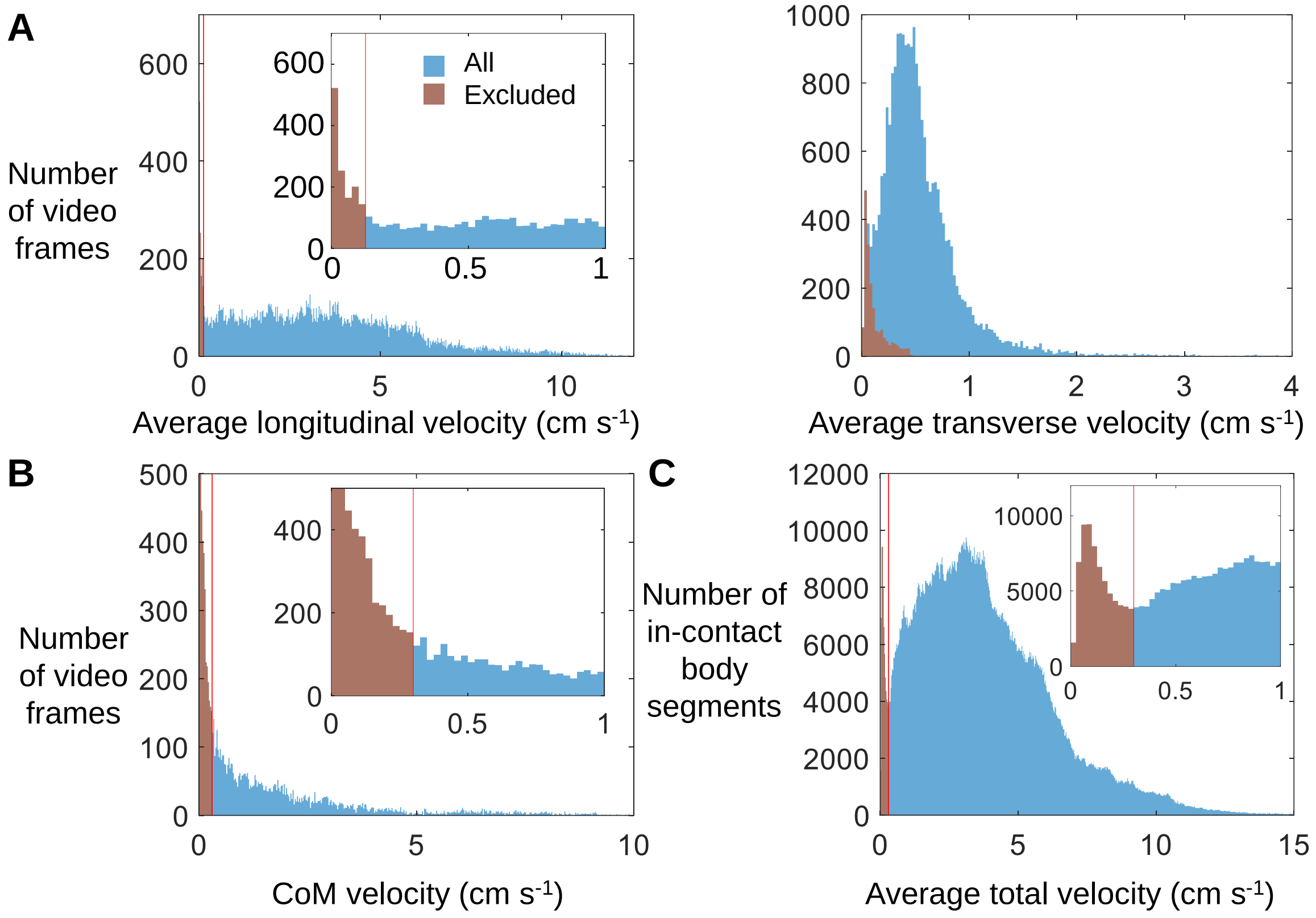

**Figure 34. Histograms of velocities used when selecting thresholds to exclude data. (A)** Histograms of average longitudinal (left) and average transverse (right) velocities (averaged along the entire reconstructed body) in all video frames pooled from all trials of all individuals (blue) and the excluded phases in which the snake stops (i.e., video frames with a small (< 0.125 cm $s^{-1}$) average longitudinal velocity) (brown). **(B)** Histogram of center of mass (CoM) velocity in all video frames (excluding phases in which the snake stops) pooled from all trials of all individuals (blue) and the excluded video frames in which CoM velocity is small (< 0.3 cm $s^{-1}$) (brown). **(C)** Histogram of average total velocity of the body segments in contact with the terrain pooled from all video frames (excluding phases in which the snake stops or CoM velocity is small) in all trials of all individuals (blue) and the excluded body segments in contact with the terrain which has a small (< 0.3 cm $s^{-1}$) total velocity (brown). Red line shows thresholds of exclusion. Inset shows close-up views at small velocities.

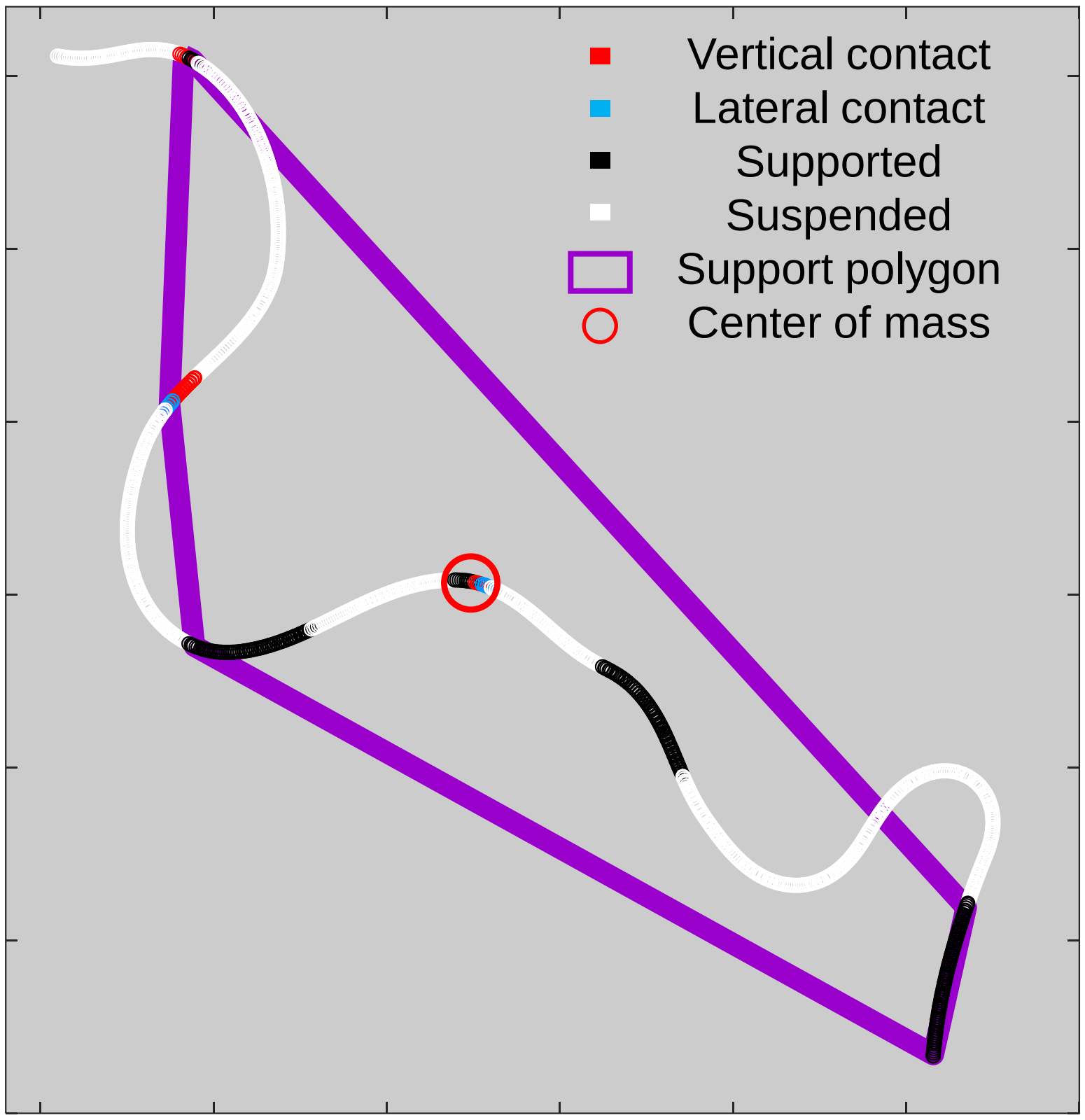

**Figure 35. Static stability analysis.** Thick curve shows body segments color-coded by contact types. Purple polygon shows support polygon, a convex region formed by body segments in contact with horizontal surfaces. Red circle shows center of mass. When center of mass is inside support polygon, the snake is statically stable.

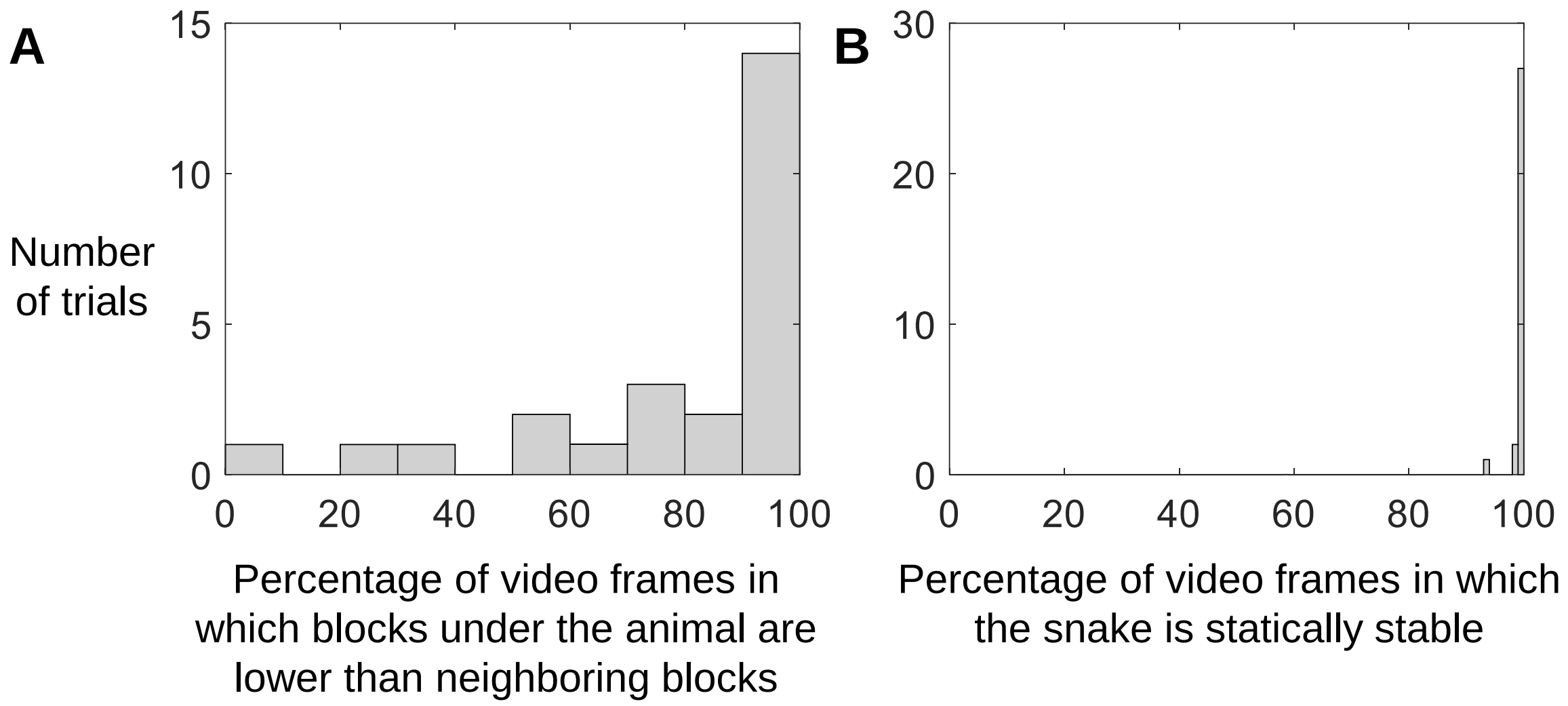

**Figure 36. Histograms of percentage of video frames in which the snake tends to move on lower blocks and in which the snake is statically stable. (A)** Histogram of the percentage of video frames in which the average height of blocks under the animal body is smaller than that of neighboring blocks. **(B)** Histogram of the percentage of video frames in which the center of mass falls within the support polygon formed by body segments in contact with horizontal surfaces (i.e., when the snake is statically stable).

# Chapter 4

# Snake robot traversing large obstacles using vertical bending reveals importance of contact feedback for propulsion generation

This chapter was previously posted on arXiv as a preprint entitled *Snake robot traversing large obstacles using vertical bending reveals importance of contact feedback for propulsion generation* authored by Qiyuan Fu and Chen Li (Fu and Li, 2021). This work has also been submitted to *Bioinspiration and Biomimetics* and is in revision at the time of submission of this dissertation.

## 4.1 Author contributions

Qiyuan Fu designed study, developed robot, performed experiments, analyzed data, and wrote the paper; Chen Li designed and oversaw study and revised the paper.

## 4.2 Summary

Snakes can bend their elongate bodies in various forms to traverse many types of environments. We understand well how snakes use lateral body bending to push against asperities on flat ground to generate propulsion, and snake robots are already effective in doing so. Recent studies revealed that snakes also use vertical bending to push against uneven terrain of large height variation for propulsion and they can adjust vertical body bending to adapt to novel terrain, presumably using mechano-sensing feedback control. Although some snake robots can move through uneven terrain, few have deliberately used vertical bending for propulsion, largely because how to control vertical bending to generate propulsion against the environment is poorly understood. To make progress, here we systematically studied a snake robot instrumented with force sensors traversing large obstacles using vertical bending. To test whether and how well sensory feedback control helps adapt to perturbations, we compared a feedforward controller and three feedback controllers that generate distinct bending patterns and body-terrain interactions. To test whether and how well each controller adapts to various perturbations, we further challenged the robot with various perturbations that break its contact with the terrain. Using feedforward propagation of a vertical bending shape, the robot could generate large propulsion when its bending shape conformed to the terrain. However, when contact was lost under perturbations, the robot easily lost propulsion and overloaded its motors. Contact feedback control helped the robot maintain propulsion by regaining contact. Unlike propulsion generation using lateral bending, during propulsion generation using vertical bending body weight can help maintain contact with the environment but may also overload actuators. Our findings advanced our understanding of how snakes slither in the three-dimensional world and will help snake robots better traverse a diversity of complex 3-D terrain.

## 4.3 Introduction

With their slender and highly flexible body, snake robots hold the promise as a versatile platform to traverse a diversity of environments (Walker et al., 2016), especially complex 3-D terrain with large obstacles (Sanfilippo et al., 2017; Tadokoro, 2019) that challenge wheeled and legged robots. Similar to snakes (Gray and Lissmann, 1950; Kano et al., 2012; Schiebel et al., 2020b), many snake robots have been developed to use lateral bending to push against vertical structures on the lateral sides (hereafter referred to as lateral push points) to move on flat surfaces (Date and Takita, 2005; Holden et al., 2014; Kamegawa et al., 2014; Kano and Ishiguro, 2020; Liljebäck et al., 2011; Sanfilippo et al., 2016). However, the real world is rarely flat but often three-dimensional. Snakes often traverse 3-D terrain with large height variations but lacking lateral push points, such as climbing over large boulders and fallen trees (Byrnes and Jayne, 2014; Gart et al., 2019; Jurestovsky et al., 2021). In contrast, snake robots in 3-D environments are still inferior to snakes in versatility and efficiency. One possible reason is the lack of understanding of whether and how vertical bending can generate propulsion by pushing against uneven terrain of large height variation below the body (hereafter referred to as vertical push points). Many snake robots use actuated wheels or treads for propulsion (Walker et al., 2016). Some snake robots form a rolling loop in the vertical plane to traverse small obstacles (Jing et al., 2017; Ohashi et al., 2010), which can be highly unstable due to the narrow base of support. A few snake robots traverse 3-D environments using geometric bending patterns designed based on geometry of limited terrain such as steps or pipes (Fu and Li, 2020; Kurokawa et al., 2008; Lipkin et al., 2007; Takemori et al., 2018b; Tanaka and Tanaka, 2013). Vertical bending is only used to connect body sections performing distinct movement patterns. Several snake robots can adapt to unstructured uneven terrain but suffer from severe slipping (Takemori et al., 2018b; Travers et al., 2018; Wang et al., 2020a). Although their body interacting with the terrain can bend vertically using controlled compliance to accommodate height variations, these robots generate propulsion primarily from gaits designed for flat surfaces with and without vertical protrusions, such as a sidewinding-like gait or lateral undulation. Only a few robots deliberately use vertical bending to push against vertical push points for propulsion, but they lack the ability to automatically adapt to novel terrains (Jurestovsky et al., 2021; Kano et al., 2014; Takanashi et al., 2022; Takemori et al., 2018a) or external forces (Date and Takita, 2005).

Recent studies revealed that snakes can traverse uneven terrain with large height variations using vertical bending to push against vertical push points. A corn snake traverses a row of horizontal cylinders by propagating a vertical wave down the body to push against the cylinders (**Figure 37**A) (Jurestovsky et al., 2021). The snake appears to coordinate contact forces in the vertical plane from multiple cylinders for propulsion or braking, shown as a variable fore-aft force on one cylinder (**Figure 37**A, bottom). When in a narrow channel with a wedge, a corn snake initially uses a concertina gait (**Figure 37**B, red, tightly bent body section) to brace against vertical walls to slowly (3.4 cm/s) move other parts of the body forward (Jurestovsky et al., 2021). But once it gains substantial contact with the wedge, it transitions to only propagating a vertical bending shape posteriorly to push against the wedge (**Figure 37**B, blue) to move forward more rapidly (7 cm/s). A corn snake can also traverse a 3-D uneven terrain by propagating a 3-D bending down the body (**Figure 38**) (Fu et al., 2022). During traversal, a similar number of potentially propulsive contact points are formed by lateral and vertical bending. For example, lateral bending (**Figure 38**, blue bands) can potentially push against vertical edges of higher blocks lateral to the body (yellow squares), while vertical bending (**Figure 38**, red bands) can potentially push against horizontal edges of blocks below the body (dark blue squares) (Fu et al., 2022). This combination may provide snakes with substantially more push points to coordinate contact forces to improve stability, maneuverability, and efficiency in complex 3-D environments than using a single mode.

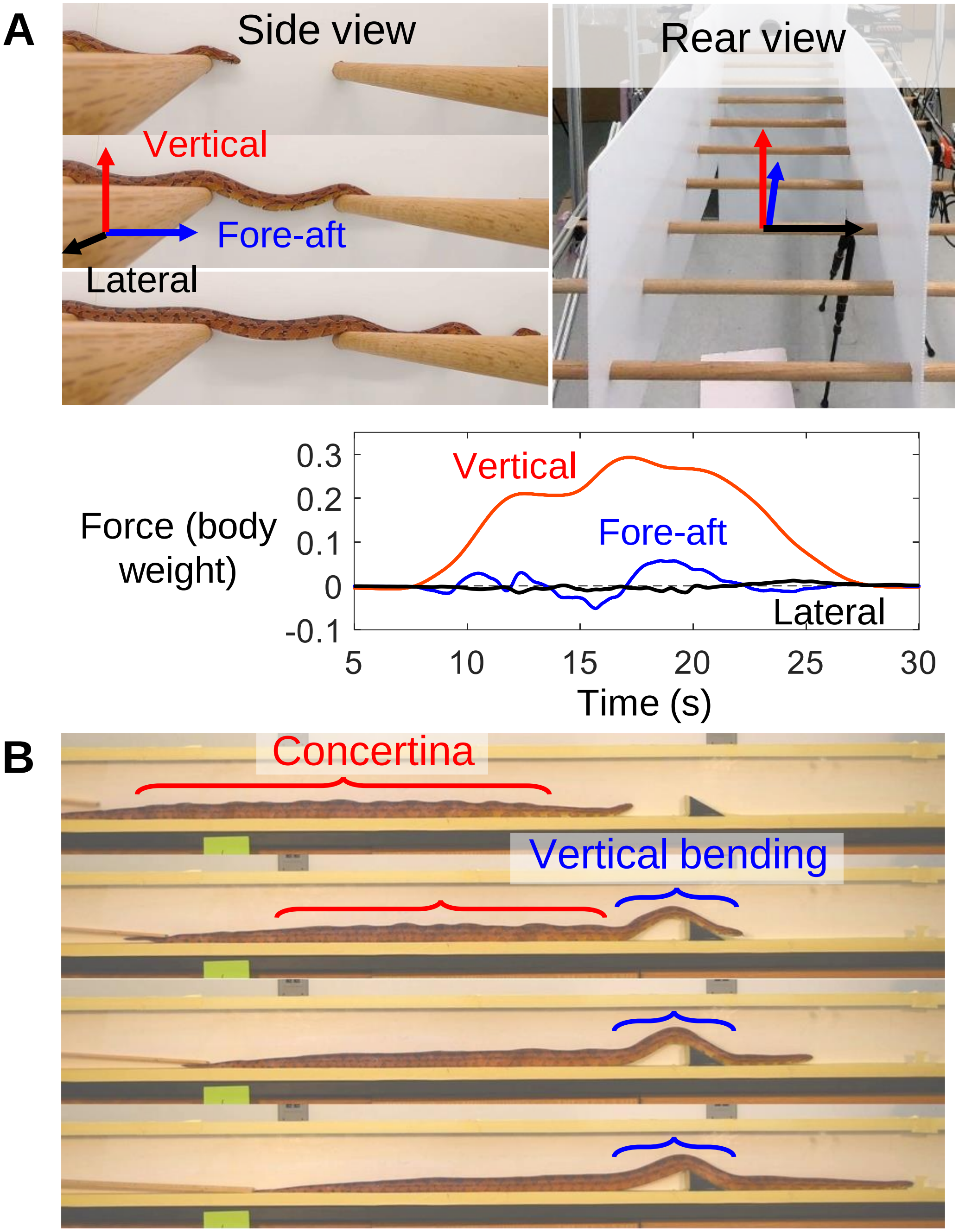

**Figure 37. Snake traversing a horizontal ladder or a wedge in a narrow tunnel. (A)** A snake traversing a horizontal ladder containing a row of horizontal cylinders by propagating a vertical

wave. Top left: representative side view snapshots. Top right: rear view of setup. Bottom: Measured terrain reaction forces on a cylinder along vertical (red), fore-aft (blue), and lateral (black) directions. **(B)** Side view of a snake traversing a wedge in a narrow tunnel (between the white back wall and a transparent front wall) using vertical bending. Snake transitions from only using concertina gait (red, evidenced by tightly bent body) to only using vertical bending (blue) once it has sufficient contact with wedge. Reproduced from (Jurestovsky et al., 2021).

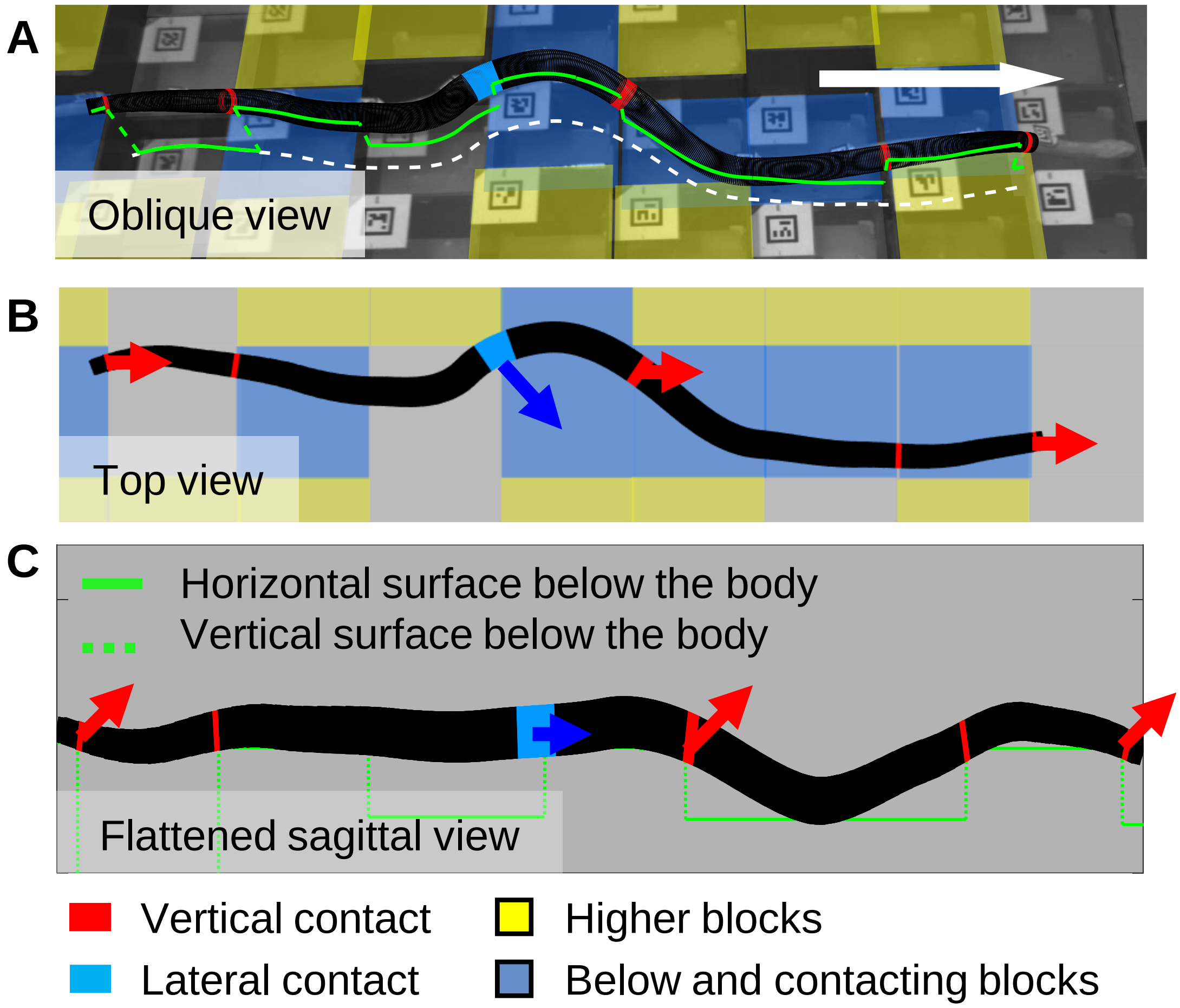

**Figure 38. Snake traversing uneven terrain.** Oblique **(A)**, top **(B)**, and flattened sagittal **(C)** view of a snake traversing a 3-D uneven terrain combining lateral and vertical bending. Snake can potentially generate propulsion by bending laterally against lateral push points (blue band with arrow) or by bending vertically against vertical push points (red band with arrow). White arrow shows direction of movement. Reproduced from (Fu et al., 2022).

This shape propagation is likely sensory-modulated, considering a generalist snake's ability to adjust bending patterns to adapt to novel terrain. For example, when encountering vertical structures with various spatial configurations on a flat surface, generalist snakes adjust lateral body bending patterns to maintain pushing against these lateral push points (Kano et al., 2012; Schiebel et al., 2020b). Snakes likely use multiple senses to guide this modulation. Most snakes likely use vision (Gans, 1975) to plan their path (Gripshover and Jayne, 2020). For example, when traversing uneven terrain with large height variation, the corn snake steers its head laterally and dorsoventrally more frequently, presumably using vision to select a path, while the rest of its body simply follows its path (Fu et al., 2022). In addition, snakes have exteroceptors within the skin to sense contact forces (Von Düring and Miller, 1979), proprioceptors within the muscles and tendons to sense stretch (Crowe, 1992), and gravity sensation (Wassersug et al., 2005). These internal or external force sensing may provide additional cues for the modulation. For example, snakes can use lateral bending to traverse a narrow tunnel on a horizontal plane where vision is unlikely modulating bending after the head, and snake robots can do so with force feedback control (Date and Takita, 2007; Gray and Lissmann, 1950; Kano and Ishiguro, 2020). In addition, a snake can modulate the lateral bending pattern when the body slips out of the path of its head or to push harder against existing contact points (Kano et al., 2012; Schiebel et al., 2020b), presumably with force feedback control.

Decoding control principles by measuring and manipulating neural activities is challenging for snakes that have complex sensory nervous systems yet to be understood (Astley, 2020a). To understand how to modulate body bending to effectively push against the environment for propulsion using sensory feedback, a snake robot is an amenable physical model that allows controlled variation of feedback control mechanisms and repeatable experimental validations (Aguilar et al., 2016). Many robot studies have investigated how sensory feedback control can help improve lateral bending to effectively push against lateral push points (Kano and Ishiguro, 2020; Sanfilippo et al., 2017). But it is unclear how to control vertical bending to generate propulsion against the environment and whether and how sensory feedback control helps it adapt to perturbations. Understanding this question will allow snake robots to exploit more terrain surfaces for propulsion generation to better traverse a diversity of complex 3-D terrain similar to snakes.

To tackle this problem, here we used a snake robot instrumented with force sensors as a physical model and drove it to traverse a track with large height variations using vertical bending. To understand whether and how sensory feedback control helps vertical bending to effectively produce propulsion, we compared four controllers that use different feedback signals and controlled states (**Figure 40**) to produce different bending patterns and terrain interactions. To test whether and how well each controller adapts to various perturbations, we tested the robot's success rates of traversal in five cases with a large bump as an initial push point. Cases (i-iii) use different additional backward loads (zero/small/large) to test how large propulsion can be continuously generated to overcome various potential longitudinal resistance in locomotion, such as large ground friction or contact forces from obstacles in the front (Schiebel et al., 2019). Case (iv) started with poor contact with the initial push point and case (v) added another unknown bump in the front. They were designed to test how well each controller can help adapt to changes in terrain geometry. We hypothesized that the robot can traverse the uneven terrain in case (i) by propagating a vertical bending shape posteriorly. If contact is maintained, the propulsion generated may also increase to accommodate additional resistance, which was evaluated in cases (ii-iii). We also hypothesized that the robot with feedforward control will struggle more when terrain geometry is changed in cases (iv-v) than in cases (i-iii). In contrast, contact feedback control can likely enable higher success rates than feedforward control in cases (iv-v). Finally, we compare our findings with previous lateral bending strategies and discuss where the differences emerge from.

## 4.4 Methods

### 4.4.1 Robotic physical model

We used a snake robot (1.18 m long, 3.0 kg) with 9 pitch and 9 yaw joints alternating as the physical model (**Figure 39**A). The alternating joint structure is common to produce 3-D motions similar to snakes (Liljebäck et al., 2012b; Mori and Hirose, 2001; Nakajima et al., 2018; Takemori et al., 2021; Wang et al., 2020a). However, because we only study vertical bending, the yaw joints were fixed to be zero. All the pitch joints were actuated by high-torque servo motors with a gear ratio of 353.5:1 (Dynamixel XM430-W350-R, ROBOTIS, Lake Forest, CA, USA). Because daisy chaining motors would limit the total current supplied to the robot, we connected the two power lines of each motor to two cables that directly drew power from a DC power supply (TekPower TP3005DM, Tektronix, Beaverton, OR, USA) at 14 V. The servo motors were connected to an Ubuntu desktop computer via an RS485 bus. Each pitch joint motor sent present joint angle and motor current to the Robot Operating System (ROS Noetic) on the computer and received goal angle or goal current commands at a frequency of 31 Hz. All the motors disabled torque output automatically when overloaded.

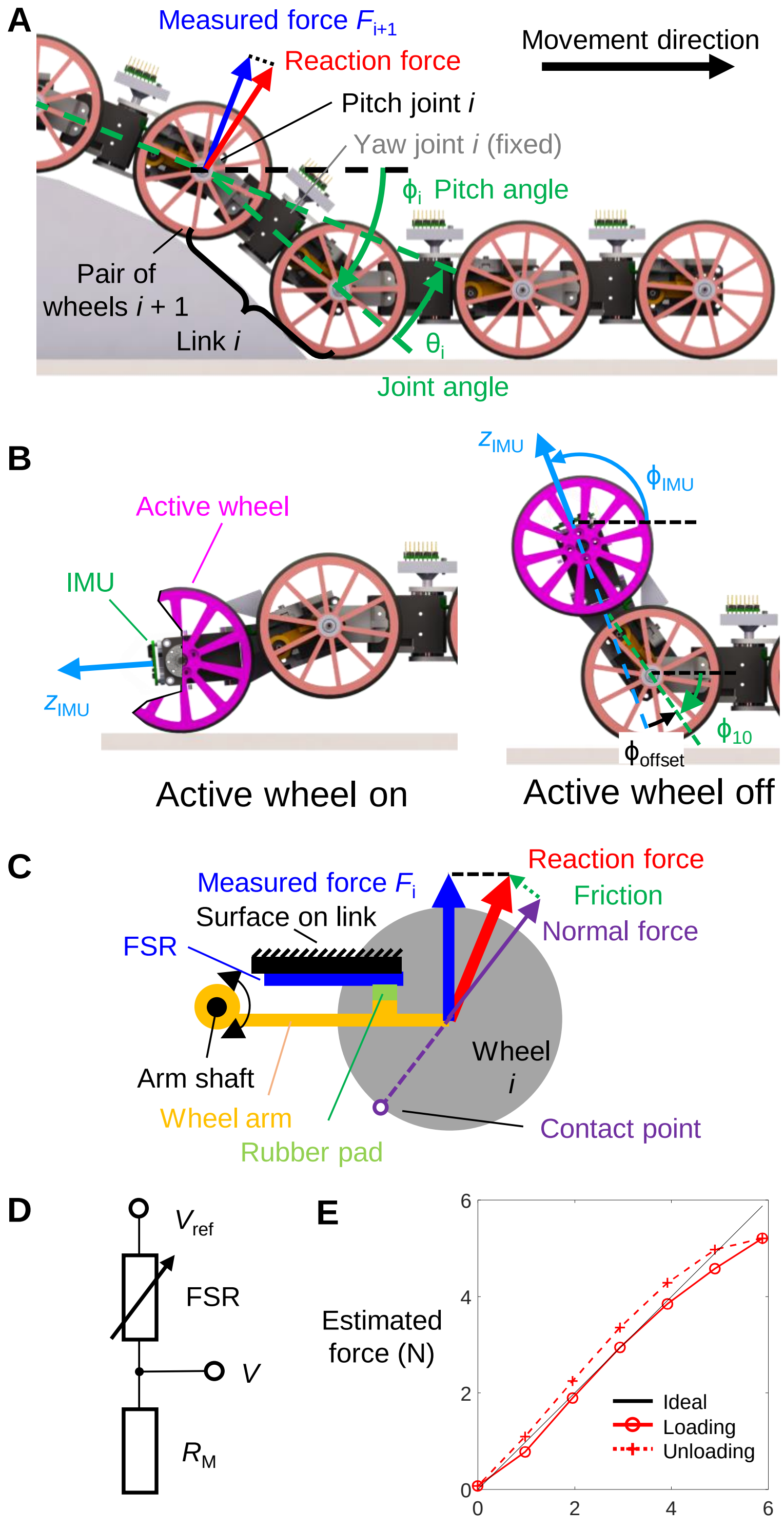

**Figure 39. Design of the robotic physical model. (A)** Overall structure. Each link contains a pitch joint motor and a fixed yaw joint motor. A pair of passive wheels (red) are installed at each end of each link. Blue vector shows measured force on wheels installed to pitch joint *i*, which is normal to link *i* + 1. Red vector shows terrain reaction force on pitch joint *i*. **(B)** Structure of active wheel module. Active wheel (magenta) is turned on and pushed against ground initially (left). It is turned off and lifted before a trial started (right). Broken out section shows IMU (green) and its *z*-axis (cyan arrow). **(C)** Wheel and force sensing resistor (FSR) installation. Wheel (gray) has its shaft attached to wheel arm (dark yellow), which has a rubber pad (green rectangle) to push on FSR (blue rectangle). Wheel arm freely rotates around an arm shaft (black circle) on link. FSR is fixated to a flat surface (black rectangle) on link. Blue, red, green, purple vectors represent measured force, terrain reaction force, friction, and normal contact force, respectively. **(D)** Measurement circuit of FSR. FSR is serially connected to a constant resistor $R_M$. Circuit is supplied with a constant voltage $V_{ref}$ and outputs a voltage $V$. **(E)** Example result of FSR calibration. Estimated force is plotted versus applied force during loading (red solid) and unloading (red dashed) in calibration.

Each robot link was defined to consist of one pitch joint motor and one yaw joint motor (**Figure 39**A). To reduce the number of contact points for easier contact force sensing, a pair of passive wheels with ball bearings (**Figure 39**A, red) were installed on the left and right sides at each end of each link. All the passive wheels had a diameter of 87 mm. The passive wheels also reduced the fore-aft friction, which allowed us to test a larger range of backward load that the robot can overcome. We measured the fore-aft friction coefficient $\mu = 0.14 \pm 0.00$ (mean ± s.d. of 3 trials) between the robot and the ground by dragging the robot longitudinally with a constant load and measuring its acceleration (**Figure 47**A).

To push the robot up the bump to the initial position before a trial started, we added an active wheel to the tail of the robot (**Figure 39**B, magenta). We did not use vertical bending in this process because no vertical push point was available for propulsion. Snakes often use lateral bending in such scenarios, such as concertina when inside a narrow tunnel (Jurestovsky et al., 2021) or lateral oscillation when on a flat ground (Gart et al., 2019). However, these approaching motions are often accompanied by variable lateral slipping and are thus not suitable for repeatable experiments with identical initial conditions. The diameter of the active wheel was the same as the passive wheels. To push the robot forward, the active wheel was pushed against the ground by the last pitch joint and rotated by a servo motor (Dynamixel XM430-W210-R, ROBOTIS, Lake Forest, CA, USA) (**Figure 39**B, left). After the robot reached the desired initial position, the active wheel stopped spinning and was lifted to reduce friction (**Figure 39**B, right; also see Chapter 4.4.3 for details). The motor spinning the active wheel was connected to the power supply and the computer in the same way as the other motors. However, it only rotated at a constant speed of 0.72 rad/s or stayed idle.

To sense ground reaction forces, we installed a force sensing resistor (FSR; FSR-400 short, Interlink Electronics, Camarillo, CA, USA; **Figure 39**C, blue) between each wheel and the corresponding robot link. To fixate and protect the sensor, we installed it on a flat surface (**Figure 39**C, black rectangle) on the link parallel to the centerline of the link. Each wheel was installed on an arm (**Figure 39**C, dark yellow) that can freely rotate around a shaft on the link (**Figure 39**C, black circle). The center of each wheel coincided with the nearest pitch joint when the wheel was pressed against the ground (**Figure 39**A). A rubber pad (**Figure 39**C, green) was added to the wheel arm to evenly distribute the exerted force over the force-sensitive area of each FSR and absorb collisional impact. Each FSR and a serially connected resistor $R_M$ = 10 kΩ were supplied with a constant voltage $V_{ref}$ = 3.3 V from a microcontroller board (Mega 2560, Arduino, Turin, Italy) to form a measurement circuit (**Figure 39**D). Four 4-channel analog-to-digital converters (ADCs) (ADS1015 breakout, SparkFun, Boulder, CO, USA) and four analog input pins on the microcontroller board were used to collect the voltage output *V* of all the 20 measurement circuits. To sense gravity, which affects vertical bending, we installed an inertial measurement unit (IMU; BNO055 breakout, Adafruit, New York, NY, USA) to the tail (**Figure 39**B). The readings from the four ADC breakouts and the IMU were first collected by the Arduino board and then sent to the computer at a sampling frequency of 23 Hz. The pitch angle of the IMU $\phi_{IMU}$ was used together with motor position feedback to estimate the pitch angle $\phi_i$ of each link in ROS: $\phi_i = \phi_{IMU} - \pi + \phi_{offset} + \sum_{j=i+1}^{10} \theta_j$, where $\phi_{offset}$ is the angle between the *z*-axis of the IMU and the centerline of the last pitch joint motor (**Figure 39**B right, cyan and green lines, respectively).

To convert the voltage reading *V* to the measured force *F*, the resistance of each force sensor *R* was first calculated using the equation $R = (V_{ref}/V - 1) \cdot R_M$. Because we observed a near linear relationship between the force *F* and the resistance *R* in the logarithmic scale (**Figure 47**C), the measured force *F* was empirically calculated by $\log(F) = k_{FSR} \cdot \log(R) + \log(g) + b_{FSR}$, where $k_{FSR}$ and $b_{FSR}$ were fitted in calibration, *g* is the local gravitational acceleration (9.81514 $m/s^2$). All the force sensors were calibrated after installation when the motors were directly fixed to an 8020 beam using 3-D printed clamps so that none of the wheels was contacting the ground. Each wheel was pushed against the FSR via a pulley system (**Figure 47**B). The force *F* applied to the wheel increased from 0 to 5.88 N (loading, **Figure 39**E, red solid) and then decreased back to 0 (unloading, **Figure 39**E, red dashed) with a step size of 0.98 N. The constants $k_{FSR}$ and $b_{FSR}$ were calculated by fitting a line to the collected values of log(*F*) and log(*R*) (**Figure 47**C). During experiments, if the measured force was negative because of fitting errors in the calibration, it was set to 0 instead. If the measured force was larger than 20 N, the maximum force-sensitive range of the FSR, it was bounded at this value.

### 4.4.2 Controller design

We implemented four controllers to propagate a vertical bending shape down the body to test the effects of different feedback usage (**Figure 40**). (1) Controller 1 is a feedforward controller propagating a pre-determined initial shape down the body. (2) Controller 2 senses contact force at the head, controls the head to maintain this contact force, and propagates the shape changes down the body. (3) Controller 3 also uses contact feedback to modulate shape propagation. Unlike controller 2 that modulates the head bending only, it also uses the sensed contact along the body and the sensed gravity direction to control the body to conform to terrain below. (4) Controller 4 is similar to controller 2 but controls internal force (joint torque) instead of the bending shape, which may allow adaptation to complex terrain without sensing the contact conditions (Date and Takita, 2005).

Note that we did not fine-tune parameters to optimize their performances; instead, we focused on analyzing how different feedback usage affects bending patterns and in turn modulates terrain interaction. The analyses can help us understand how vertical bending can produce effective interaction and how sensory feedback control can help achieve it. Aside from contributing to snake robot control, the analyses will also inform animal behaviors or sensory mechanisms that inspire these controllers. As an extreme end that does not react to any environmental changes, controller 1 served as a control group in the comparison. Controllers 2 and 4 mimic the follow-the-leader behaviors of a snake traversing uneven terrain accompanied by potential exploration behaviors of the head (Fu et al., 2022; Jurestovsky et al., 2021). Controllers 3 uses multiple sensory information that can be sensed by the snakes (Crowe, 1992; Von Düring and Miller, 1979) and generates the whole-body adaption to the terrain, similar to generalist snakes change bending shape to maintain contact with push points (Schiebel et al., 2020b).

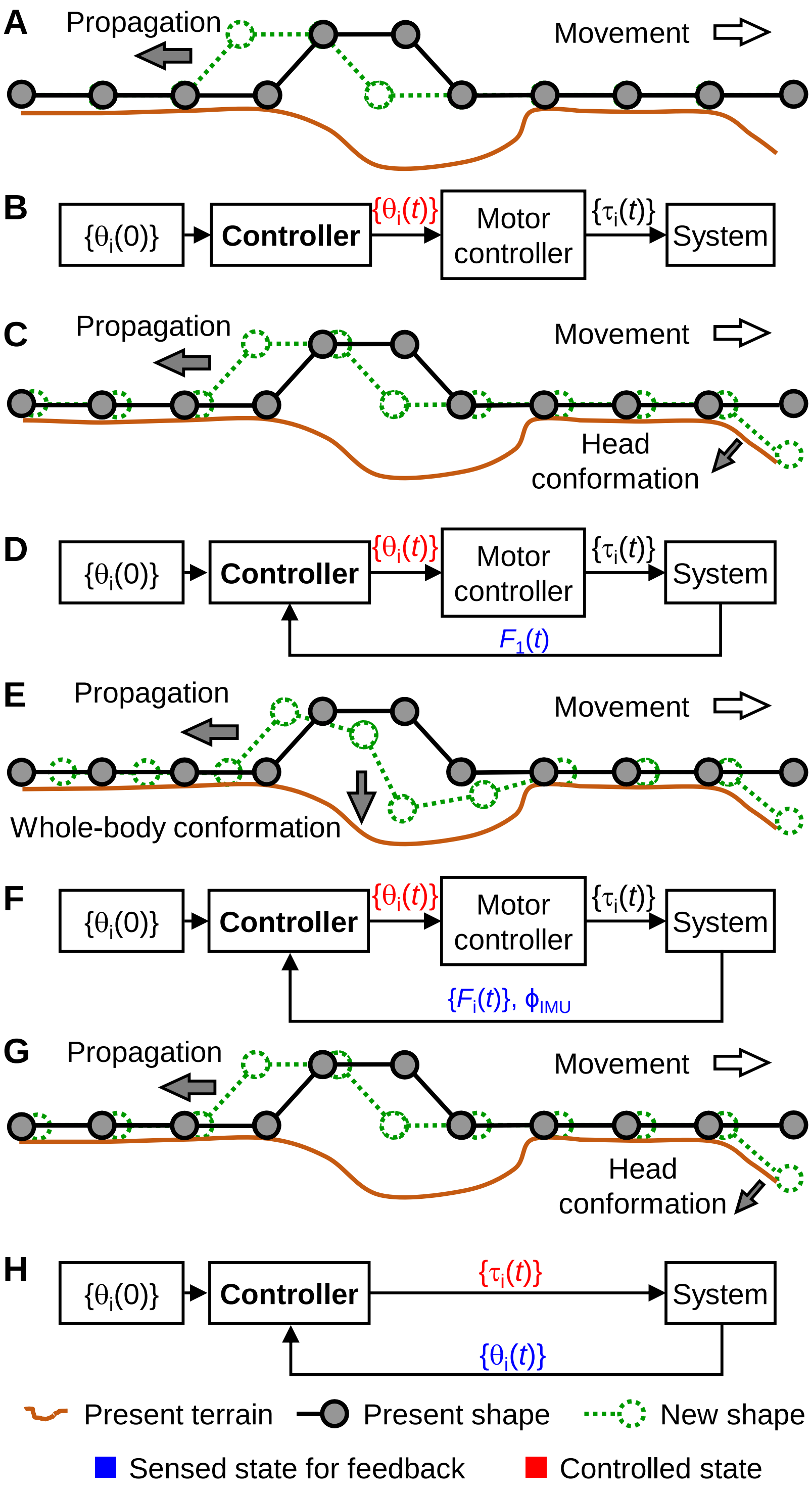

**Figure 40. Comparison of four controllers. (A-B)** Controller 1: Feedforward shape propagation. **(C-D)** Controller 2: Propagation with head conformation. **(E-F)** Controller 3: Propagation with whole-

body conformation. **(G-H)** Internal torque-based propagation with head conformation. Schematic on top of each group shows present (black solid) and new (green dashed) robot shapes and present terrain (brown solid). New shape is right aligned to present shape and only represents expected change of shape using each controller instead of actual shape and position. Diagram at bottom of each group shows how controllers use initial shape and sensed states (blue) to send controlled states (red) to embedded motor controllers and robot system.

All four controllers were implemented in ROS and ran at 100 Hz, much higher than that of measuring contact force (23 Hz), tail orientation (23 Hz), and joint angles and current (31 Hz). To accommodate the measurement and control frequencies and protect the robot, propagation speed and gains of the sensory feedback were set conservatively.

### 4.4.2.1 Feedforward shape propagation

To realize shape propagation (**Figure 41**, **Figure 40**A, B), this controller controls the angle of each pitch joint $\theta_i$ to approach that of the anterior pitch joint $\theta_{i-1}$ (Hirose, 1993) :

$$\theta_i(t) = \theta_{i-1}(nT) \cdot \frac{t-nT}{T} + \theta_i(nT) \cdot \frac{(n+1)T-t}{T}, nT \leq t < (n+1)T, i \geq 2 \quad (1)$$

where $t$ is the current time, $T$ = 8 s is the time taken to shift one joint angle down one link (hereafter referred to as the period), $n$ is the number of periods that have passed, and *i* is the index of the joint counting from the head.

The first pitch joint angle is controlled to linearly decay to 0 (the first two links being straight) and stay at zero after one period:

$$\theta_1(t) = \begin{cases} \theta_1(0) \cdot \frac{t}{T}, t < T \\ 0, t \geq T \end{cases} \quad (2)$$

With this controller, the initial shape was only propagated down the body once before the robot became completely straight excluding the active wheel module (**Figure**

**41**). Because the entire bending pattern solely depends on the pre-determined initial shape, controller 1 is essentially feedforward despite having the terms $\theta_{i-1}(nT)$ and $\theta_i(nT)$ in Equation (1).

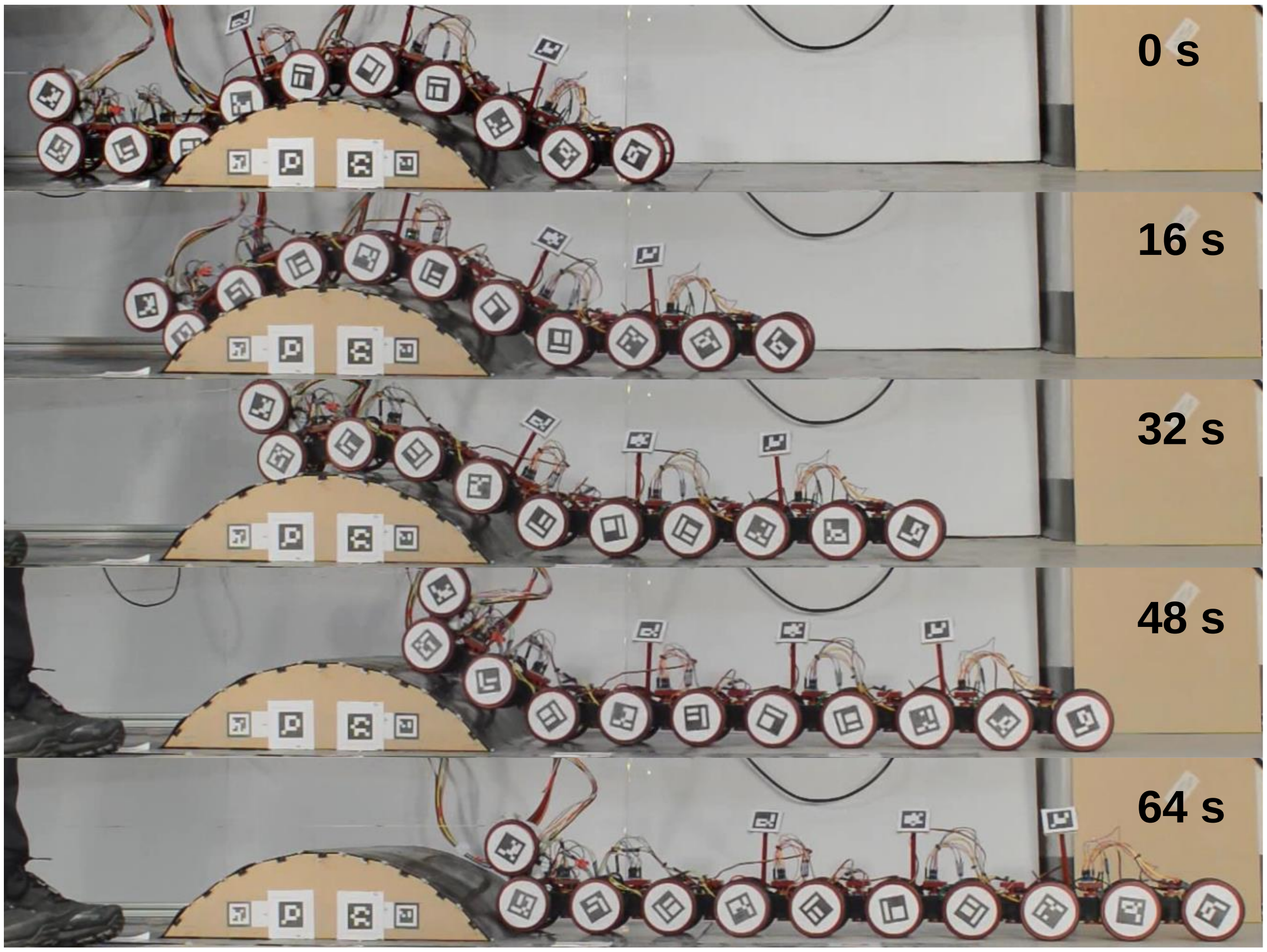

**Figure 41. Robot traversing a large bump using feedforward shape propagation.** Robot propagates initial shape posteriorly by one link length every 8 seconds. Eventually, robot becomes completely straight except active wheel module.

### 4.4.2.2 Propagation with head conformation

To propagate a shape, controller 2 controls the pitch joints after the head using Equation (1), the same as controller 1 (**Figure 40**C, D). For the head, instead of using manual input (Kano et al., 2014; Takanashi et al., 2022) or distance feedback that can be easily obstructed in complex environments (Date and Takita, 2005), this controller controls the head to automatically adapt to the ground by maintaining the contact force exerted by the head:

$$\dot{\theta}_1(t) = \begin{cases} k_1(F_1 - \bar{F}), |F_1 - \bar{F}| < \Delta F_{max} \\ k_1 \Delta F_{max} \cdot sgn(F_1 - \bar{F}), |F_1 - \bar{F}| \geq \Delta F_{max} \end{cases} \tag{3}$$

where $k_1$ = 0.4 rad·s$^{-1}$·N$^{-1}$ is a constant gain that controls the bending speed, $F_1$ is the sum of forces detected by the first pair of FSRs, $\bar{F}$ = 0.25 N is the desired contact force, and $\Delta F_{max}$= 0.25 N is the threshold to limit the bending speed. We use a small $\bar{F}$ so that the head does not push against the terrain too hard to avoid large backward resistance.

### 4.4.2.3 Propagation with whole-body conformation

To allow simultaneous shape propagation and adaptation to the ground, we used a backbone method (Liljebäck et al., 2014b) to guide the deformation (**Figure 40**E, F, **Figure 42**). The method fits the robot's discrete body (**Figure 42**A, blue line segments) to a continuous virtual curve (backbone curve; **Figure 42**A, red curves), which goes through a series of virtual shape control points (SCPs; **Figure 42**A, curves with red outlines) that can be manipulated to deform the curve.

SCPs are initially placed at the endpoints (**Figure 42**A, red solid circles) of all the robot links (**Figure 42**A, blue solid). The relative positions of SCPs are calculated using joint angle feedback and forward kinematics. A cubic spline (Kluge, 2022) is then fitted to all the SCPs and used as the backbone curve (**Figure 42**A, solid red curve). In each control cycle, the controller updates the joint angles as follows (**Figure 42**B): (1) Identify the lifted wheels (**Figure 42**A, blue dashed) with the terrain by checking whether the measured force $F_i$ equals to 0. (2) Move the SCP corresponding to each lifted wheel toward the ground: (A) If a pitch joint angle already exceeds a threshold of 60°, move the SCP normal to the local backbone curve toward the concave side by $3\Delta_{SCP}$. This is to avoid self-collision. (B) Otherwise, if an SCP belongs to a suspended section of the robot with only one end contacting the ground (**Figure 42**A, left most red solid circle), move the SCP downward and normal to the line segment connecting it and the nearest SCP whose wheel is contacting the ground (**Figure 42**A, second solid red circle from the left). (C) Otherwise, move the SCP vertically downward (**Figure 42**A, the right three green arrows). The displacement of each moved SCP in (B) and (C) is $\Delta_{SCP}$ times the number of links between the SCP and the closest wheel contacting the ground (**Figure 42**A, green arrows). (3) Fit a new backbone curve to the updated SCPs (**Figure 42**A, red dashed). (4) To propagate the shape down the body, move the most anterior SCP forward along the new backbone curve by $\Delta s$ (**Figure 42**A, right most red circle with yellow filling), then move the other SCPs along the new backbone curve such that each line segment connecting two adjacent SCPs (**Figure 42**A, blue dashed) has the same length as a robot link. (5) Calculate the new joint angles using the angles between the adjacent line segments fitted in (4) (**Figure 42**A, blue dashed). In this study, we used $\Delta_{SCP}$ = 0.1 mm, $\Delta s$ = 0.2 mm in a control cycle of 0.01 s. In this controller, force sensors were used as on-off switches that only detected whether each wheel contacted the ground or not.

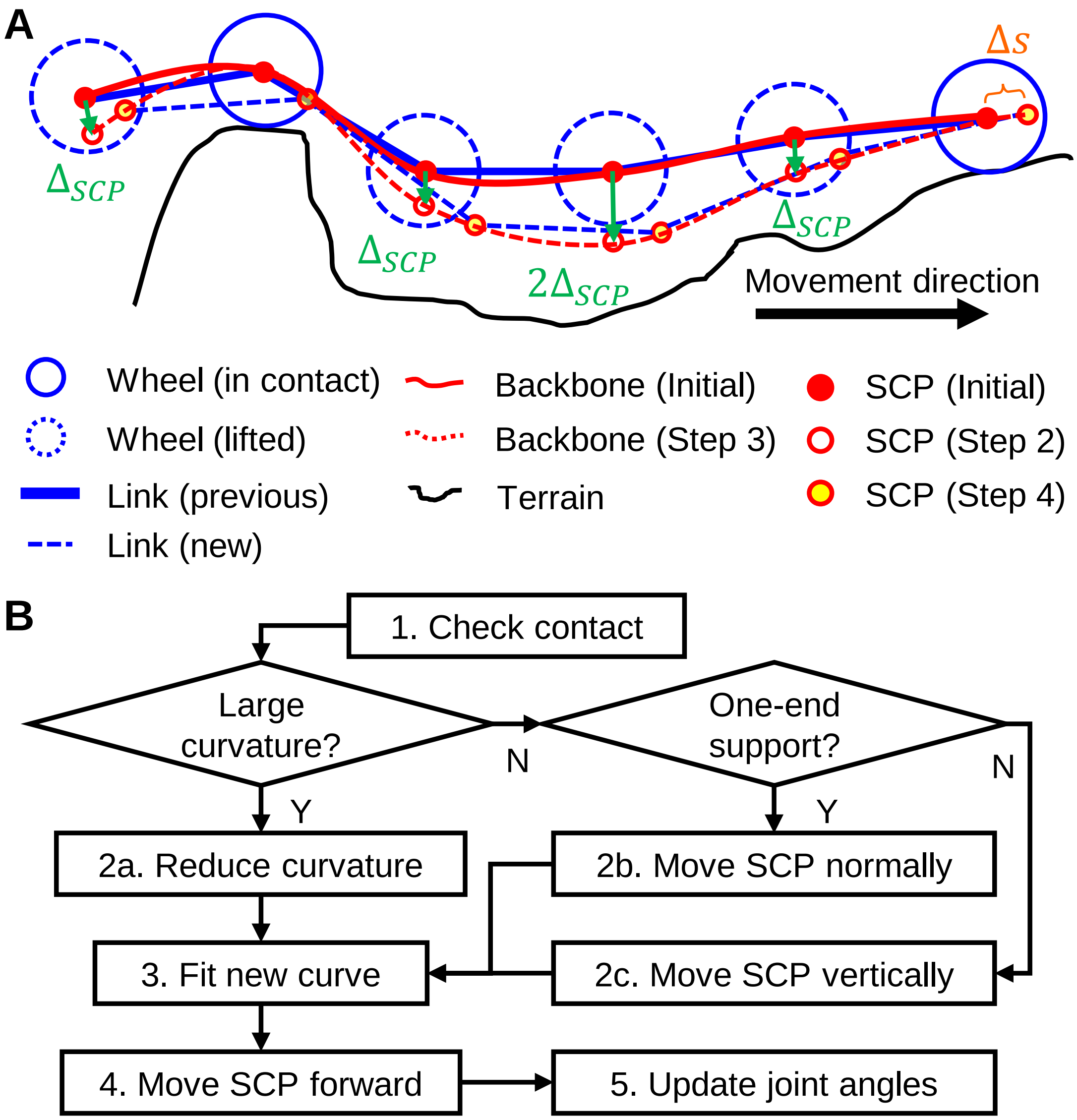

**Figure 42. Implementation of Controller 3: Propagation with whole-body conformation. (A)** Definition of backbone curve and SCPs. This controller controls robot deformation by fitting robot links (blue) to a virtual spline curve (backbone curve, red) that is controlled by virtual shape control points (SCPs, circles with red outlines). **(B)** Flowchart of a control cycle.

### 4.4.2.4 Internal torque-based propagation with head conformation

Controller 4 is adapted from a controller derived by optimizing a cost function that increases with joint torques in a theoretical continuous model traversing an arbitrary, smooth 3-D terrain (Date and Takita, 2005). Assuming that joint torque is linearly proportional to current, it is simplified as follows to use only the joint angles as feedback:

$$I_i = \begin{cases} I_{head}, i = 1 \\ K_P(\theta_i - \theta_{i-1}), i \geq 2 \end{cases} \tag{7}$$

where $I_{\text{head}}$ = 0.016 A is a constant current to bend the head toward the ground and $K_{\text{P}}$ = $8.07 \times 10^{-4}$ A·rad$^{-1}$ is a constant gain. This equation controls the head to push against the ground with a constant torque while bending each of the other pitch joints toward the angle of its anterior pitch joint (**Figure 40**G, H).

### 4.4.3 Experimental setup & protocol

To test their performance, we challenged the robot to traverse a 4-m-long track using each of the 4 controllers in each of the 5 cases with 5 trials, resulting in a total of 100 trials. A bump with a cylindrical upper surface was used as an initial vertical push point (**Figure 43**A). It was 0.49 m long, 0.12 m high, with a radius of 0.3 m, and fixed to the ground (**Figure 43**A). It was made by gluing together laser-cut 6.35 mm thick wooden sheets (McMaster-Carr, Elmhurst, IL, USA). The obstacle and the ground were covered by a rubber sheet (EPDM 60A 1.6 mm thick rubber sheet, Rubber-Cal, Fountain Valley, CA, USA) for uniform surface condition. During experiments, cables powering and controlling the robot were routed by a wheel carriage so that they always stayed vertically above the tail under manual control to minimize their effects on propulsion.

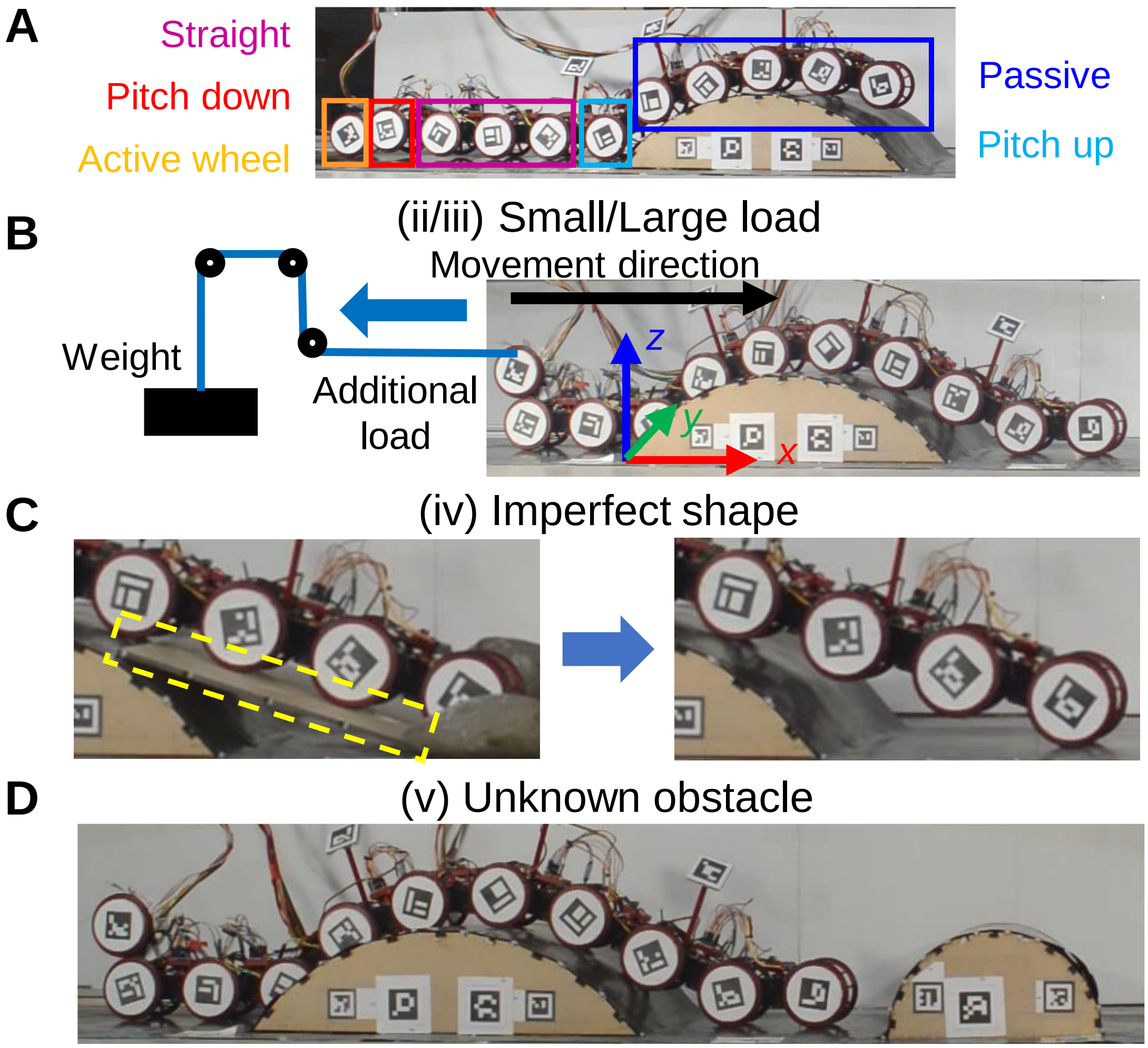

**Figure 43. Experimental setup. (A)** Initial climbing process using an active wheel. Torques are off for pitch joints in blue box and on for others. As robot is propelled forward by active wheel in orange box and hits bump using wheel in cyan box, all boxes shift down body by one link length. **(B)** Traversal with an additional backward load (cases (ii, iii)). Additional load is applied by a weight via a pulley system. **(C)** Traversal with an initial shape that has poor contact with terrain (case (iv)). An additional acrylic plate (yellow dashed box) is placed under robot before initial body shape is determined and removed before a trial starts. Stones are used to hold plate in place. **(D)** Traversal with an unknown bump in front (case (v)).

Before each trial, the robot was initially placed horizontal and straight behind the bump. Initially, the robot pushed its active wheel against the ground (**Figure 39**B, right; **Figure 43**A, orange) by bending the last pitch joint (**Figure 43**A, red) down by 20°. Meanwhile, all the links between the last pitch joint and the pitch joint closest to the obstacle were held straight (**Figure 43**A, purple) to increase the contact force exerted by the active wheel. As the spinning active wheel pushed the robot forward, the joint closest to the obstacle pitched up (**Figure 43**A, cyan) to reduce the resistance from pushing the bump, while the pitch joints anterior to it turned torque off (**Figure 43**A, blue) to passively conform to the bump under gravity. When the wheel in the cyan box started pushing against the bump, the angle and torque status of all the pitch joints were propagated down the robot by one link. This process was manually controlled until the third pair of passive wheels from the tail contacted the obstacle (**Figure 43**B).

Then, the last joint pitched up by 75° to lift the active wheel. All the other pitch joint torques were disabled to let the robot conform to the terrain under gravity, which kept the initial shape for all the robot controllers the same in each case. The initial joint angles were recorded before motor torques were enabled. A trial started when the vertical bending controller started. A trial was ended if the robot detached the bump(s), stopped bending after a controller finished propagation, failed from motor stalling, or got stuck at the same position for over 15 seconds and after two cycles of periodic behaviors. If any part broke from impact, the trial was rejected and re-collected after replacing the part.

In cases (ii) and (iii), we pulled the tail of the robot backward (similar to a drawbar test in terramechanics (Wong, 2009)), by hanging a weight using a string through a pulley system (**Figure 43**B). The pulley system was placed 1 m away from the robot tail at the beginning of a trial, such that the direction of the backward force remains horizontal (with a variation of around 3° in each trial) despite the up-and-down displacement of the robot tail. Weights of 150 and 300 g were used to generate a small and large load (1.5 and 2.9 N, or 36% and 71% of the fore-aft frictional drag of the robot lying straight on the flat ground, respectively). The load was applied to the robot in each trial after the initial shape was recorded and torque was enabled but before the controllers started working.

In case (iv), we added a 0.3 m long acrylic plate below the robot initially (**Figure 43**C, left). The robot climbed over it and passively conformed to it under gravity before the initial joint angles were recorded. The plate was then removed after the torque was enabled (**Figure 43**C, right) but before a trial began.

In case (v), we added an additional 0.13 m high, 0.25 m long half-cylindrical bump 0.29 m in front of the main bump (**Figure 43**D). The additional obstacle was also covered with the same rubber sheet to ensure the same surface condition.

To track the positions of the wheels, we attached an ArUco marker (Garrido-Jurado et al., 2014) to each wheel. To track the terrain geometry, we attached two ArUco markers to the main obstacle, one ArUco marker to the additional obstacle, and five ArUco markers to the horizontal ground. Four synchronized cameras (N-5A100-Gm/CXP-6-1.0, Adimec, Eindhoven, The Netherlands) recorded each trial at 60 frame/s with a resolution of 2592 × 2048 pixels. To correct lens distortion, we calculated the distortion parameters of each lens using a checkerboard and the MATLAB Camera Calibrator application. Snapshots and videos recorded by the four cameras were undistorted before being used for calibration and tracking. To calibrate the cameras for 3-D reconstruction, we placed a 61 × 66 cm calibration grid made of DUPLO bricks (The Lego Group, Billund, Denmark) with BEEtag markers in the setup before experiments (Crall et al., 2015) and tracked 2-D coordinates of their centers in each camera view. The coordinates were used to calculate intrinsic and extrinsic camera parameters necessary for 3-D reconstruction using direct linear transform (Hedrick, 2008). The ArUco markers were tracked in each camera view and reconstructed for their 3-D positions and orientations using the calibrated camera parameters. Bad tracking results of the ArUco markers attached to the wheels were rejected by checking whether the angle between the marker plane and the vertical plane (**Figure 43**B, *x*-*z* plane) was larger than 30°. Then, the missing 3-D positions and orientations of each ArUco marker were filled temporally using a linear interpolation of the Lie algebra (Fu et al., 2021).

All the forces, IMU orientation, motor present angles and current, and motor goal angles and current data were recorded in ROS. To synchronize the ROS-recorded data with the camera tracking data, the active wheel was turned on and off at before and after each trial. The corresponding data frames from the two data resources were aligned for synchronization. A digital camera (D3200, Nikon, Tokyo, Japan) was used to record the experiments from the side view at 25 frame/s with a resolution of 1920 × 1080 pixels.

### 4.4.4 Data analyses

To reconstruct the 3-D terrain profile, we first obtained the geometry of the bumps and the ground, then used the tracked ArUco markers to locate the terrain relative to the cameras in each trial. To obtain the geometry, we digitized 3-D positions of 20 and 24 vertices of the main and the additional obstacles, respectively, and the corner points of all the ArUco markers attached to the terrain in all four camera views at one moment using DLTdv8 (Hedrick, 2008). The ground was then represented by a plane (**Figure 43**B, *x*-*y* plane) fitted to all the digitized points on the ground. Each obstacle was represented by a polyhedron using the digitized vertices. The terrain geometry containing one plane and two polyhedrons was then obtained. If no markers were tracked because of occlusion or bad lighting, additional corner points were manually digitized using DLTdv8.

To evaluate the performance of each controller in each case, we calculated the success rate. To evaluate how well the robot adapted to the terrain, we calculated the lift-off height of each wheel in the vertical plane (**Figure 43**B, *x*-*z* plane) in each video frame and averaged them spatiotemporally for each trial. The lift-off height was calculated as the closest distance between the wheel center and the terrain (Yoshpe, 2015) deducted by one wheel radius. We noticed that the wood sheets used to construct the obstacles were sometimes deformed, but the effect was small (2 mm for every 10 N of normal force) compared to the total values (> 4 mm on average for all passive wheels, **Figure 45**B). When calculating the lift-off height, the point on the terrain closest to each wheel center was recorded as a contact point. Thus, if multiple contact points were present for one wheel, only one was recorded. This is inaccurate for wheels contacting both the bump and the ground, but its effect on results was minor because such contacts were momentary unless the robot was stuck in this state.

To evaluate the net propulsion generated by pushing against the terrain, we summed the horizontal components (positive along the +*x*-axis in **Figure 43**B) of terrain reaction forces acting on the wheels (**Figure 39**C, red). To obtain the direction of the reaction force, we first obtained the direction of normal force (**Figure 39**C, purple vector), which pointed from the contact point (**Figure 39**C, purple point) to the tracked center of the wheel. Frictional force (**Figure 39**C, green dashed) was then calculated using the friction coefficient $\mu = 0.14$, which was perpendicular to the normal force and opposite to the velocity of this wheel center relative to the contact point. We set friction to be 0 if this velocity was smaller than 2 mm/s (i.e., with little relative motion to the terrain and likely produce unpredictable static friction). The total ground reaction force (**Figure 39**C, red) was then along the vector sum of the normal force and the friction. If the angle between the reaction force (**Figure 39**C, red) and the measured force (**Figure 39**C, blue) was larger than 85°, the force data on this wheel was rejected to avoid excessively large reaction force values. Terrain reaction forces were averaged temporally using a moving window size of 0.11s. The average net propulsion of a trial was then calculated by averaging the sum of the horizontal components of the reaction forces across all the video frames of each trial.

To evaluate the electrical energy cost by the robot to traverse the bumps using vertical bending, we first calculated the energy consumed by each pitch joint motor in each video frame by multiplying its present current, voltage, and video frame duration. We then summed the total energy cost by all the pitch motors in all the video frames in each successful trial and calculated their average.

## 4.5 Results and discussion

The tests with backward load and terrain variations revealed differences in how different controllers helped the robot generate bending patterns under unexpected challenges. These behaviors in turn affected the interaction with the terrain that influenced success rates. Analyzing the results helped us test our hypotheses: (1) Propagation of a vertical bending shape posteriorly can generate propulsion to traverse the obstacle in case (i) and accommodate additional loads in cases (ii-iii). (2) Feedforward control will struggle more under terrain variations in cases (iv-v) than in cases (iv-v). (3) Contact feedback control will achieve higher success rates than feedforward control in cases (iv-y). Below, we first describe the performance of different controllers in each test, then discuss failure modes, and finally elaborate on findings based on the observations and analyses of failures.

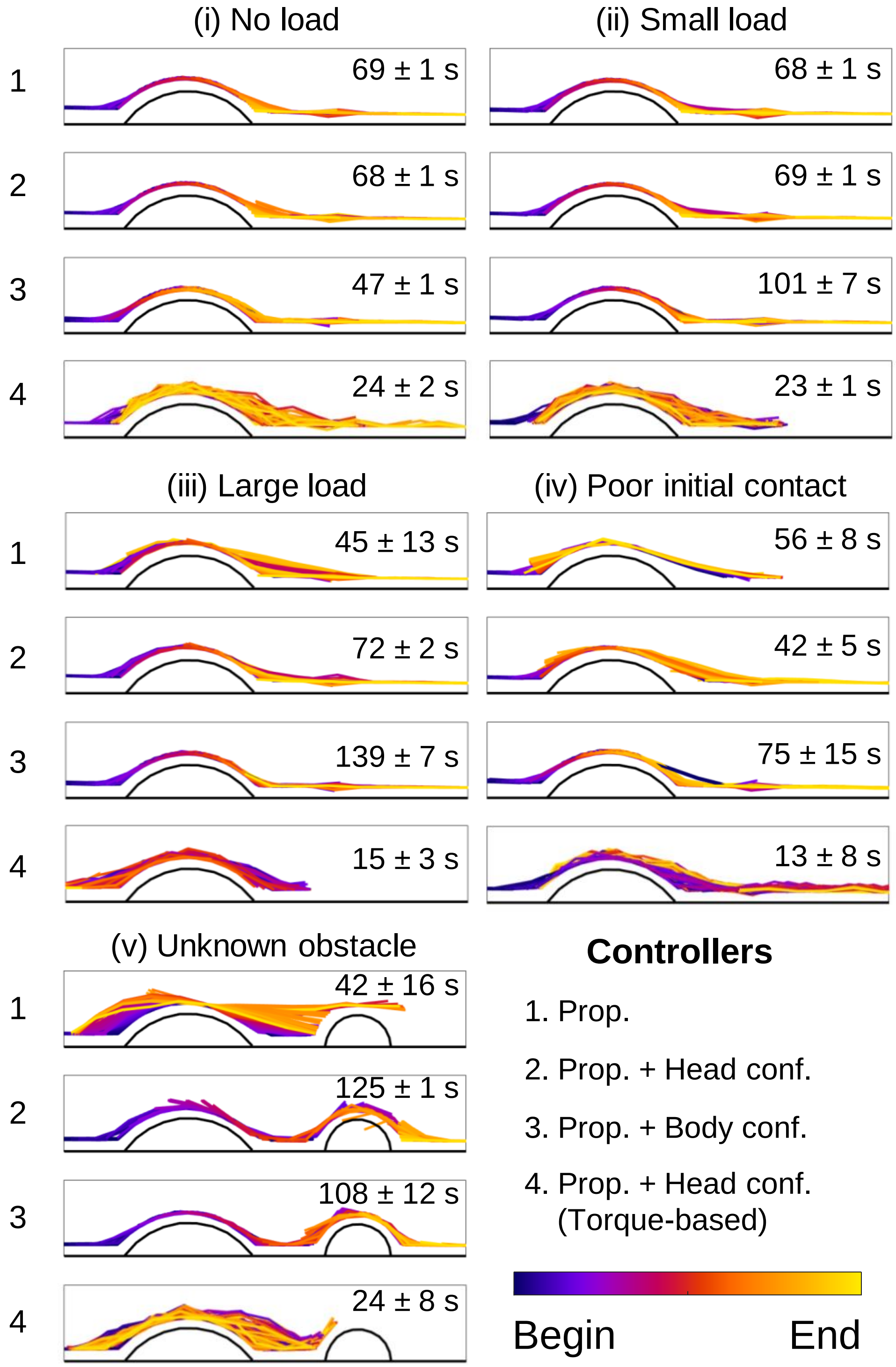

**Figure 44. Side views of robot traversing bumps.** Reconstructed robot centerlines in all 5 trials are overlaid for each controller in each case. Centerline color changes from purple to yellow with elapse of time from the beginning to the end of each trial. Black lines show terrain geometry. Numbers show mean ± s.d. values of total time of one trial, either successful or not, for each controller in each case.

## 4.5.1 Performance of different controllers in each case

### 4.5.1.1 Success rates and bending patterns

#### 4.5.1.1.1 Managing increasing load

In cases (i-iii) with only one bump and different additional loads, the robot using controller 1 succeeded in all the trials when the load was zero/small (**Figure 45**A) and maintained good contact with the terrain (**Figure 44**(i-ii), **Figure 45**B). However, it failed in 80% of the trials when the load was large (**Figure 45**A) after a large section of the body was lifted off the ground and stalled the motors (**Figure 44**(iii)).

Using controllers 2 and 3, the robot succeeded in all the trials (**Figure 45**A) and maintained good contact with the terrain for all these three cases (**Figure 44**(i-iii), **Figure 45**B).

Using controller 4, the robot failed in all but 1 trial when the load was zero (**Figure 45**A). When the load was zero or small, it initially moved forward rapidly and kept accelerating due to a lack of velocity regulation in Equation (7) (Date and Takita, 2005). However, it stopped moving forward after the last passive wheel contacted the bump, except in one trial because of the inertia from the initial acceleration. When the load was large, the robot slipped backward after the controller started (Movie 7). Controller 4 always caused large up and down oscillations at each pitch joint (**Figure 44**(i-iii), Movie 7) and a larger average lift-off height (**Figure 45**B).

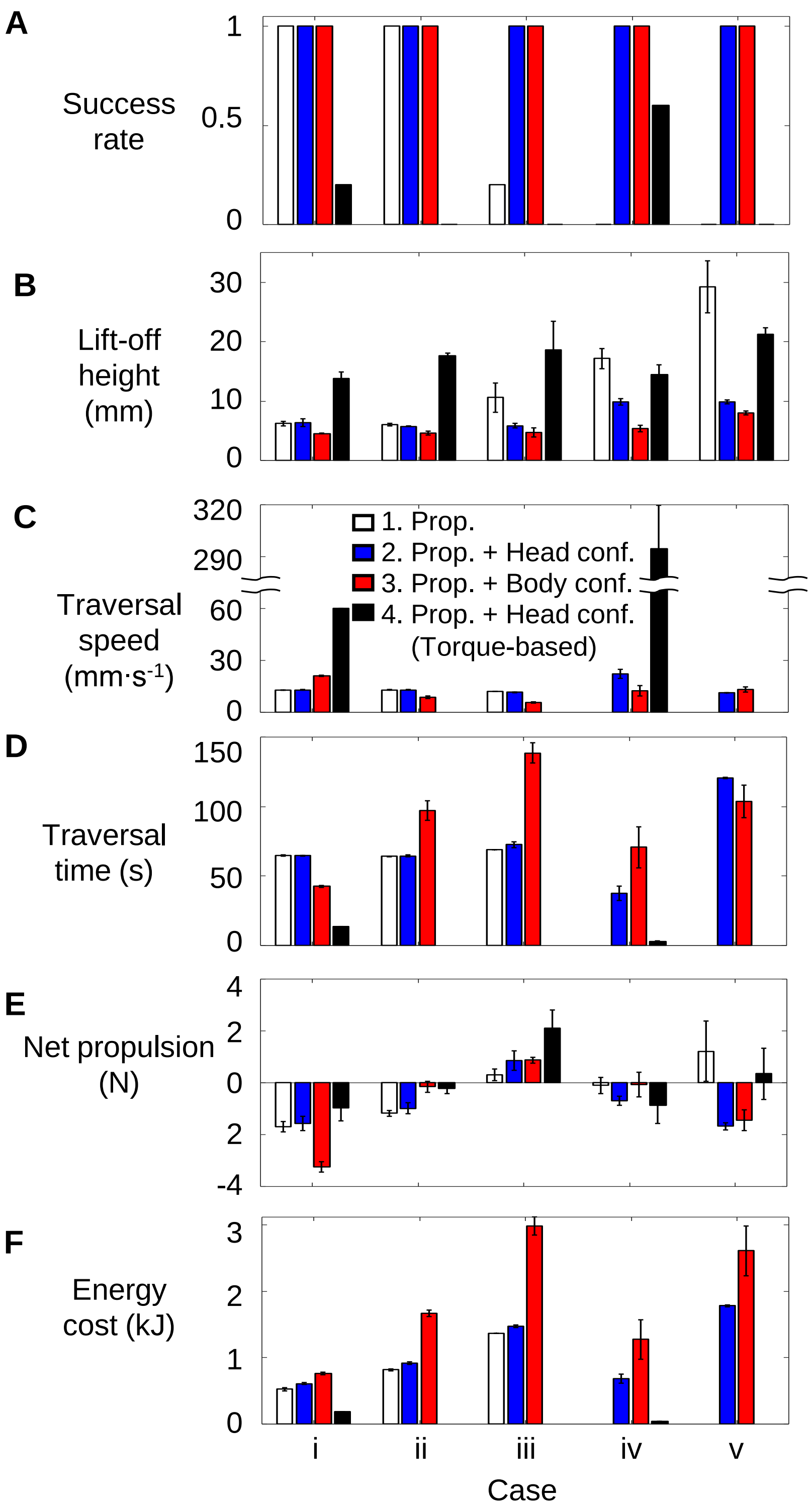

**Figure 45. Comparison of performance. (A)** Success rate. **(B)** Average lift-off height of all passive wheels. **(C)** Traversal speed of successful trials. **(D)** Traversal time of successful trials. **(E)** Net

propulsion generated. **(F)** Electrical energy cost of successful traversal. Cases include: (i-iii) with a single bump and a zero/small/large additional load, (iv) with poor initial contact, and (v) with an additional unknown bump.

#### 4.5.1.1.2 Overcoming terrain variations

In case (iv) with an initial shape that had poor contact with the terrain, the robot succeeded in all the trials using controllers 2 and 3, in 60% of the trials using controller 4, and in none of the trials using controller 1 (**Figure 45**A). When using controller 1, the robot did not regain the lost contact with the front side of the obstacle and was stuck on the obstacle (**Figure 44**(iv), **Figure 46**B), followed by motor stalling in 80% of the trials (Movie 7). Controllers 2 and 3 helped the robot regain contact with the terrain (**Figure 44**(iv)), resulting in smaller lift-off heights (**Figure 45**B). When using controller 4, the robot regained contact with the terrain immediately and gained an initial forward speed from the impact (**Figure 44**(iv), Movie 7). Afterwards, its behavior was similar to when using controller 4 in case (i).

In case (v) with an additional unknown bump in the front, the robot succeeded in all the trials using controllers 2 and 3 but failed in all the trials using controllers 1 and 4 (**Figure 45**A). Controller 1 failed from loss of contact (**Figure 44**(v), **Figure 46**A) with large lift-off heights (**Figure 45**B), followed by motor stalling similar to that in case (iv) (Movie 7). Controllers 2 and 3 helped the robot adapt to the additional unknown obstacle (**Figure 44**(v)), resulting in smaller lift-off heights (**Figure 45**B). Controller 4 failed with similar behaviors as in cases (i), (ii), and (iv) (**Figure 44**, Movie 7).

### 4.5.1.2 Traversal speed and time

The traversal speed in successful trials (**Figure 45**C) remained the same using controllers 1 and controller 2 in cases (i-iii). The traversal time in these trials were roughly the time taken to finish propagation down the entire body (64 s; **Figure 45**D). Because of the extra time spent to regain contact, traversal speed using controller 3 decreased with load in cases (i-iii). Traversal speed using controller 2 was larger in case (iv) than in case (i), because of a sudden slipping down the obstacle near the end that resulted from a different shape being propagated (Movie 7). Traversal speed using controller 4 was substantially faster in case (iv) than in case (i) because of the initial forward speed gained from the impact.

### 4.5.1.3 Net propulsion

The average net propulsion in different cases (**Figure 45**E) deducting the additional loads applied (1.5 and 2.9 N in cases (ii) and (iii), respectively) was in the range of static friction, which was approximately 0 ± 4.2 N (fore-aft kinetic friction). This was likely because the robot moved quasi-statically in most of the trials. The deviation from 0 may come from the inaccuracy of our estimation of the force direction (e.g., when friction was ignored if the local relative velocity was smaller than 2 mm/s) or from the errors of contact force measurements (**Figure 39**E).

### 4.5.1.4 Electrical energy cost

Despite higher success rates and better conformation to the terrain, the robot consumed more electrical energy to complete a successful traversal when using the feedback controllers 2 and 3 than when using the feedforward controller 1 (**Figure 45**F). For each controller, the energy also increased with additional loads in cases (ii-iii) or when additional adaptation was needed in cases (iv) and (v) compared to in case (i). The energy consumption of the robot (32~3171 J) was substantially higher than the mechanical work needed to drag a straight robot forward on a flat ground covered by the same rubber sheet (6 J for a similar displacement of 1.4 m without additional load).

### 4.5.2 Failure modes

The failure when using controller 1 can be classified into 2 modes: getting stuck at the same position (**Figure 46**A) and motor stalling (**Figure 46**B). Both modes resulted from loss of contact. In cases (iv) and (v), after losing contact with terrain surfaces steep enough (**Figure 46**A), the horizontal components (red) of the normal contact forces (yellow) on other terrain surfaces were insufficient for overcoming resistance such as friction. When a substantial part of the body was suspended in the air (**Figure 46**B), the large weight (cyan) required a large torque at the joint lifting this section (red), which often stalled the motor. This happened frequently when the large additional load applied to the tail in cases (iii) caused a small contact force on the head (yellow).

The failure when using controller 4 resulted from insufficient joint torque to propagate the shape down the body. To propagate the shape down the body, the pitch joints on the front side of the obstacle (**Figure 46**C, red) should bend as if lifting several links in front of them (**Figure 46**C, from blue solid to blue dashed). However, controller 4 was not able to generate sufficient torque to realize this motion, resulting in a deformed shape (**Figure 46**C, purple). The maximal joint torque generated by controller 4 was smaller than that of other controllers (0.52 N·m versus at least 1.16 N·m; **Figure 48**). We tried increasing the gain $K_P$ in Equation (7) to increase the torque. Despite an increased probability of successful propagation, this induced larger up and down oscillations and faster initial acceleration that damaged the robot frequently. We could not eliminate the oscillations by introducing integral and derivative terms in the feedback controller. We suspect that the poor performance came from three reasons: (1) The inaccurate torque estimation using current for our servo motors with a large gear ratio. (2) The slow sampling frequency of the joint angles (31 Hz). This had a larger effect on controller 4 which kept bending the robot with a constant torque profile in each time step than other controllers that only changed the robot shape with a small, limited increment. (3) The derivation of the controller did not consider gravity, longitudinal friction, and additional backward load (Date and Takita, 2005). In addition, the controller was designed to minimize a cost function that increases with torque. Thus, it was not expected to generate larger torques.

### 4.5.3 Shape propagation down the body in the vertical plane is propulsive if the shape matches the terrain profile

Our results supported the hypothesis that vertical bending can generate large propulsion if: (1) the robot continuously propagates a vertical bending shape down the entire body, and (2) the shape matches the current terrain profile well. The robot traversed the entire terrain successfully in all the trials when the two requirements were satisfied. These included the trials using controllers 2 and 3 in all the cases and those using controller 1 in cases (i) and (ii) (**Figure 45**A, B, **Figure 49**). The propulsion was large enough to overcome the large additional load when using controllers 2 and 3 in case (iii). This was achieved by creating an asymmetry in the body-terrain interaction: as the shape was propagated posteriorly, the body continuously pushed against the vertical push points for propulsion while detaching the back side of terrain surfaces to reduce resistance.

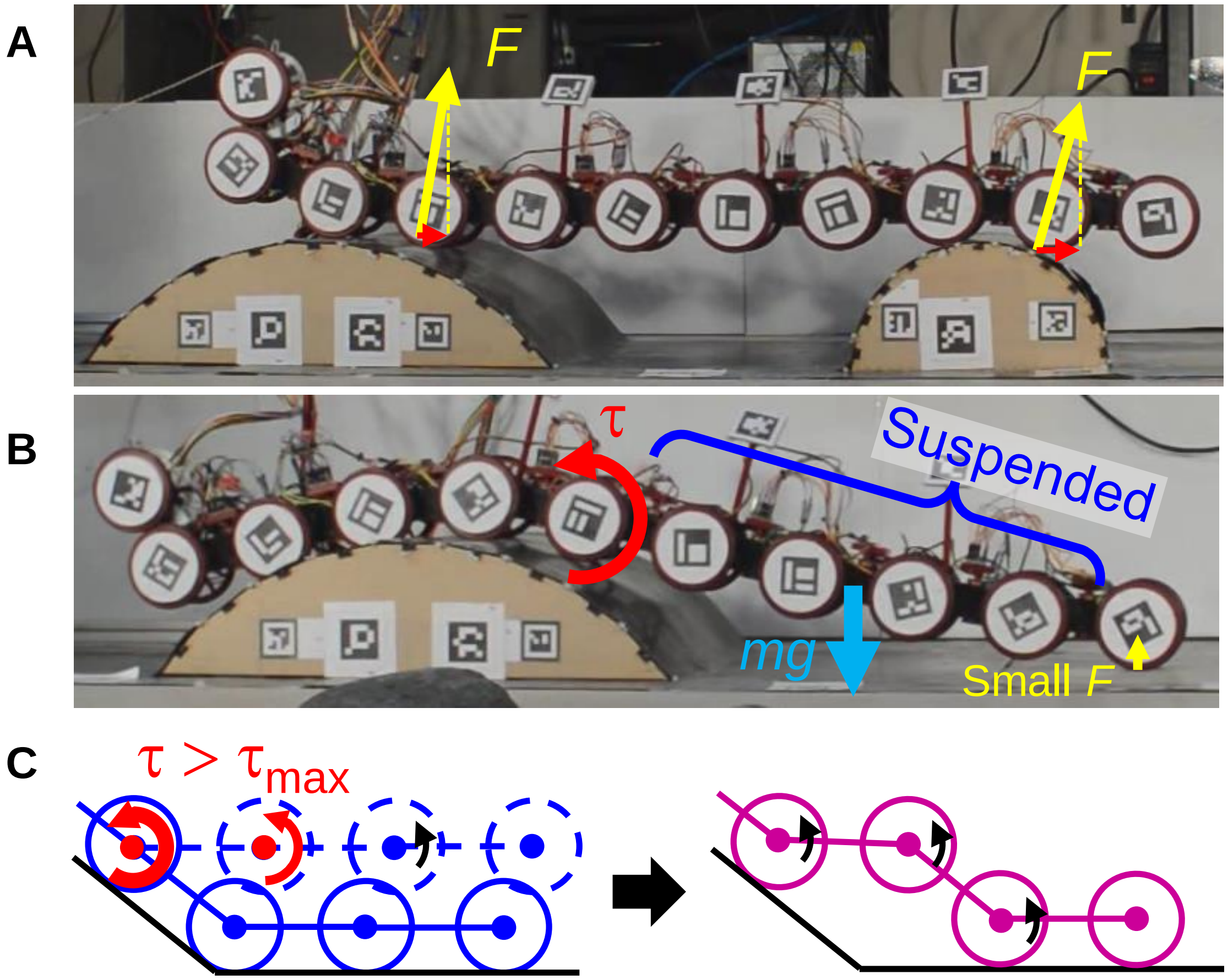

**Figure 46. Typical failure modes. (A)** Insufficient propulsion after loss of contact. Limited propulsion can be generated from horizontal components (red) of normal contact forces (yellow). **(B)** Motor stalling after loss of contact. A large torque (red) is required to balance weight of suspending body section (cyan), especially when limited support is provided by other end (yellow). **(C)** Insufficient joint torque to propagate shape when using controller 4. Blue solid shows present shape of robot, blue dashed shows expected shape after propagation, and purple shows real shape after propagation. Red and black show expected torque beyond and within controller capacity, respectively.

The robot stopped moving forward whether or not the current shape was contacting vertical push points once the shape propagation down the body stopped. For example, controller 4 failed to generate sufficient torque to propagate a shape (**Figure 46**C).

If the shape did not match the terrain profile well, the robot also failed easily, whether or not a propagation was completed. (1) When losing contact with the steep front sides of terrain surfaces, the robot could not generate large propulsion from the horizontal components of contact forces on the other surfaces (**Figure 46**A). Controller 1 failed in cases (iv) and (v) because of this reason. (2) When a long section of the robot lost contact, the motors easily stalled from the large torques required to lift this section (**Figure 46**B). Examples included the trials using controller 1 in cases (iii) and (v).

### 4.5.4 Feedforward propagation struggles with loss of contact while contact feedback can help by improving contact

The decrease of success rates of controller 1 in cases (iii-v) (**Figure 45**A) demonstrated that feedforward vertical bending control cannot handle unexpected changes such as loss of contact, which is often followed by loss of propulsion or motor stalling.

Had the controller been able to sense the loss and adjust the shape in time to regain contact, its success rate could increase dramatically. Indeed, this was realized by controllers 2 and 3 by introducing contact feedback as demonstrated in cases (iii-v) (**Figure 45**A, B).

However, feedback control did not always improve performance. Controller 4 failed in most of the trials because of a reduction of the asymmetry in body-terrain interaction. While the asymmetry can be provided by proper shape propagation, controller 4 reduced the propulsion due to the limited torque that compromised the bending shape.

### 4.5.5 Comparison between propulsion generation using propagation of a lateral and a vertical bending shape

Most of the previous propulsion generation strategies used propagation of a lateral bending shape (Jayne, 2020; Kano and Ishiguro, 2020; Sanfilippo et al., 2017). Similar to them, propagation of a vertical bending shape generates propulsion by pushing against obstacles and generating contact force components along the movement direction. Thus, the angle between the normal directions of such a contact surface and movement direction has to be small such that normal pushing forces can be effectively converted to propulsion. The maximum angle depends on the resistance the locomotor needs to overcome during slithering, such as friction (Zhang et al., 2022) or external load (cases (ii, iii)). If no such suitably oriented surfaces are available, other strategies that rely on alternating static contact points are needed, such as concertina, rectilinear, and sidewinding (Jayne, 2020).

Aside from this similarity, propagation of a vertical bending shape has two major differences in propulsion generation with propagation of a lateral bending shape.

(1) Difference in push points. To generate propulsion, propagation of a lateral bending shape relies on push points lateral to the body, such as walls, pillars, or trunks. In contrast, propagation of a vertical bending shape relies on push points below or above the body, such as uneven ground and horizontal branches. While lateral push points often appear at both sides of the body, vertical push points usually only occur below the body because gravity always pulls the body downward against the ground.

Different environments have different availability of lateral or vertical push points, which requires choosing strategies accordingly. When on large boulders or over horizontal tree branches that lack vertical extrusions, the body may need to slip over a large distance to reach one lateral push point available for lateral bending to push against, which creates difficulty for continuous propulsion generation and for maintaining heading. When on a flat, rigid ground with plant stems, the body can barely find a vertical push point for vertical bending to push against. But often, both push points are available and allow a combination of both bending (Fu et al., 2022), such as on rocky beaches or forest floors.

(2) Difference in force balance between environmental forces. Because of a difference in the primary bending plane, snakes (and snake robots) need to balance different external forces when using vertical and lateral bending strategies.

Lateral bending strategies mainly need to balance lateral forces in the horizontal plane, including contact forces from lateral push points and lateral friction from the ground. Because lateral push points often exist on both sides of the body, a snake or snake robot can adjust leftward and rightward contact forces to counteract friction and avoid yawing or lateral slipping (Gans, 1962). However, lateral bending can easily push similarly hard against push points on both sides of the body and cause overloading or large friction. This has led to many studies on how sensory feedback can help resolve such a jamming issue (Liljebäck et al., 2011; Liljebäck et al., 2014b; Wang et al., 2020a). If push points are only on one side of the body, it is instead challenging to maintain heading by balancing pushing forces and friction that can vary with multiple factors such as velocity and substrate properties. Although snakes can balance it well (Gasc et al., 1989; Gray and Lissmann, 1950; Moon and Gans, 1998), the principles are not well studied and not applied to any robots yet.

Vertical bending mainly needs to balance forces in the vertical direction. It is less critical to consider jamming along this direction because vertical push points usually only occur below the body, except in very narrow pipes (Sawabe et al., 2019) or in subterranean environments (Ozkan-Aydin et al., 2022). However, gravity needs to be considered. It pulls on every part of the body and can be used to push the body against the ground for contact forces. However, it can also overload actuators easily if a long body section is suspended in the air (**Figure 46**B). It is impossible to modulate the magnitude of weight similar to increasing contact forces by changing the bending patterns (Kano and Ishiguro, 2020). In addition, failure to balance weight with contact forces when supporting points are limited can cause loss of stability and falling which can severely damage the robot (Fu and Li, 2020).

## 4.6 Summary

By comparing a feedforward and three feedback controllers that generate distinct bending patterns and body-terrain interactions under different perturbations on a robophysical model, we conclude that: (1) Vertical bending can generate propulsion to overcome friction and other backward resistance by propagating a vertical bending shape to push against height variations if the shape conforms to the terrain. (2) Feedforward shape propagation fails from loss of contact easily under perturbations such as unknown terrain geometry or excessive external forces that cause slipping or lifting. Contact feedback control can help maintain propulsion by sensing the change and adapting bending shapes to regain contact while preserving the propagation. (3) Unlike propulsion generation using lateral bending, a snake robot using vertical bending for propulsion needs to consider the differences in push point positions and external forces. Body weight can help maintain contact with the environment but may also overload actuators.

Our study is an important step toward understanding how to slither in the 3-D world and will help snake robots better traverse a diversity of complex 3-D terrain. Propagation of a vertical bending shape can provide a new method to generate propulsion by pushing against vertical push points on uneven terrains. Contact feedback control further enhances the robustness of this strategy against unexpected loss of contact. The expansion of push points will allow more redundancy to adjust distribution of contact forces along the body. This will help improve snakes' and snake robots' stability, maneuverability, and efficiency and contribute to their superior versatility in various 3-D environments.

## 4.7 Limitations and future work

The performance of our robot and the controllers can be further improved. First, many control parameters were selected conservatively to protect the robot and resulted in slow speeds. Secondly, the robot can likely adjust its bending shape during propagation to push harder against existing push points for larger propulsion than simply conforming to the terrain (Kano et al., 2014). Thirdly, the accuracy of force sensing can be improved. FSRs can be better calibrated using a model that considers the hysteresis behavior of the sensors (**Figure 39**E). Alternatively, sensors like load cells with better linearity can be used (Liljebäck et al., 2014a). The torque sensing and control accuracy of joint actuators also need to be improved by using back-drivable motors or introducing additional torque sensors, which can improve the performance of torque-based control (Date and Takita, 2007; Rollinson et al., 2014) and better facilitate contact sensing (Ordonez et al., 2013; Wang et al., 2020b).

Difference in mechanical structures can also contribute to the performance gap between robots and snakes, which remain to be investigated. A snake has a highly flexible body with hundreds of vertebrae (Jayne, 1982), a smooth body surface, and more distributed mechanoreceptors over the body (Von Düring and Miller, 1979). This reduces the chance of getting caught on terrain asperities and allows continuous contact force sensing. Snake robots with a smoother body rather than wheels (Borenstein and Hansen, 2007; Haraguchi et al., 2005; Liljebäck et al., 2014a; Ramesh et al., 2022; Takemori et al., 2018b; Wright et al., 2012) and with more force sensors distributed over the body (Liljebäck et al., 2014a; Ramesh et al., 2022) likely share the same advantages in complex environments. In addition, unlike snake robots that typically actuate each joint with a dedicated motor, snakes have muscle-tendon groups that often span dozens of vertebrae. The effects of such musculoskeletal structure on body bending remain to be investigated and may inspire the design of new robot actuation mechanisms (Kojouharov et al., 2023; Schiebel et al., 2020c; Wang et al., 2023).

It remains to be understood how a snake processes such an excessive amount of sensory information efficiently and controls such a large degree of freedom agilely, especially during fast motion or when the substrate (such as thin branches) deforms fast unexpectedly. In this study, the control frequency was significantly limited by the ability of the central controller to collect information from sensors and send commands to actuators. A combination of centralized and decentralized control can likely save communication bandwidth and computational cost for faster response speed (Holmes et al., 2006). This is because decentralized control can process detailed sensory feedback information to generate control signals locally, which can be applied to generate shape propagation and local reflexes (Kano and Ishiguro, 2020; Thandiackal et al., 2021).

Although we used five different cases to simulate different environmental conditions, the natural environment can generate even more complex challenges for vertical bending strategies. For example, it remains to be investigated how a controller can handle dynamic environmental changes such as collapse of deformable substrate like loose sand and compliant tree branches. A robot should also avoid excessive adaptation that can cause falling, such as when crawling on a horizontal ladder (**Figure 37**A) or when bridging a deep gap.

More broadly, we need to better understand the principles of how to combine vertical, lateral, or obliquely-oriented bending to fully exploit available environmental surfaces for propulsion stably (Fu et al., 2022). For example, how to choose the proper bending in three dimensions in response to different terrain features? How to coordinate and transition between different bending strategies at different body sections? How can a snake or snake robot plan a path with sufficient push points by combining mechanical sensing and vision?

Aside from further improving robots and using them as amenable physical models to test hypotheses about how snakes generate propulsion using vertical body bending, it is also possible to design animal experiments to deepen the understanding despite the limited knowledge about their neuromechanical structures. For example, by observing how snakes react to variation of body-terrain contact with reduced tatitle stimulus response after anesthesia (Strahl-Heldreth et al., 2019), we can better confirm the role of contact feedback control and reveal other potential mechanisms that can facilitate adaptation without feedback control (Kojouharov et al., 2023; Schiebel et al., 2020c; Wang et al., 2023).

## 4.8 Appendix

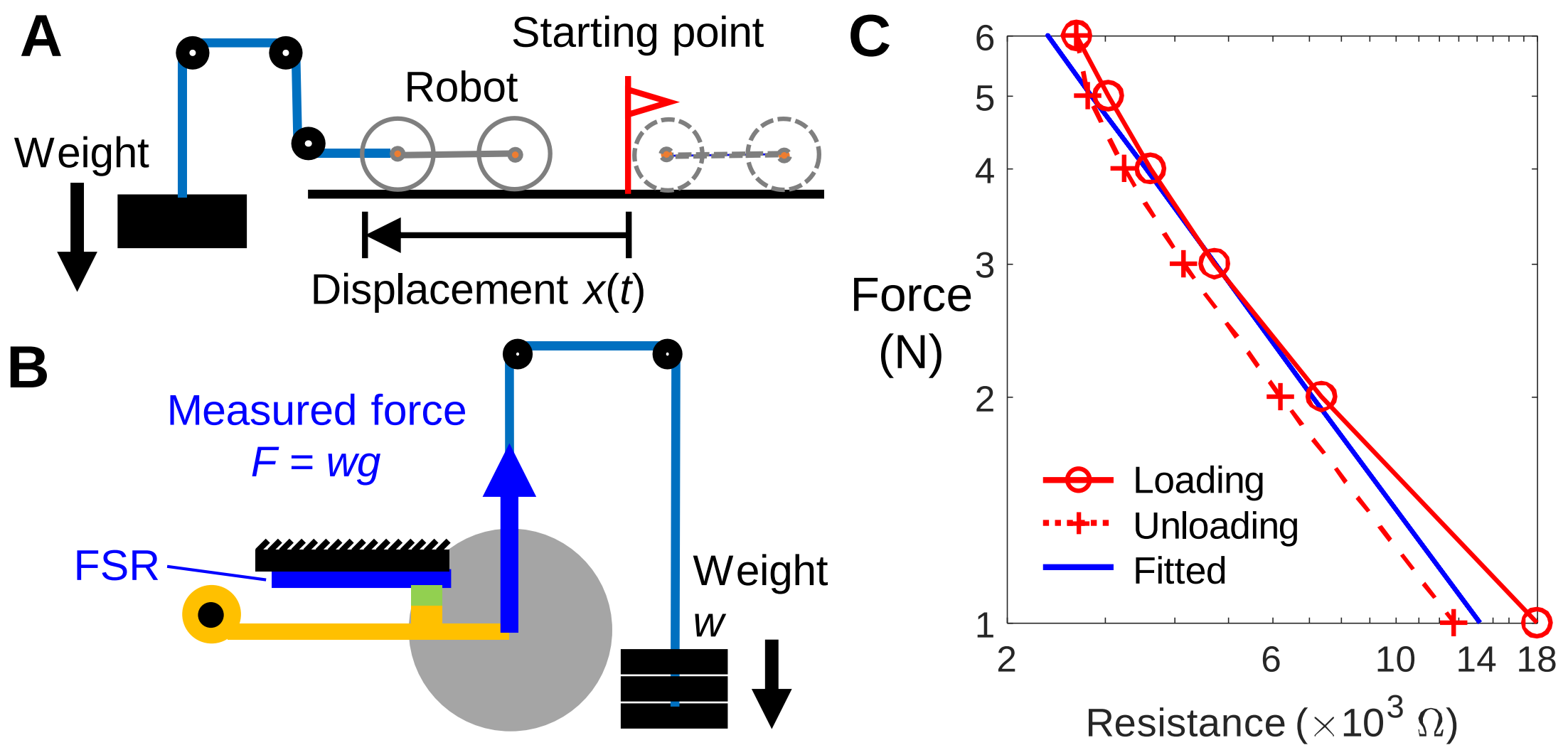

**Figure 47. Additional experimental setup. (A)** Setup to measure fore-aft friction coefficient. Robot is initially placed straight on a flat ground covered by same rubber sheets used in experiments. A weight is then added and drag robot longitudinally by a weight using a string through a pulley system. We track ArUco markers attached to robot and obtain robot displacement *x* as a function of time *t*. We then calculate acceleration of robot *a* by fitting a quadratic function to robot displacement $x(t) = 0.5at^2$. Friction coefficient is calculated as $\mu = (m_2g - (m_1+m_2)a)/(m_1g)$, where $m_1$ is mass of robot, $m_2$ is mass of weight, *g* is local gravitational acceleration (9.81514 m/s$^2$). **(B)** Setup to calibrate force sensing resistor. Wheel (gray) and its wheel arm (dark yellow) is pushed against force sensing resistor (green) by a slotted weight set *w* through a pulley system. Weight increases from 0 to 5.88 N then decreased to 0 with a 0.98 N increment. Values of force $F = wg$ and resistance of force sensing resistor *R* are collected 5 seconds after a change of weight. **(C)** Force as a function of resistance. Red solid and dashed show forces applied during loading and unloading in calibration, respectively. Blue shows line function $\log(F) = k_{FSR} \cdot \log(R) + \log(g) + b_{FSR}$ fitted to collected values of $\log(F)$ and $\log(R)$. Measurements when $F = 0$ are not used for fitting because force is too small to actuate sensor.

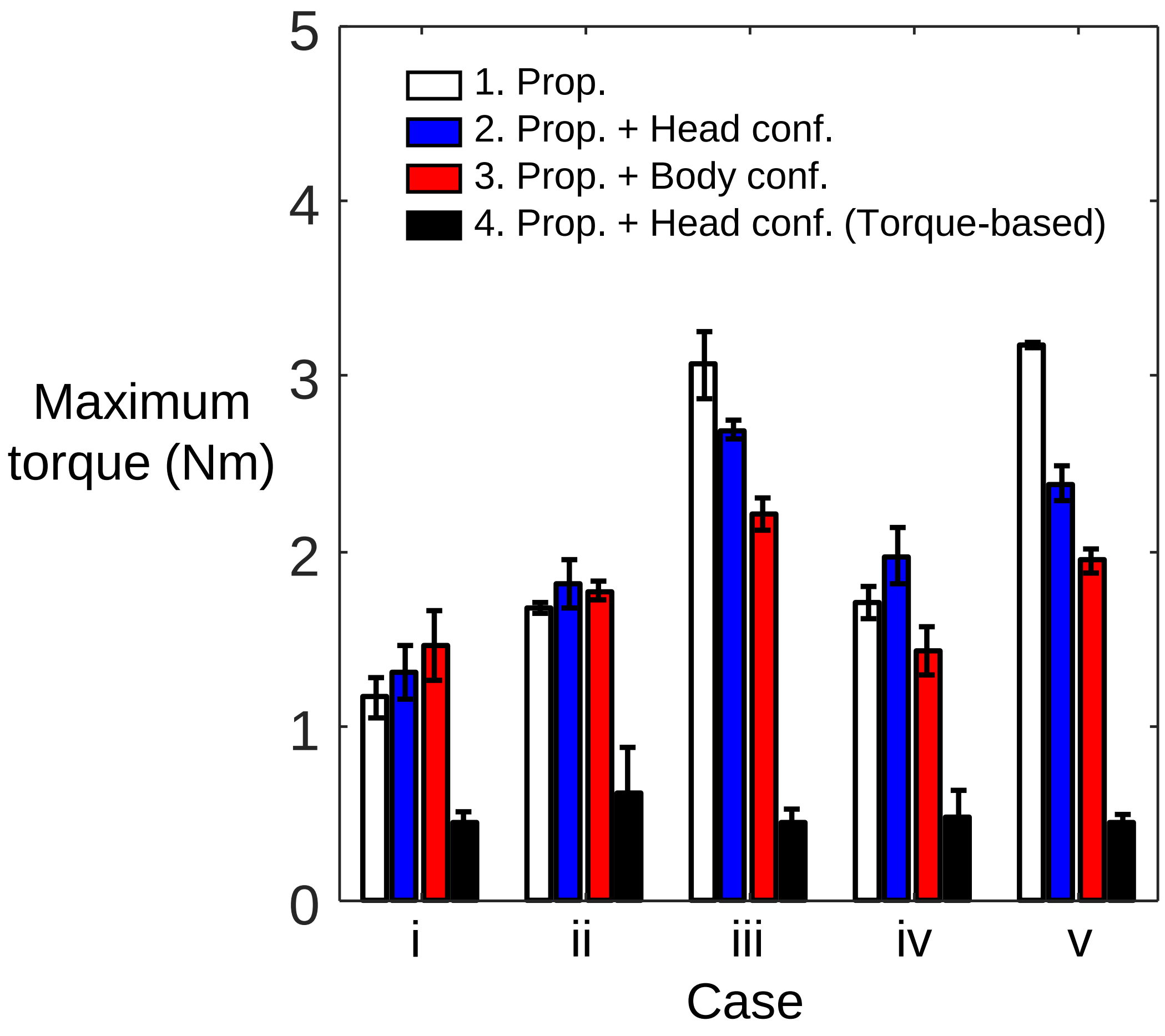

**Figure 48. Maximum joint torque generated in each trial.** Test cases include: (i) with only a single obstacle, (ii) with a small additional load, (iii) with a large additional load, (iv) with an initial shape that has poor contact with terrain, and (v) with another unknown obstacle in front.

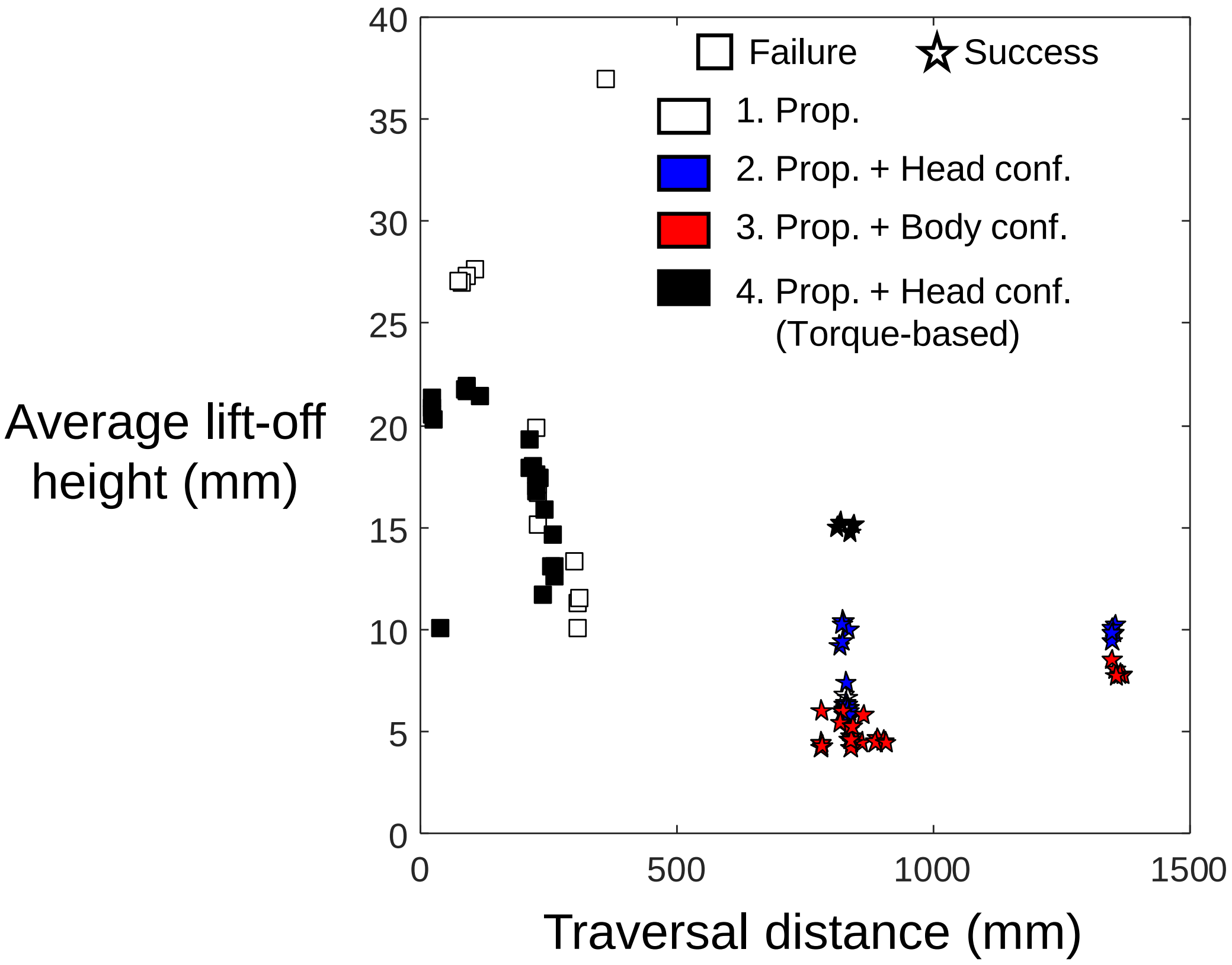

**Figure 49. Average lift-off height compared with traversal distance.**

# Chapter 5

# Conclusions

## 5.1 General remarks

In this dissertation research, we combined biological and robotic studies to investigate how a snake or a snake robot can traverse complex 3-D terrain versatilely and stably using 3-D body bending (**Figure 12**). By focusing on body-terrain contact that connects different mechanisms (such as body compliance and 3-D body bending control) with system performance (such as stability and propulsion), we discovered several principles of how to use vertical body bending for robust propulsion generation while reducing the instability induced by it. Our findings shed light on how snakes traverse complex 3-D environments versatilely and efficiently using 3-D body bending. The mechanisms discovered and the idea of improving performance by modulating physical interactions also provided inspiration to design snake robots. For example, by integrating passive mechanisms and body bending control with sensory feedback, the robots can potentially traverse a diversity of 3-D terrain beyond flat surfaces versatilely and efficiently with minimum human-in-the-loop control.

## 5.2 Specific accomplishments

### 5.2.1 Challenges of using vertical body bending when traversing complex, 3-D terrain

- Confirmed that roll instability increases with the amplitude of vertical body bending when traversing a large smooth step using a partitioned gait that combines lateral body undulation and vertical cantilevering (Chapter 2, **Figure 12** left red arrow).
- Confirmed that feedforward body bending easily loses contact with the terrain under perturbations such as dynamic movement, terrain geometry variation, or excessive external forces that cause slipping or lifting (Chapters 2 and 4).
- Discovered that body weight can overload actuators when a long section of the body loses contact with the terrain (Chapter 4).

### 5.2.2 Mechanisms that can improve system performance in 3-D terrain by modulating body-terrain contact

- Discovered that body compliance can improve surface contact statistically and reduce roll instability induced by large vertical body bending (Chapter 2, **Figure 12** purple arrows).
- Confirmed that vertical bending is used by corn snakes to push terrain as frequently as lateral body bending during traversal of uneven terrain (**Figure 12** top red and blue arrows). The combination of lateral and vertical bending in 3-D may drastically expand the range of natural surfaces available for propulsion generation (**Figure 12** bottom red and blue arrows) in all but the smoothest environments, by allowing each part of the bending body to push against its nearby terrain surfaces (Chapter 3).

- Confirmed that vertical bending can generate large propulsion to accommodate friction and other backward resistance by propagating a vertical bending shape to push against height variations if the shape conforms to the terrain (Chapter 4, **Figure 12** right red arrows).
- Discovered that contact feedback control can help vertical body bending to maintain propulsion generation by modulating bending patterns to regain lost contact (Chapter 4, **Figure 12** magenta arrows).

## 5.3 Future directions

While this dissertation research reveals several principles that potentially contribute to the versatile and efficient locomotion of snakes and snake robots in complex 3-D environments, much more work is needed to better understand how a snake achieves this by integrating multiple components of its complex neuromechanical system and how we can further shorten the performance gap between snake robots and snakes. Below are some of the potential directions that can help advance our understanding.

### 5.3.1 Measurement of 3-D terrain reaction forces

In this dissertation research, we used 3-D geometry (Chapters 2-4) and 1-D force measurement (Chapter 4) to quantify body-terrain interaction. However, to fully understand the physical interaction between a snake or a snake robot and the terrain, it is essential to measure 3-D terrain reaction forces distributed along the body. This is challenging at present because current commercial high-fidelity 3-D force sensors are expensive (Han et al., 2021; Jurestovsky et al., 2021), whereas low-cost, customizable force sensors typically only measure 1-D forces, have low precision, or are large (Fu and Li, 2021; Kalantari et al., 2012; Liljebäck et al., 2012b; Shimojo et al., 2007; Sundaram et al., 2019). Development of precise, compliant, small-sized, and affordable 3-D force sensors that can cover arbitrary 3-D surfaces (Zhou et al., 2022) and can be sampled at a reasonably high frequency may enable such measurement in the future. The force measurement can inform not only the dynamics of 3-D body movement, but also the correlation between body bending and terrain reaction forces when compared with simultaneous measurement of body kinematics, or sensory feedback control principles when analyzed together with neuromotor signals.

### 5.3.2 Measurement of neuromotor activities of moving snakes

In addition to measuring the kinematics of 3-D body bending in Chapter 2, measurements of muscle activities and neural signals are needed to understand how snakes control such body bending. The measurement can help us further confirm whether snakes generate propulsion from active propagation of muscular activation (Jayne, 1988; Moon and Gans, 1998) or also from skin and scale movement not reflected by the kinematics of the backbone (Marvi and Hu, 2012; Newman and Jayne, 2018). The understanding can also reveal potential mechanisms that can improve energy efficiency or simplify body motion control, such as passive body bending under gravity (Jorgensen and Jayne, 2017) or bilateral muscle activation that can passively conform to obstacles with feedforward control (Kojouharov et al., 2023; Schiebel et al., 2020c; Wang et al., 2023).

### 5.3.3 Comparison across snake species and limbless clades

This dissertation research and the previous studies that are directly related (Gart et al., 2019; Jurestovsky et al., 2021) used two generalist terrestrial species, corn snakes [*Pantherophis guttatus* (Utiger et al., 2002)] and variable kingsnakes [*Lampropeltis mexicana thayeri* (Garman, 1883)]. Although many other snake species or other limbless clades such as nematodes (Dorgan, 2015; Kwon et al., 2013) and elongate fish (Ekeberg et al., 1995; Gidmark et al., 2011; Tatom-Naecker and Westneat, 2018) also use 3-D body bending for locomotion, it is unclear whether different species share the same principles of controlling the body bending to interact with the environment. The bending patterns can depend on specific neuromechanics such as body bending capacity (Jurestovsky et al., 2020; Kelley et al., 1997), skin and scale deformation (Marvi and Hu, 2012; Newman and Jayne, 2018), muscular force capability (Astley, 2020b; Long Jr, 1998), and sensing and neural control capacity (Sulston et al., 1983; White et al., 1986). Environmental factors can also result in differences in locomotion strategies, such as push point density (Majmudar et al., 2012), friction (Dorgan et al., 2013), substrate deformability (Gu et al., 2017), and terrain heterogeneity (Mitchell and Soga, 2005).

### 5.3.4 Integration of passive mechanics, decentralized reflexes, and centralized modulation

The elongate body provides a snake or a snake robot with the benefits of extra flexibility and many contact points with the terrain. Despite the additional redundancy to improve robustness, maneuverability, and efficiency, the additional complexity also poses significant challenges to control. An animal deals with different control challenges using neuromechanical components with different degrees of centralization and different levels of feedback (Holmes et al., 2006). For example, passive mechanics such as body compliance in Chapter 2 can modulate body-terrain interaction robustly and fast without adding computational cost, but need dedicated mechanical structures and are less variable. A centralized feedback controller like the ones discussed in Chapter 4 can produce more diverse reflective behaviors coordinated along the body, but is computationally expensive and has limited bandwidth to communicate with sensors and actuators. Decentralized controllers such as coupled central pattern generators and local reflexes can produce adaptive periodic movement more variable than using passive mechanics (Kano and Ishiguro, 2020; Thandiackal et al., 2021), but have latency in coordination between distant body segments. A hierarchical structure integrating these components can likely contribute to successful snake and snake robot locomotion in complex environments by capitalizing on the strength of each.

## 5.4 Final thoughts

It was fun working on snakes, experts in moving through almost any environment using a seemingly simple limbless body. The more I worked on the project, the more mysteries I realized that need to be understood in these natural creatures. I am lucky to have had the opportunity to learn how to solve the mysteries using an integrative approach and how to interpret the findings for broader applications in science and engineering. I look forward to hearing about relevant work and working along this path together with the fantastic people I have met and will meet in the future.

# References

## Appendices

### List of supplementary movies

The movies are available through the YouTube links.

Movie 1. Mechanical design of snake robot. Link: https://youtu.be/3K89FT0Zhws.

Movie 2. Comparison of large step traversal between rigid and compliant body snake robot. Link: https://youtu.be/VmLP_0lPtHE.

Movie 3. Comparison of large step traversal between rigid and compliant body snake robot. Link: https://youtu.be/9Rj5VSo5huA.

Movie 4. Adverse events of snake robot traversing a large step. Link: https://youtu.be/xUkAD268k3w.

Movie 5. A representative trial of a snake utilizing vertical bending to traverse an uneven terrain. Link: https://youtu.be/U8UXmh6MfYQ.

Movie 6. A representative trial showing little transverse motion with reconstructed midline overlaid at different time instances. Link: https://youtu.be/98PT8YKAFvY.

Movie 7. Representative trials of robot using different controllers in different cases. Link: https://youtu.be/GWcGQ2FP5I8.

## Unpublished results

Below are some of the preliminary results that were not included in the previous chapters or other publications, but can be potentially useful in future research or inspire new ideas that contribute to science and technology.

Kirigami scales with anisotropic friction

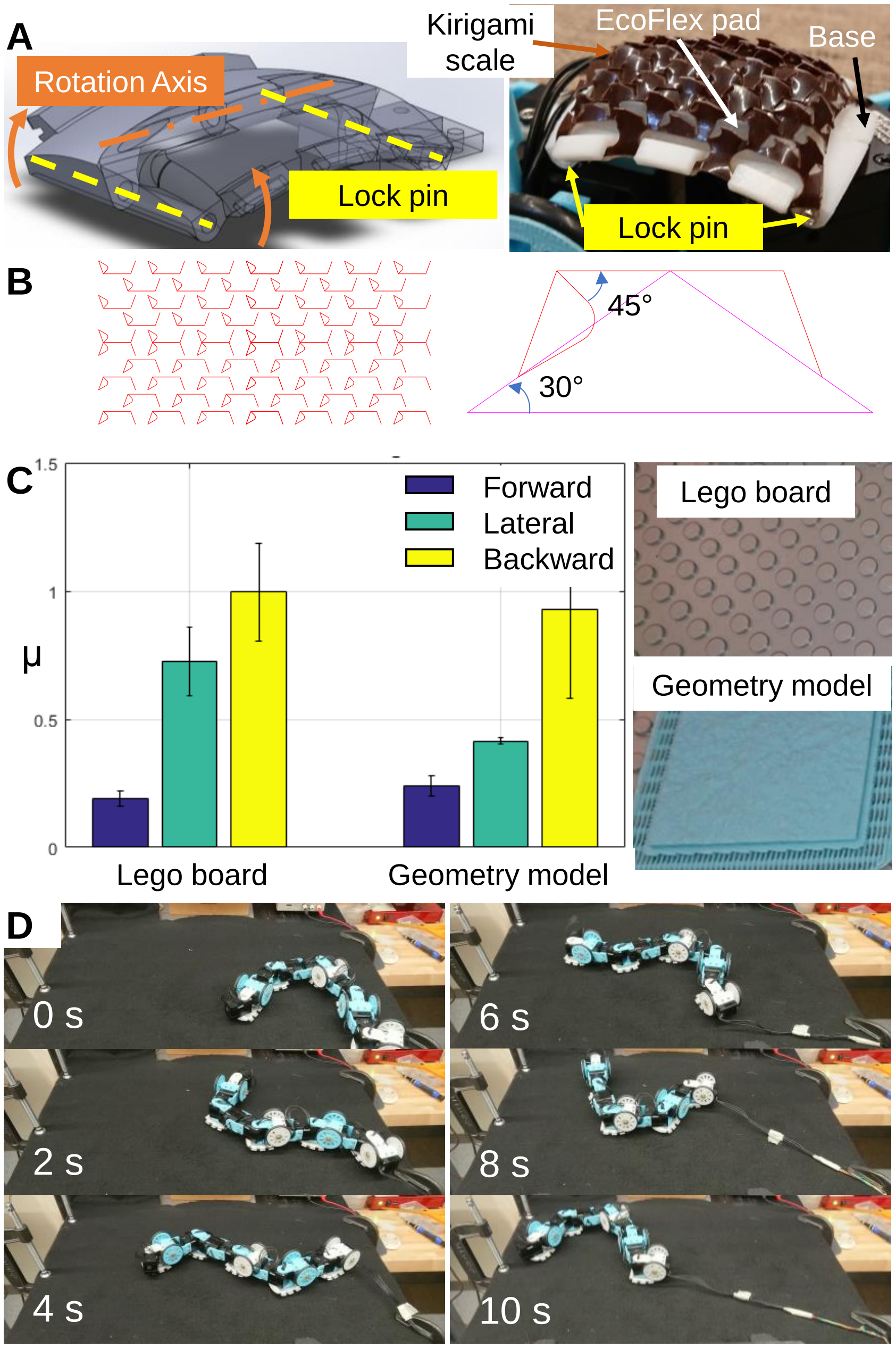

**Figure 50. Kirigami scales with anisotropic friction. (A)** Assembly of a module with a 3-D printed curved base (top), a stretched Kirigami scale (brown), and an EcoFlex pad in between. Kirigami scale is made by laser cutting a 0.01-inch-thick polyester sheet (McMaster-Carr, Elmhurst, IL, USA). **(B)** Laser cut pattern of kirigami scale inspired by (Rafsanjani et al., 2018). Red shows laser cuts

and magenta shows design auxiliary line. **(C)** Measured friction coefficient μ of kirigami scale sliding against a Lego pad (left) and a 3-D printed rough terrain model (right) using slope tests (Hu et al., 2009). **(D)** Snapshots of a robot equipped with kirigami scale moving forward on black felt using lateral undulation. A scale module is installed to each link between every two adjacent yaw joints. This work was later improved by Xiangyu Peng but his results were not included here.

# Vita

Qiyuan Fu was born in Taihe, China on March 7, 1996 (January 18 of the Chinese Calendar) to Hailin Fu and Xiaoli Guo. He grew up in Wuxi, China and went to Beijing, China to attend Tsinghua University in 2013. He received a Bachelor of Engineering degree in Mechanical Engineering there in 2017. He then went to Baltimore, MD, USA in August 2017 to pursue a doctoral degree at Johns Hopkins University under the supervision of Prof. Chen Li. During this program, he received a Master of Science in Engineering degree in Robotics in 2020. His work has been published in journals such as *Royal Society Open Science*, *Integrative and Comparative Biology*, *Journal of Experimental Biology*, and *Bioinspiration and Biomimetics*.